\documentclass[11pt]{article}

\usepackage{amsmath}
\usepackage{amssymb}
\usepackage{latexsym}
\usepackage{xspace}
\usepackage{epsfig}

\newcommand{\be}{\begin{itemize}}
\newcommand{\ee}{\end{itemize}}
\newcommand{\bn}{\begin{enumerate}}
\newcommand{\en}{\end{enumerate}}

\newcommand{\bp}{\begin{proposition}}
\newcommand{\ep}{\end{proposition}}
\newcommand{\bl}{\begin{lemma}}
\newcommand{\el}{\end{lemma}}
\newcommand{\bco}{\begin{corollary}}
\newcommand{\eco}{\end{corollary}}
\newcommand{\bt}{\begin{theorem}}
\newcommand{\et}{\end{theorem}}
\newcommand{\bpr}{\begin{proof}}
\newcommand{\epr}{\end{proof}}
\newcommand{\bd}{\begin{definition}}
\newcommand{\ed}{\end{definition}}

\newcommand{\beqn}{\begin{centeqn}}
\newcommand{\eeqn}{\end{centeqn}}

\newcommand{\bleqn}[1]{\begin{centlabeqn}{#1}}
\newcommand{\eleqn}{\end{centlabeqn}}

\newcommand{\bc}{\begin{center}}
\newcommand{\ec}{\end{center}}

\renewcommand{\ss}{\smallskip}

\newcommand{\bfg}{\begin{figure}}
\newcommand{\efg}{\end{figure}}

\newenvironment{centeqn}	
   {{\ms\\ \hspace*{\fill}}} 
   {\hspace*{\fill}\ms\\}

\newsavebox{\EqnLabel}
\newenvironment{centlabeqn}[1]		
   {\sbox{\EqnLabel}{#1}
    {\ss\\ \hspace*{\fill}}
   } 
   {\hfill{\makebox[0in][r]{\usebox{\EqnLabel}}}\ms\\}

\newenvironment{centeqn-nbsp} 
   {{\ms\\ \hspace*{\fill}}}
   {\hspace*{\fill}}

\newcommand{\intrdef}{\emph}	
\newcommand{\intr}{\emph}	

\newcommand{\empi}[1]{\textit{#1\/}}

\newcommand{\INT}{\bigcap}

\newcommand{\UN}{\bigcup}

\newcommand{\choice}{\mbox{$[\hspace*{-0.5pt}]$}}
\newcommand{\df}{\mbox{$\:\stackrel{\rm df}{=\!\!=}\:$}}

\newcommand{\ex}{\exists}
\newcommand{\fa}{\forall}
\newcommand{\false}{\mathit{false}}

\newcommand{\imp}{\Rightarrow}		
\newcommand{\ind}{\hspace*{3.0em}}
\newcommand{\ints}{\cap}
\renewcommand{\l}{\ell}

\newcommand{\pl}{\!\parallel\!}

\newcommand{\sat}{\models}

\newcommand{\sub}{\subseteq}
\newcommand{\true}{\mathit{true}}
\newcommand{\un}{\cup}
\newcommand{\union}{\cup}
\newcommand{\up}{\raisebox{0.2ex}{$\upharpoonright$}}

\newcommand{\flag}{\mathit{flag}}

\newcommand{\lla}[2]{\mbox{$\, \stackrel{#1}{\longrightarrow}_{#2} \,$}}

\newcommand{\GREEN}{\mathsf{Green}}
\newcommand{\RED}{\mathsf{Red}}
\newcommand{\G}{\mathsf{G}}
\newcommand{\R}{\mathsf{R}}
\newcommand{\pair}[2]{\lpb #1, #2 \rpb}
\newcommand{\lpb}{\langle \hspace{-0.34em} \langle \hspace{-0.34em} \langle}	
\newcommand{\rpb}{\rangle \hspace{-0.34em} \rangle \hspace{-0.34em} \rangle}

\newcommand{\acts}{\mathit{acts}}
\newcommand{\bad}{\mathit{bad}}
\newcommand{\al}{\alpha}		
\newcommand{\bottom}{\mathit{bottom}} 
\newcommand{\cat}{\mbox{$^\frown$}}
\newcommand{\clos}[1]{\widehat{{#1}}}	

\newcommand{\execs}{\mathit{lexecs}}
\newcommand{\lexecs}{\mathit{lexecs}}
\newcommand{\iexecs}{\mathit{execs}^\omega}
\newcommand{\ext}{\mathit{ext}}

\newcommand{\fault}{\mathit{fault}}
\newcommand{\fstate}{\mathit{fstate}}
\newcommand{\ga}{\gamma}		
\newcommand{\good}{\mathit{good}}

\renewcommand{\int}{\mathit{int}}
\newcommand{\lattices}{\mathit{lattices}}

\newcommand{\lprec}{\prec}
\newcommand{\lpreceq}{\preceq}
\newcommand{\lstate}{\mathit{lstate}}

\newcommand{\pr}{\mathit{pair}}
\newcommand{\sig}{\mathit{sig}}

\newcommand{\start}{\mathit{start}}
\newcommand{\states}{\mathit{states}}
\newcommand{\steps}{\mathit{steps}}
\newcommand{\subs}{\lhd}		
\newcommand{\suc}{\mathit{succ}}	
\newcommand{\topp}{\mathit{top}}
\newcommand{\trace}{\mathit{trace}}
\newcommand{\traces}{\mathit{traces}}

\newcommand{\lpreorder}{\sqsubseteq_\l}	
\newcommand{\spreorder}{\sqsubseteq_s}	
\newcommand{\simu}{\leq}

\newcommand{\lpsim}{\simu_{\l F}}	
\newcommand{\lpfs}{\simu_{\l F}}		
\newcommand{\lprm}{\simu_{\l R}}		
\newcommand{\lpbs}{\simu_{\l B}} 	
\newcommand{\ilpbs}{\simu_{i\l B}} 
\newcommand{\lphr}{\simu_{\l H}}		
\newcommand{\lppr}{\simu_{\l P}} 	
\newcommand{\ilppr}{\simu_{i\l P}} 

\newcommand{\nonalwayssilent}{\mathit{nonalwayssilent}}

\newcommand{\ESDSAlg}{\mbox{\textit{ESDS-Alg}}}
\newcommand{\ESDSI}{\mbox{\textit{ESDS-I}}}
\newcommand{\ESDSII}{\mbox{\textit{ESDS-II}}}
\newcommand{\MI}{\mbox{\textit{M-I}}}
\newcommand{\MII}{\mbox{\textit{M-II}}}
\newcommand{\SpReq}{\mathit{SpReq}}
\newcommand{\SpStab}{\mathit{SpStab}}
\newcommand{\ImpReq}{\mathit{ImpReq}}
\newcommand{\ImpStab}{\mathit{ImpStab}}
\newcommand{\Users}{\mbox{\textit{Users}}}
\newcommand{\Frontend}{\mathit{Frontend}}
\newcommand{\Replica}{\mathit{Replica}}
\newcommand{\Channel}{\mathit{Channel}}

\newcommand{\chan}{\mathit{channel}}
\newcommand{\client}{\mathit{client}}

\newcommand{\pndg}{\mathit{pending}}
\newcommand{\req}{\mathrm{``request\mbox{''}}}
\newcommand{\resp}{\mathrm{``response\mbox{''}}}
\newcommand{\strict}{\mathit{strict}}

\newcommand{\remove}[1]{}

\newcommand{\ea}{\Diamond \Box}

\newcommand{\iof}{\Box \Diamond} 

\newcommand{\always}{\Box}
\newcommand{\eventually}{\Diamond}

\newtheorem{theorem}{Theorem}
\newtheorem{lemma}[theorem]{Lemma}
\newtheorem{proposition}[theorem]{Proposition}
\newtheorem{corollary}[theorem]{Corollary}

\newtheorem{definition}{Definition}

\newtheorem{assumption}{Assumption}

\newcommand{\case}[2]{\vspace{1.5ex} \noindent \textit{Case} #1: #2.}

\newcommand{\ms}[1]{%
        \relax\ifmmode
                \mathord{\mathcode`\-="702D\it #1\mathcode`\-="2200}%
        \else
                {\it #1}%
        \fi
}

\def\ioisize{\footnotesize}

\newcommand{\ioi}[3]{\bgroup\ioisize \begin{tabbing}
XX\=XX\=    \kill           
Input:\+ \\                 
  #1 \-  \\
Output: \+ \\
  #2 \- \\
Internal: \+\\
  #3 \-
\end{tabbing} \egroup} 

\newcommand{\io}[2]{\bgroup\ioisize \begin{tabbing}
XX\=XX\=    \kill           
Input:\+ \\                 
  #1 \-  \\
Output: \+ \\
  #2 \- 
\end{tabbing} \egroup} 

\newcommand{\ioti}[4]{\bgroup\ioisize \begin{tabbing}
XX\=XX\=    \kill           
Input:\+ \\                 
  #1 \-  \\
Output: \+ \\
  #2 \- \\
Time-passing: \+ \\
  #3 \- \\
Internal: \+\\
  #4 \-
\end{tabbing} \egroup} 

\newcommand{\ei}[2]{\bgroup\ioisize \begin{tabbing}
XX\=XX\=    \kill           
External:\+ \\              
  #1 \-  \\
Internal: \+\\
  #2 \-
\end{tabbing} \egroup}

\newcommand{\eonly}[1]{\bgroup\ioisize \begin{tabbing}
XX\=XX\=    \kill           
External:\+ \\              
  #1 \-  \\
\end{tabbing} \egroup}

\def\iocodesize{\scriptsize}
 
\newcommand{\iocodeonecol}[1]{\begin{center}\iocodesize
\begin{minipage}[t]{.95\linewidth}
\hbox to\linewidth{}
#1
\end{minipage}
\end{center}}

\newcommand{\iocode}[2]{\vspace{-15pt}\begin{center}\iocodesize
\begin{minipage}[t]{.475\linewidth}
\hbox to\linewidth{}
#1
\end{minipage}
\hspace{.01\linewidth} 
\begin{minipage}[t]{.475\linewidth}
\hbox to\linewidth{}
#2
\end{minipage}
\end{center}}

\newcommand{\ef}[2]{
\begin{tabbing}         
\= XXX\=XX\=XX\=XX\=XX\=   \kill
\protect #1\\
\>Eff: \+ \+ \>
   #2  \- \-
\end{tabbing}}

\newcommand{\prcef}[3]{\begin{tabbing}        
\= XXX\=XX\=XX\=XX\=  \kill
\protect #1\\
\>Pre: \+ \+ \>
  #2  \-  \- \\ 
\>Eff: \+  \+ \>
  #3 \- \- 
\end{tabbing}}

\newcommand{\prc}[2]{\begin{tabbing}        
\= XXX\=XX\=XX\=XX\=  \kill
\protect #1\\
\>Pre: \+ \+ \>
  #2  \-  \- 
\end{tabbing}}

\newlength\Mwidth
\newlength\jsawidth

\def\statedef#1{\def\auxstatedef{#1}%
\def\auxempty{}%
\settowidth{\Mwidth}{M}
\jsawidth=\textwidth
\advance\jsawidth by -7cm
\advance\jsawidth by -\Mwidth
\advance\jsawidth by -10\tabcolsep
\advance\jsawidth by -6\arrayrulewidth
\begin{trivlist}\item[]%
\begin{tabular}{|p{2.5cm}|c|p{2.2cm}|p{2.3cm}|p{\jsawidth}|}
\hline
{\bf Variable} & \phantom{M} & {\bf Type} & {\bf Initially} & {\bf Description}\\
\hline\hline}
\def\endstatedef{\\ 
\ifx\auxstatedef\auxempty
  \hline
\else%
  \hline\hline
   \multicolumn{5}{|l|}{\auxstatedef}\\\hline
 \fi
\end{tabular}%
\end{trivlist}}

\def\statetext{\begin{minipage}[t]{\linewidth}}
\def\endstatetext{\end{minipage}\vspace{2pt}}

\newcommand{\action}[1]{\textsf{#1}}
\newcommand{\statevar}[1]{\ms{#1}}

\newcommand{\automatontitle}[1]{\vspace{2ex} \noindent 
     \textbf{Automaton #1}\\}

\newcommand{\ioautomatontitle}[1]{\vspace{2ex} \noindent 
     \textbf{I/O Automaton #1}\\}

\newenvironment{signature}[1][Signature]{%
\noindent\textbf{#1}\iocodesize\vspace{0ex}}{}

\newenvironment{statevarlist}[1][State]
   {\noindent\textbf{#1}\iocodesize\vspace{0ex}
    \begin{list}{}{\topsep=2ex \itemsep=-0.5ex \leftmargin=0ex \labelwidth=0ex}}
   {\end{list}}

\newenvironment{actionlist}[1][Actions]{%
\vspace{1ex}\noindent\textbf{#1}\iocodesize}{}

\newcommand{\externalaction}[3]{\prcef{\textbf{External} #1}{#2}{#3}}

\usepackage{calc}       

\renewcommand{\iocode}[3][.475]{\vspace{-15pt}\begin{center}\iocodesize
\begin{minipage}[t]{#1\linewidth}
\hbox to\linewidth{}
#2
\end{minipage}
\hspace{.01\linewidth} 
\begin{minipage}[t]{.95\linewidth-#1\linewidth}
\hbox to\linewidth{}
#3
\end{minipage}
\end{center}}

\newcommand{\dbspec}{\textit{DB-Spec}\xspace}
\newcommand{\dbimp}{\textit{DB-Imp}\xspace}

\newcommand{\Op}{{\cal O}}

\newcommand{\request}{\action{request}}
\newcommand{\response}{\action{response}}

\newcommand{\send}{\action{send}}
\newcommand{\receive}{\action{receive}}
\newcommand{\doit}{\action{do\_it}}

\newcommand{\enter}{\action{enter}}
\newcommand{\calc}{\action{calculate}}

\newcommand{\stabilize}{\action{stabilize}}
\newcommand{\addcons}{\action{add\_constraints}}

\newcommand{\wait}{\statevar{wait}}
\newcommand{\rept}{\statevar{rept}}

\newcommand{\channel}{\statevar{channel}}
\newcommand{\pending}{\statevar{pending}}
\newcommand{\rcvd}{\statevar{rcvd}}
\newcommand{\done}{\statevar{done}}
\newcommand{\stable}{\statevar{stable}}

\newcommand{\ops}{\statevar{ops}}
\newcommand{\po}{\statevar{po}}
\newcommand{\newpo}{\statevar{new-po}}
\newcommand{\stabilized}{\statevar{stabilized}}

\newcommand{\id}{\ms{id}}

\newcommand{\prev}{\ms{prev}}

\newcommand{\idof}[1]{{#1}.\ms{id}}

\newcommand{\prevof}[1]{{#1}.\ms{prev}}
\newcommand{\strictof}[1]{{#1}.\ms{strict}}

\newcommand{\mkmsg}[1]{\ensuremath{\langle #1 \rangle}}
\newcommand{\reqmsg}[1]{\mkmsg{\text{``request''}, #1}}		
\newcommand{\respmsg}[1]{\mkmsg{\text{``response''}, #1}}	
\newcommand{\gossipmsg}[1]{\mkmsg{\text{``gossip''}, #1}}	

\newcommand{\msgset}{\mathcal{M}}
\newcommand{\reqmsgset}{\msgset_{\ms{req}}}
\newcommand{\respmsgset}{\msgset_{\ms{resp}}}
\newcommand{\gossipmsgset}{\msgset_{\ms{gossip}}}

\newcommand{\lbl}{\statevar{label}}

\newcommand{\valset}{\ms{valset}}

\newcommand{\CSC}{\mathit{CSC}}

\newcommand{\opdescs}{\mathcal{O}}
\newcommand{\labels}{\mathcal{L}}

\newcommand{\idset}{\mathcal{I}}
\newcommand{\reps}{\mathcal{R}}

\newcommand{\requested}{\statevar{requested}}
\newcommand{\responded}{\statevar{responded}}

\newcommand{\datavals}{V}

\newcommand{\setinsert}[2]{#1 \gets #1 \cup \{#2\}}
\newcommand{\setdelete}[2]{#1 \gets #1 -  \{#2\}}
\newcommand{\spn}[1]{\mathit{span}(#1)}
\newcommand{\set}[1]{\{#1\}}
\newcommand{\logicor}{\lor}

\newcommand{\LC}{\mathit{lc}}
\newcommand{\intersect}{\cap}
\newcommand{\Intersect}{\bigcap}
\newcommand{\horline}{\rule{\textwidth}{1pt}}

\newenvironment{proof}{\vspace{-1.0ex}\textit{Proof.} }
                      {\hfill{$\Box$}}

\begin{document}

\bc
\textbf{\Large On the Refinement of Liveness Properties of Distributed Systems}\footnote{Some of the 
results in this paper appeared in the eighteenth
ACM Symposium on Principles of Distributed Computing, (PODC'99), under the title
``Liveness-preserving Simulation Relations''.}\\[0.2in]

{\large Paul C. Attie} \\[0.05in]
\vskip 0.05in
{\large Department of Computer Science}\\
{\large American University of Beirut}\\
{\large and}\\
{\large Center for Advanced Mathematical Sciences}\\
{\large American University of Beirut}\\
\texttt{paul.attie@aub.edu.lb}

\today


\ec


\begin{abstract}

We present a new approach for reasoning about liveness properties of
distributed systems, represented as automata. Our approach is based
on simulation relations, and requires reasoning only over finite
execution fragments. Current simulation-relation based methods for
reasoning about liveness properties of automata require reasoning
over entire executions, since they involve a proof obligation of the
form: if a concrete and abstract execution ``correspond'' via the
simulation, and the concrete execution is live, then so is the
abstract execution.

Our contribution consists of (1) a formalism for defining liveness
properties, (2) a proof method for liveness properties based on that
formalism, and (3) two expressive completeness results: firstly, our
formalism can express any liveness property which satisfies a natural
``robustness'' condition, and secondly, our formalism can express any
liveness property at all, provided that history variables can be used.

To define liveness, we generalize the notion of a complemented-pairs
(Streett) automaton to an infinite state-space, and an infinite number
of complemented-pairs. Our proof method provides two main techniques:
one for refining liveness properties across levels of abstraction, and
the other for refining liveness properties within a level of
abstraction. The first is based on extending simulation relations so
that they relate the liveness properties of an abstract (i.e., higher
level) automaton to those of a concrete (i.e., lower level)
automaton. The second is based on a deductive method for inferring new
liveness properties of an automaton from already established liveness
properties of the same automaton. This deductive method is
diagrammatic, and is based on constructing ``lattices'' of liveness
properties. Thus, it supports proof decomposition and separation of
concerns. 

\end{abstract}

\section{Introduction and Overview}

One of the major approaches to the construction of correct distributed
systems is the use of an operational specification, e.g., an \intr{automaton}
or a \intr{labeled transition system}, which is successively refined, via
several intermediate levels of abstraction, into an
implementation. The implementation is considered correct if and only if each of
its externally visible behaviors, i.e., \intr{traces}, is also a trace of the
specification. 
This ``trace inclusion'' of the implementation in the specification is
usually established transitively by means of establishing the trace
inclusion of the system description at each level of abstraction in
the system description at the next higher level.  When reasoning at
any particular level, we call the lower level the concrete level, and
the higher level the abstract level.

The correctness properties of a distributed system are classified into
\intr{safety} and \intr{liveness} \cite{Lam77}: 
safety properties state that ``nothing bad happens,'' for example,
that a database system never produces incorrect responses to queries, 
while
liveness properties state that ``progress occurs in the system,'' for
example, every query sent to a database system is eventually responded to.
Safety properties are characterized by the fact that they are violated
in finite time: e.g., once a database has returned an incorrect
response to an external user, there is no way to recover to where the
safety property is satisfied. Liveness properties, on the other
hand, are characterized by the fact that there is always the
possibility of satisfying them: the database always has the
opportunity of responding to pending queries.
Thus, an operational specification defines the required safety properties by
means of an automaton, or labeled transition system. The reachable
states and transitions of the automaton are the ``good''
states/transitions, whose occurrence does not violate safety. Any
unreachable states, if present, are ``bad,'' i.e., they represent a
violation of the safety properties, e.g., due to a fault.
The occurrence of such a ``bad'' state is something that happens in
finite time, and so constitutes the violation of a safety property.
The liveness properties are specified by designating a subset of the
executions of the automaton as being the ``live'' executions, leading
to the notion of \intr{live execution property}.
These are the executions along which eventually, all the necessary
actions are executed, e.g., the actions that respond to pending queries.
To express the idea that there is always the possibility of
satisfying a liveness property, this subset of the executions must have the property
that any finite execution can be extended to an execution in the
subset \cite{AL91}. 

Distributed systems consist of many sequential processes which execute
concurrently. To reason effectively about such large systems,
researchers have proposed the use of \emph{compositional reasoning}:
global properties of the entire system are inferred by first
deducing local properties of the constituent processes or subsystems, and then
combining these local properties to establish the global properties.
In particular, we desire that refinement is compositional: when a
particular process $P_i$ is refined to a new process $P'_i$, we wish
to reason only about whether $P'_i$ is a correct refinement of $P_i$,
without having to engage in global reasoning involving all of the
other processes in the system. The need for compositional reasoning,
as well as notions such as behavioral subtyping \cite{LW94} and
information hiding, motivated the development of the notion of
\emph{externally visible behavior}, e.g., the set of traces of an
automaton, where a trace is a sequence of ``external'' actions,
visible at the interface, which the automaton can engage in.
Typically, a trace is obtained by taking an execution and removing all
the \emph{internal} information, i.e., the states and the internal actions.

The notion of externally visible behavior then leads naturally to
notions of external safety and liveness properties, which are
specified over the traces of an automaton, rather than over the
(internal) states and executions.
The external safety property is the set of all traces,
since this is the external ``projection'' of all the executions, which
define the reachable states and transitions, which in turn give us the
safety properties, as discussed above.
The external liveness property is obtained by taking the 
traces of all the live executions. These are called the \emph{live
traces}, and the set of all live traces is a \intr{live trace
property}.

Trace inclusion usually means that every trace of the concrete
automaton is a trace of the abstract automaton.  Thus, trace inclusion
deals with safety properties: every safety property of the set of
traces of the concrete automaton is also a safety property of the set
of traces of the abstract automaton. Thus, external
safety properties are preserved by the refinement from the abstract to
the concrete.  Trace inclusion does not address liveness properties,
however.
The appropriate notion of inclusion for external liveness properties is
\intr{live trace inclusion} \cite{GSSL93,GSSL98}: every live trace of the
concrete automaton is a live trace of the abstract automaton.

Consider again the database example, with the external liveness property that
every query submitted is eventually processed.  Let $B$ be a
high-level specification of such a system.  By using state variables
that record requests and responses, this property can be easily stated
in terms of the executions of $B$, which results in a live execution
property.  The set of traces of the live executions then gives the
corresponding live trace property. Provided that the state variables
which record requests and responses are updated correctly, the live
trace property will only contain traces in which every input of a
query to the database (e.g., from an external ``user'') is eventually
followed by an output of a response from the database (to the user).

Let $A$ be an implementation of $B$. The live executions of $A$ are
defined by the liveness properties that typically can be guaranteed by
reasonable implementations, e.g., ``fair scheduling''
\cite{Fr86}---every continuously enabled action (or process) is
eventually executed, and fair polling of message channels---every
message sent is eventually received\footnote{We do not address
fault-tolerance for the time being, thus messages are always received
along a live execution. See Section~\ref{sec:fault-tolerance} for a
discussion of how the techniques presented in this paper can be
applied to fault-tolerance.}.
The set of traces of the live executions then gives the 
live trace property corresponding to this action/process fairness and
reliable message delivery in the underlying execution behavior.
However, the live trace property that we wish to verify for $A$ is not
this property per se, but the same live trace property which $B$ has, namely that
every input of a query to
the database is eventually followed by an appropriate output from the database.
This paper addresses the problem of verifying such liveness properties
for an implementation  $A$.

It is clear that verifying that the live traces of $A$ are contained
in the live traces of $B$ immediately yields the desired conclusion,
namely that $A$ has the desired live trace property.
Thus, live trace inclusion applied to the above example implies that
every trace of an execution of $A$ in which all messages sent are
eventually received, and all continuously enabled actions (processes)
are eventually executed, i.e., a live trace of $A$, is also a live
trace of $B$, i.e., a trace in which all queries receive a response. This
is exactly what is required, since the liveness properties of $A$
along executions where, for example, messages sent are not received,
are not of interest. Conversely, a live execution of $A$, in which all
messages sent are received, and scheduling is fair, should produce an
external behavior which has the desired liveness properties:
every query receives a response.
More generally, live trace inclusion 
implies that external liveness properties are
preserved by the refinement from the specification $B$ to the implementation
$A$.

One of the main proof techniques for establishing trace inclusion is
that of establishing a \textit{simulation} \cite{LV95} or
\emph{bisimulation} \cite{Mil99} between the concrete and the abstract
automata.  A simulation (or bisimulation) establishes a certain
correspondence (depending on the precise type of simulation)
between the states/transitions of the concrete automaton and
the states/transitions of the abstract automaton, which then implies trace
inclusion.
An important advantage of the simulation-based approach is that it
only requires reasoning about individual states and finite execution
fragments, rather
than reasoning about entire (infinite) executions.  Unfortunately, the
end-result, namely the establishment of trace inclusion, does not, as
we establish in the sequel, imply live trace inclusion, since 
the set of live traces is, in general, a proper subset of the set of traces.
\remove{
Reasoning about liveness has traditionally been associated with
reasoning over entire executions, e.g., using well-foundedness arguments, 
variant functions, fairness etc. 
}

\paragraph{Our contributions.}
In this paper, we show how to use simulation relations to reason about
liveness.
\remove{This is therefore a significant methodological advance over current methods.}
Our approach uses a state-based technique to specify live execution 
properties: a \intr{liveness condition} is given as a (possibly infinite) set
of ordered pairs $\pair{\RED_i}{\GREEN_i}$, where $\RED_i$, $\GREEN_i$
are sets of states.
An execution is considered to satisfy a single pair
$\pair{\RED}{\GREEN}$ iff whenever it contains infinitely many states
in $\RED$, then it also contains infinitely many states 
in $\GREEN$.  An execution is live iff it satisfies all the pairs
in the liveness condition. A trace is live iff it is the trace of some
live execution.
Our notion of liveness condition is akin to the acceptance condition
of a \intr{complemented-pairs (or Streett) automaton}
\cite{AH95,EL85,GL94}, except that we allow an infinite number of pairs,
and also our automata can have an infinite number of states and
transitions. 
We then present the notion of \intr{liveness-preserving simulation
relation}, which appropriately relates the states mentioned in the
concrete automaton's liveness condition to those mentioned in the
abstract automaton's liveness condition.  This is done in two stages.
The first stage refines the liveness condition of the abstract
automaton into a ``derived'' liveness condition of the concrete
automaton.  This derived condition may contain complemented-pairs that
are not directly specified in the liveness condition of the concrete
automaton.  The second stage then proves that the derived condition is
implied by the directly specified liveness condition of the concrete
automaton (using a ``lattice'' construction).  
The use of such a derived liveness condition allows us to
break down the refinement problem at each level into two simpler
subproblems, since the derived liveness condition of the concrete
automaton can usually be formulated to better match with the
liveness condition of the abstract automaton.
Establishing a liveness-preserving simulation relation then allows us
to conclude that every live trace of the concrete automaton is also a
live trace of the abstract automaton.
As discussed above, our method can be applied to multiple levels of
abstraction, where the specification is successively refined in
stages, producing several intermediate descriptions of the specified
system, until a description that is directly implementable on the
desired target architecture and has adequate performance and
fault-tolerance properties is derived.
Thus, we address the problem of preserving liveness properties in the
successive refinement of a specification into an implementation,
which contributes to making the method scalable, as our 
extended example in Section~\ref{sec:example} shows.

We establish two expressive completeness results for 
complemented-pairs liveness conditions. The first shows that any live execution
property which satisfies a natural ``robustness'' condition can be
specified by a complemented-pairs liveness condition. The second shows that any
live execution property whatsoever can be specified by a
complemented-pairs liveness condition, provided that history variables can be
used.

The paper is organized as follows.
Section~\ref{sec:background} provides technical background on automata
and simulation relations from
\cite{GSSL93} and \cite{LV95}.  Section~\ref{sec:live-automata} gives
our key technical notion of a live automaton, i.e., an automaton
equipped with a liveness condition, and also defines live executions,
live traces, and derived liveness properties.
Section~\ref{sec:sim-rels} presents our definitions for
liveness-preserving simulation relations, and shows that
liveness-preserving simulation relations
imply live trace inclusion.  Section~\ref{sec:lattices}
shows how a derived liveness condition can be deduced from the directly
specified condition. Together, these two sections give our
method for refining liveness properties.
Section~\ref{sec:example} applies our
results to the eventually-serializable data service of \cite{FGLLS99,LLSG92}.
Section~\ref{sec:discussion} examines some
alternative choices for expressing liveness, shows that our method
can also be applied to fault-tolerance properties, and briefly discusses the
mechanization of our method. 
Section~\ref{sec:expressiveness} discusses the expressiveness of complemented-pairs
for liveness properties, and presents two relative completeness results.
Section~\ref{sec:related} discusses related work.  Finally,
Section~\ref{sec:conc} presents our conclusions and discusses avenues
for further research.
Appendix~\ref{app:simulation-relations} gives some background on
simulation relations,
Appendix~\ref{app:TL} gives some background on temporal logic, and
Appendix~\ref{app:esds} presents I/O automaton pseudocode for the 
eventually-serializable data service of \cite{FGLLS99,LLSG92}.

\section{Technical Background}
\label{sec:background}

The definitions and theorems in this section are taken from \cite{GSSL93} and
\cite{LV95}, to which the reader is referred for details and proofs.

\subsection{Automata}
\label{sec:automata}

\bd[Automaton]
An \emph{automaton} $A$ consists of four components:
\bn
\item a set $\states(A)$ of states,
\item a nonempty set $\start(A) \sub \states(A)$ of start states,
\item an action signature $\sig(A) = (\ext(A), \int(A))$ where
$\ext(A)$ and $\int(A)$ are disjoint sets of external and internal
actions, respectively $($let $\acts(A)$ denote the set $\ext(A) \un
\int(A)$$)$, and
\item a transition relation $\steps(A) \sub \states(A) \times
\acts(A) \times \states(A)$.
\label{def:automaton}
\en
\ed

Let $s,s',u,u',\ldots$ range over states and $a,b,\ldots$ range over
actions. Write $s \lla{a}{A} s'$ iff $(s,a,s') \in \steps(A)$. We say
that $a$ is \intr{enabled} in $s$.  An execution fragment $\al$ of
automaton $A$ is
an alternating sequence of states and actions $s_0 a_1 s_1 a_2 s_2
\ldots$ such that $(s_i, a_{i+1}, s_{i+1}) \in \steps(A)$ for all $i
\geq 0$, i.e., $\al$ conforms to the
transition relation of $A$. Furthermore, if $\al$ is finite then it
ends in a state. 
If $\al$ is an execution fragment, then
$\fstate(\al)$ is the first state along $\al$, and if 
$\al$ is finite, then
$\lstate(\al)$ is the last state along $\al$.
If $\al_1$ is a
finite execution fragment, $\al_2$ is an execution fragment, and
$\lstate(\al_1) = \fstate(\al_2)$, then $\al_1 \cat \al_2$ is the
concatenation of $\al_1$ and $\al_2$ (with $\lstate(\al_1)$ repeated
only once). 
Let $\al = s_0 a_1 s_1 a_2 s_2 \ldots$ be an execution fragment. Then
the length of $\al$, denoted $|\al|$, is the number of actions in
$\al$. $|\al|$ is infinite if $\al$ is infinite, 
and $|\al| = 0$ if $\al$ consists of a single state.
Also, $\al|_i \df  s_0 a_1 s_1 \ldots a_i s_i$.
If $\al$ is a prefix of $\al'$, we write $\al \leq \al'$. We also
write $\al < \al'$ for $\al \leq \al'$ and $\al \neq \al'$.

An execution of $A$ is an execution fragment that begins with a state
in $\start(A)$. 
The set of all executions of $A$ is denoted by $\mathit{execs}(A)$, and
the set of all infinite executions of $A$ is denoted by $\iexecs(A)$.
A state of $A$ is \emph{reachable} iff it occurs in some execution of
$A$.
The trace $\trace(\alpha)$ of execution fragment $\alpha$ is obtained
by removing all the states and internal actions from $\al$.
The set of traces of an automaton $A$ is defined as the set of traces
$\beta$ such that $\beta$ is the trace of some execution of $A$.  It
is denoted by $\traces(A)$.
If $\varphi$ is a set of executions, then
$\traces(\varphi)$ is the set of traces $\beta$ such that $\beta$ is the
trace of some execution in $\varphi$. 
If $a$ is an action, then we define $\trace(a) = a$ if $a$ is external, and
$\trace(a) = \lambda$ (the empty sequence) if $a$ is internal.
If $a_1 a_2 \cdots a_n$ is a sequence of actions, then 
$\trace(a_1 \cdots a_n) = \trace(a_1) \trace(a_2) \cdots \trace(a_n)$,
where juxtaposition denotes concatenation.

If $R$ is a relation over $S_1 \times S_2$ (i.e., $R \subseteq S_1
\times S_2$) and $s_1 \in S_1$, then we define
$R[s_1] = \{ s_2 ~|~ (s_1,s_2) \in R\}$.
We use $\up$ to denote the restriction of a mapping to a subset of its domain.

\subsection{Simulation Relations}
\label{sec:simulation-relations}

We shall study five different simulation relations: 
forward simulation, refinement mapping, backward simulation,
history relation, and prophecy relation.
These relations all preserve safety properties.
In Section~\ref{sec:sim-rels}, we extend these simulation relations so
that they preserve liveness as well as safety.
A forward simulation requires that (1) each execution of
an external action $a$ of $A$ is matched by a finite execution
fragment of $B$ containing $a$, and all of whose other actions are
internal to $B$, and (2) each execution of an internal action of $A$
is matched by a finite (possibly empty) execution fragment of $B$ all
of whose actions are internal to $B$ (if the fragment
is empty, then we have $u \in f[s']$, i.e., $u$ and $s'$ must be related by
the simulation).
It follows that forward simulation implies trace inclusion
(also referred to as the \intr{safe preorder} below), i.e., 
if there is a forward simulation from $A$ to $B$, then
$\traces(A) \sub \traces(B)$.
Likewise, the other simulation relations all imply trace inclusion (the backward
simulation and prophecy relation must be image-finite) for similar reasons.
See Lemma 6.16 in \cite{GSSL93} for a formal proof of this result.

We use $F, R, iB, H, iP$ to denote forward simulation, refinement
mapping, image-finite backward simulation, history relation,
image-finite prophecy relation, respectively. 
Thus, when we write $X \in \{F, R, iB, H, iP\}$, we mean that $X$ is
one of these relations.
We write $A \simu_F B$ if there exists a forward simulation from $A$
to $B$ w.r.t. some invariants, and 
$A \simu_F B$ via $f$ if $f$ is a forward simulation from
$A$ to $B$ w.r.t. some invariants. Similarly for the other simulation relations.
Appendix~\ref{app:simulation-relations} gives formal definitions for
all of these simulation relations.

\subsection{Execution Correspondence}

Simulation relations induce a correspondence between the executions of
the concrete and the abstract automata. This correspondence is
captured by the notion of $R$-relation.
If $\alpha' = u_0 b_1 u_1 b_2 u_2 \cdots$ is an execution of automaton
$B$, then define 
$\trace(\al',j,k)$ to be $\trace(b_j \cdots b_k)$ if $j \le k$, and
to be $\lambda$ (the empty sequence) if $j > k$.

\bd[$R$-relation and Index Mappings]
Let $A$ and $B$ be automata with the same external actions and let $R$
be a relation over $\states(A) \times \states(B)$. Furthermore, let
$\alpha$ and $\alpha'$ be executions of $A$ and $B$, respectively:
\ms\\
\ind $\alpha  = s_0 a_1 s_1 a_2 s_2 \cdots$ \\
\ind $\alpha' = u_0 b_1 u_1 b_2 u_2 \cdots$ \ms\\
Say that $\alpha$ and $\alpha'$ are \intrdef{$R$-related}, written
$(\alpha, \alpha') \in R$, if there exists a total, nondecreasing mapping
$m: \{0,1,\ldots,|\alpha|\} \mapsto \{0,1,\ldots,|\alpha'|\}$
such that:
\bn
   \item \label{clause:index-mapping:init} $m(0) = 0$,

   \item \label{clause:index-mapping:corr}
         $(s_i, u_{m(i)}) \in R$ for all $i$, $0 \leq i \leq |\alpha|$,

   \item \label{clause:index-mapping:trace}
         $\trace(\al', {m(i-1)+1}, {m(i)}) = \trace(a_i)$ for all 
               $i$, $0 < i \leq |\alpha|$, and

   \item \label{clause:index-mapping:cofinal}
         for all $j, 0 \leq j \leq |\alpha'|$, there exists an $i$, 
             $0 \leq i \leq |\alpha|$, such that $m(i) \geq j$.
\en
The mapping $m$ is referred to as an \intrdef{index mapping} from
$\alpha$ to $\alpha'$ with respect to $R$.
Write $(A,B) \in R$ if for every execution $\alpha$ of $A$, there
exists an execution $\alpha'$ of $B$ such that $(\alpha,\alpha') \in
R$.
\label{def:index-mapping}
\ed

\bt [Execution Correspondence Theorem]
Let $A$ and $B$ be automata with the same external actions.
Suppose $A \simu_X B$ via $S$, where $X \in \{F, R, iB, H, iP\}$.
Then $(A,B) \in S$.
\label{thm:execution-correspondence}
\et

\bl  
Let $A$ and $B$ be automata with the same external actions and let $R$
be a relation over $\states(A) \times \states(B)$. If $(\alpha,
\alpha') \in R$, then $\trace(\alpha) = \trace(\alpha')$.
\label{lem:traces}
\el

Theorem~\ref{thm:execution-correspondence} and Lemma~\ref{lem:traces} appear
in \cite{GSSL93} as Theorem 6.11 and Lemma 6.15, respectively.

\subsection{Linear-time Temporal Logic}

We use the fragment of linear-time temporal
logic consisting of the $\always$ (always) and
$\eventually$ (eventually) operators over state assertions \cite{Pn77,MP92}.
In particular, we use the ``infinitary'' operators $\iof$ (infinitely often)
and $\ea$ (eventually always).
We specify state assertions as a set of states, the state in question
satisfying the assertion iff it belongs to the set.

For example, if $U$ is a set of states, then
$\alpha \sat \iof U$ means ``$\alpha$ contains infinitely many states from $U$,''
and
$\alpha \sat \ea U$ means ``all but a finite number of states of
                          $\alpha$ are from $U$.'' 
These operators can be combined with propositional connectives
$(\neg, \land, \lor, \imp)$ so that, for example,
$\alpha \sat \iof U' \imp \iof U''$ means ``if $\alpha$ contains
infinitely many states from $U'$, then it also contains 
infinitely many states from $U''$,
and
$\alpha \sat \ea \neg U$ means ``all but a finite number of states of
                               $\alpha$ are not from $U$.'' 

Appendix~\ref{app:TL} provides a formal definition of the syntax and
semantics of the temporal logic that we use.

\section{Live Automata}
\label{sec:live-automata}

We first formalize the notions of live execution property and live
trace property, that discussed in the introduction.

\bd[Live Execution Property]
\label{def:liveness-property}
Let $A$ be an automaton, and $\varphi \sub \iexecs(A)$. Then,
$\varphi$ is a \emph{live execution property} for $A$ if and only if
for every finite execution $\al$ of $A$, there exists an infinite
execution $\al'$ of $A$ such that
$\al < \al'$ and $\al' \in \varphi$.
\ed
In other words, a live execution property is a set of infinite
executions of $A$ such that every finite execution of $A$ can be
extended to an infinite execution in the set.
This requirement was proposed in \cite{AL91}, where it is called
\emph{machine closure}.

Note that we do not consider interaction with an environment in this
paper. THis is why we use automata rather than I/O automata, i.e., we
have external actions without an input/output distinction.
This issue is treated in detail in \cite{GSSL98}, where a
liveness property is defined as a set of executions (finite or
infinite) such that 
any finite execution can be extended to an execution in the set.
Thus, an extension may be finite, unlike our approach. This is because
requiring extension to an infinite execution may constrain the
environment: an execution ending in a state with no enabled internal
or output action will then require the environment to execute an
action that is an output of the environment and an input of the
automaton, so that the execution can be extended to an infinite one. 
We defer treating this issue to another occasion.

\bd[Live Trace Property]
\label{def:live-trace-property}
Let $A$ be an automaton, and $\psi \sub \traces(A)$. Then,
$\psi$ is a \emph{live trace property} for $A$ if and only if
there exists a live execution property $\varphi$ for $A$ such that 
$\psi = \traces(\varphi)$.
\ed

In \cite{GSSL93,GSSL98}, the notion of live execution property was the
basic liveness notion, and a live automaton was defined to be an
automaton $A$ together with a live execution property.
This use of an arbitrary set of executions as a liveness property,
subject only to the machine closure constraint resulted in a proof
method in \cite{GSSL93} which requires reasoning over entire
executions. Since we wish to avoid this, we take as our basic liveness
notion the \intr{complemented-pairs condition} of Streett automata,
with the proviso that we extend it to an infinite state-space and an
infinite number of complemented-pairs.
In the next section, we show that this approach to specifying liveness
entails no loss of expressiveness, provided that we can use history
variables.

Let $A$ be an automaton. We say that $p$ is a
\intrdef{complemented-pair}\footnote{When it is clear from context, we just
say ``pair''.} over $A$ iff $p$ is an ordered pair
$\pair{\RED}{\GREEN}$ where $\RED \sub \states(A)$, $\GREEN \sub states(A)$.
Given $p = \pair{\RED}{\GREEN}$, we define the selectors $p.\R = \RED$ and $p.\G =
\GREEN$. 
Let $\al$ be an infinite execution of $A$. Then, we write
$\al \sat \pair{\RED}{\GREEN}$ iff $\al \sat \iof \RED \imp \iof \GREEN$,
i.e., if $\al$ contains infinitely many states
in $\RED$, then it also contains infinitely many states in
$\GREEN$.
We also write 
$\al \sat p$ in this case. %
Our goal is a method for refining liveness properties using reasoning
over states and finite execution fragments only, in particular,
avoiding reasoning over entire (infinite) executions. We therefore formulate a
liveness condition based on states rather than executions.

\bd[Live Automaton with Complemented-pairs Liveness Condition] 
\label{def:live-automaton}
A \intrdef{live automaton} is a pair $(A,L)$ where:
\bn

\item $A$ is an automaton, and

\item $L$ is a set of pairs
$\{ \pair{\RED_A^i}{\GREEN_A^i} ~|~ i \in \eta\}$
where $\RED_A^i \sub \states(A)$ and $\GREEN_A^i \sub \states(A)$ for
all $i \in \eta$, and $\eta$ is some cardinal, which serves as an
index set,

\en

and $A$, $L$ satisfy the following constraint:

\be
\item for every finite execution $\al$ of $A$, there exists an infinite
execution $\al'$ of $A$ such that

$\al < \al'$ and $(\fa p \in L: \al' \sat p)$.

\ee
\ed
$(A,L)$ inherits all of the attributes of $A$, namely the states,
start states, action signature, and transition relation of $A$. 
The executions (execution fragments) of $(A,L)$ are the executions
(execution fragments) of $A$, respectively. 
We say that $L$ is a \emph{complemented-pairs liveness condition} over $A$.
Often we use just ``liveness condition'' instead of 
``complemented-pairs liveness condition.''

The constraint in Definition~\ref{def:live-automaton} is the machine
closure requirement, that every finite execution can be extended to a
live execution.

\bd[Live Execution]
\label{def:live-execution}
Let $(A,L)$ be a live automaton. An execution $\al$ of $(A,L)$ is a
\intrdef{live execution} iff $\al$ is infinite and $\fa p \in L: \al \sat p$.\\
We define $\execs(A,L) = \{\al ~|~ \al \in \iexecs(A) \mbox{ and } 
(\fa p \in L: \al \sat p)\}$.
\ed

Our notion of liveness condition is essentially the acceptance condition for
finite-state complemented-pairs automata on infinite strings
\cite{EL85}, with the important difference that we generalize it to an
arbitrary (possibly infinite) state space, and allow a possibly
infinite set of pairs. 
Despite the possibility that $\RED_A^i$ and
$\GREEN_A^i$ are infinite sets of states, it is nevertheless very
convenient to have an infinite number of complemented-pairs.  Using the database
example of the introduction, we can express the 
liveness property ``every query submitted is eventually processed'' as the
infinite set of pairs 
$\{ \pair{x \in \wait}{x \not\in \wait} ~|~ \mbox{$x$ is a query} \}$, 
and where $\wait$ is the set of all queries that have been submitted
but not yet processed ($x$ is removed from $\wait$ when it is processed).
Being able to allocate one pair for each query facilitates the very
straightforward expression of this liveness property.
Our extended example in Section~\ref{sec:example} also uses an
infinite number of pairs in this manner.

The above discussion applies to any system in which there are an infinite number
of \empi{distinguished} operations, e.g., each operation has a unique
identifier, as opposed to, for example mutual exclusion for a fixed
finite number of processes,, where there
are an infinite number of entries into the critical section by some
process $P_i$, but these need not be ``distinguished,'' since the single
liveness property 
$\always ( request(P_i) \implies \eventually critical(P_i) )$
is sufficient to account for all of these. 
The key point is that only a bounded number of 
outstanding requests must be dealt with ($\le$ the number of
processes) , whereas in 
a system in which there are an infinite number
of {distinguished} operations, an unbounded number of outstanding
requests must be dealt with. We conjecture that the liveness property 
``every request is eventually satisfied'' cannot even be stated using
a finite number of complemented pairs.

The safe preorder, live preorder \cite{GSSL93} embody our notions of correct
implementation with respect to safety, liveness, respectively.

\bd[Safe preorder, Live preorder] Let $(A,L)$, $(B,M)$ be live
automata with the same external actions $(\ext(A) = \ext(B))$.
We define: \ms\\
\ind Safe preorder:  $(A, L) \spreorder (B, M)$ iff
		       $\traces(A) \sub \traces(B)$\\[1.0ex]
\ind Live preorder:  $(A, L) \lpreorder (B, M)$ iff
		       $\traces(\execs(A,L)) \sub \traces(\execs(B,M))$
\label{def:preorders}
\ed
From \cite{LV95,GSSL93}, we have that simulation relations imply the safe
preorder, i.e., if
$A \simu_X B$ where $X \in \{F, R, iB, H, iP\}$, then $(A, L) \spreorder (B, M)$.

Returning to the database example of the introduction, 
if $\al$ is some live execution of the implementation $A$, 
then, along $\al$, every continuously enabled action is eventually
executed (action fairness) and every message sent is eventually received
(message fairness).
The trace $\beta$ of $\al$ is then an externally visible live
behavior of $A$: $\beta \in \traces(\execs(A,L))$.
If $A$ is a correct implementation, then we expect that the
enforcement of action fairness and message fairness in $A$ then
guarantees the required liveness properties of the specification,
namely that every query is eventually processed. Thus, 
the externally visible live
behavior $\beta$ of $A$ must satisfy the 
required liveness properties of the specification,
i.e.,  $\beta \in \traces(\execs(B,M))$.
This is exactly what the live preorder requires.

\bd[Semantic Closure of a Liveness Condition]
\label{def:closure}
Let $(A,L)$ be a live automaton. 
The semantic closure $\clos{L}$ of $L$ in $A$ is given by
$\clos{L} =  \{ \pair{\R}{\G} ~|~ \fa \al \in \execs(A,L): \al \sat \pair{\R}{\G} \}$.
\ed

$\clos{L}$ is the set of complemented-pairs over $A$ which are ``semantically
entailed'' by the complemented-pairs in $L$, with respect to the executions of $A$.
In general, $\clos{L} - L$ is nonempty.
Every pair in $\clos{L} - L$ represents a ``derived'' liveness property, since it
is not directly specified by $L$, but nevertheless can be deduced from the
pairs in $L$, when considering only the executions of $A$.

\bd[Derived Pair]
\label{def:derived-pair}
Let $(A,L)$ be a live automaton, and let $p \in \clos{L} - L$.
Then $p$ is a \intrdef{derived pair} of $(A,L)$.
\ed

\bp
$L \subseteq \clos{L}$.
\label{prop:clos}
\ep
\bpr
Let $p$ be any complemented-pair in $L$.
Hence, by definition of $\execs(A,L)$, we have 
$\fa \al \in \execs(A,L): \al \sat p$. Hence $p \in \clos{L}$.
\epr

\bp
$\execs(A,\clos{L}) = \execs(A,L)$.
\label{prop:execs-clos}
\ep
\bpr
$\execs(A,\clos{L}) \subseteq \execs(A,L)$ follows immediately from 
Proposition~\ref{prop:clos} and the relevant definitions.
Suppose $\al \in \execs(A,L)$. 
By Definition~\ref{def:closure}, $\fa p \in \clos{L}: \al \sat p$.
Hence, $\al \in \execs(A,\clos{L})$.
Hence $\execs(A,L) \subseteq \execs(A,\clos{L})$.
\epr

From Proposition~\ref{prop:execs-clos}, it follows that $(A,\clos{L})$ is
a live automaton.

\section{Refining Liveness Properties Across Levels of Abstraction: Liveness-preserving
	 Simulation Relations}
\label{sec:sim-rels}

The simulation relations given in
Section~\ref{sec:simulation-relations} induce a relationship between
the concrete automaton $A$ and abstract automaton $B$ whereby for
every execution $\al$ of $A$ there exists a corresponding, in the
sense of Definition~\ref{def:index-mapping}, execution $\al'$ of $B$.
This correspondence between executions does not however take liveness
into account. So, if we were dealing with live automata $(A,L)$ and
$(B,M)$ instead of automata $A$ and $B$, then it would be possible to
have $\al \in \lexecs(A,L)$, $\al' \not\in \lexecs(B,M)$, and
$(\al,\al') \in S$ where $S$ is a simulation relation from $A$ to $B$.
So, $\beta \in \traces(\execs(A,L))$ and $\beta \notin \traces(\execs(B,M))$,
where $\beta = \trace(\al)$, is possible.
Hence establishing 
$A \simu_X B$ via $S$, where $X \in \{F, R, iB, H, iP\}$
does not allow one to conclude 
 $\traces(\execs(A,L)) \sub \traces(\execs(B,M))$, as desired, 
whereas it does allow one to conclude 
$\traces(A) \sub \traces(B)$, \cite[Lemma 6.16]{GSSL93}.
For example, consider 
Figures~\ref{fig:simple-DB-spec} and \ref{fig:simple-DB-lossy-impl}
which respectively give a specification and a ``first level'' refinement of the
specification, for a toy database system. 
The database takes input requests of the form 
$\request(x)$, where $x$ is a query, computes a response for $x$ using
a function $val$ (which presumably also refers to the underlying
database state, we do not model this to keep the example simple), and
outputs a response $(x,v)$ where $v = val(x)$.
This behavior is dictated by the specification in 
Figure~\ref{fig:simple-DB-spec}, where received queries are placed in
the set $\requested$, and queries responded to are placed in the set
$\responded$ (this prevents multiple responses to the same query).
The first-level refinement of the specification 
(Figure~\ref{fig:simple-DB-lossy-impl}) 
is identical to the specification except that it can 
``lose'' pending requests: the $\request(x)$ nondeterministically
chooses between adding $x$ to $\requested$, or doing nothing, as
represented by $\choice skip$ in Figure~\ref{fig:simple-DB-lossy-impl}.
Despite this fault, it is possible to
establish a forward simulation $F$ from 
\dbimp to \dbspec, as follows.
A state $s$ of \dbimp and a state $u$
of \dbspec are related by $F$ if and only if
$s.\requested \sub u.\requested$ and $s.\responded = u.\responded$
(where $s.var$ denotes the value of variable $var$ in state $s$).
Now suppose we add the following liveness condition to both 
\dbimp and of \dbspec: 
$\{ \pair{x \in \requested}{x \in \responded} ~|~ \mbox{$x$ is a query} \}$.
Thus, an operation $x$ that has been requested must eventually be
responded to, since ${x \in \requested}$ is stable; once true, it is
always true, and therefore it is true infinitely often.
Now let $\al$, $\al'$ be executions of \dbimp,
\dbspec, respectively, which are related by $F$ in
the sense of Definition~\ref{def:index-mapping}. Suppose some query $x_0$ is
lost along $\al$, and no other query is lost. Let $\al$ be live, i.e., 
if a query is placed into $\requested$, and is not lost, then it will
eventually be responded to. We now see that $\al'$ cannot be
live, since $x_0 \in \requested$ holds along an infinite suffix of $\al'$, but 
$x_0 \in \responded$ never holds along $\al'$.
Hence, establishing a forward simulation from 
\dbimp to \dbspec is not sufficient to 
establish live trace inclusion from 
\dbimp to \dbspec.

\begin{figure}[htb]

\horline
\automatontitle{\dbspec}

\begin{signature}
\eonly{%
	$\request(x)$, where $x$ is a query\\
	$\response(x,v)$, where $x$ is a query and $v$ is a value
}
\end{signature}

\begin{statevarlist}

\item $\requested$, a set of received queries, initially empty
\item $\responded$, a set of computed responses to queries, initially empty

\end{statevarlist}

\begin{actionlist}

\iocode{%

\externalaction{$\request(x)$}
{%
  $\true$}
{%
  $\setinsert{\requested}{x}$
}

}{%

\externalaction{$\response(x, v)$}
{%
  $x \in \requested - \responded \land v = val(x)$}
{%
  $\setinsert{\responded}{x}$
}

}

\end{actionlist}

\horline
\caption{Specification of a simple database system}
\label{fig:simple-DB-spec}
\efg

\begin{figure}[htb]

\horline
\automatontitle{\dbimp}

\begin{signature}
\eonly{%
	$\request(x)$, where $x$ is a query\\
	$\response(x,v)$, where $x$ is a query and $v$ is a value
}
\end{signature}

\begin{statevarlist}

\item $\requested$, a set of received queries, initially empty
\item $\responded$, a set of computed responses to queries, initially empty

\end{statevarlist}

\begin{actionlist}

\iocode{%

\externalaction{$\request(x)$}
{%
  $\true$}
{%
  $(\setinsert{\requested}{x}) \ \choice\ skip$
}

}{%

\externalaction{$\response(x, v)$}
{%
  $x \in \requested - \responded \land v = val(x)$}
{%
  $\setinsert{\responded}{x}$
}

}

\end{actionlist}

\horline
\caption{First level refinement of the specification of a simple database system}
\label{fig:simple-DB-lossy-impl}
\efg

This example demonstrates that the simulation relations of 
Section~\ref{sec:simulation-relations} do not imply 
live trace inclusion. The problem is that these simulation relations do not
reference the liveness conditions of the concrete and abstract automata.
To remedy this, we augment the simulation relations so that every pair $q$ in the
abstract liveness condition $M$ is related to a pair $p$ in the
concrete liveness condition $L$. The idea is that the simulation
relation relates occurrences of states in $q.\R$, $q.\G$ in
transitions of the abstract automaton $(B,M)$ with occurrences of
states in $p.\R$, $p.\G$ in transitions of the concrete automaton
$(A,L)$. The relationship is defined so that the augmented simulation
implies that, in ``corresponding'' executions $\al$ of $(A,L)$, $\al'$
of $(B,M)$, if $\al$ satisfies $p$, then $\al'$ must satisfy $q$.

In more detail, an occurrence of a $q.\R$ state in an
abstract (live) execution $\al'$ must be matched by at least one $p.\R$
state in the corresponding concrete (live) execution $\al$, and an occurrence of
a $p.\G$ state in $\al$ must be matched by at least one $q.\G$ state in
$\al'$. Thus, 
if $\al' \sat \iof q.\R$, then $\al \sat \iof p.\R$, and 
if $\al \sat \iof p.\G$, then $\al' \sat \iof q.\G$.
Assuming $\al$ is live,
we get $\al \sat \iof p.\R \imp \iof p.\G$. This and the previous two 
implications yields
$\al' \sat \iof q.\R \imp \iof q.\G$. Hence $\al'$ is live.
Hence we can show that if an abstract execution $\al'$ and concrete
execution $\al$ correspond (according to the simulation), and $\al$ is live,
then $\al'$ is also live.  
The matching thus allows us to show that every live execution of $(A,L)$ has a
``corresponding'' live execution in $(B,M)$. Live trace inclusion follows
immediately.

Since the semantic closure $\clos{L}$ of $L$ specifies the same set of live
executions (Proposition~\ref{prop:execs-clos}), as $L$ does,
we can relax the requirement $p \in L$ to $p \in \clos{L}$.
Since $\clos{L}$ is in general a superset of $L$, this can be very helpful in
refining the abstract liveness condition. In particular, it enables us to split
the refinement task into two subtasks: refinement across abstraction levels
(which we address in this section) and refinement within an abstraction level
(which we address in the next section).

Let $(A,L)$ be a live automaton, $\al$ be a finite execution fragment
of $A$, and $p \in L$.  We abuse notation and write $\al \in p.\R$ iff
there exists a state $s$ along $\al$ such that $s \in p.\R$. $\al \in
p.\G$ is defined similarly.
The above considerations lead to the following definitions of
liveness-preserving simulation relations.

\bd[Liveness-preserving Forward Simulation w.r.t. Invariants]
\label{def:live-fwd-sim-inv}
Let $(A,L)$ and $(B,M)$ be live automata with the same external actions.
Let $I_A$, $I_B$ be invariants of $A$, $B$ respectively.
Let $f = (g,h)$ where $g \subseteq \states(A) \times \states(B)$
and $h: M \mapsto \clos{L}$ is a total mapping over $M$\footnote{That is,
$h(q)$ is defined for all $q \in M$.}.
Then $f$ is a \intrdef{liveness-preserving forward simulation} from $(A,L)$ to
$(B,M)$ with respect to $I_A$ and $I_B$ iff:

\bn

\item \label{clause:live-fwd-sim-inv:init} If $s \in \start(A)$, then
$g[s] \ints \start(B) \neq \emptyset$.

\item \label{clause:live-fwd-sim-inv:trans}
If $s \lla{a}{A} s'$, $s \in I_A$, and $u \in g[s] \ints I_B$, then
there exists a finite execution fragment
$\al$ of $B$ such that
$\fstate(\al) = u$, $\lstate(\al) \in g[s']$, and $\trace(\al) = \trace(a)$.
Furthermore, for all $q \in M$,
   \bn

   \item \label{clause:live-fwd-sim-inv:RED}
	 if $\al \in q.\R$ then	$s \in p.\R$ or $s' \in p.\R$, and

   \item \label{clause:live-fwd-sim-inv:GREEN}
	 if $s \in p.\G$ or $s' \in p.\G$ then $\al \in q.\G$,

   \en
where $p = h(q)$.

\item \label{clause:live-fwd-sim-inv:silent}
Call a transition $s \lla{a}{A} s'$ \intrdef{always-silent} iff
$s \in I_A$ and for every finite execution fragment
$\al$ of $B$ 
such that $\fstate(\al) \in g[s] \ints I_B$, $\lstate(\al) \in g[s']$, and
$\trace(\al) = \trace(a)$, we have $|\al| = 0$, i.e., $\al$ consists
of a single state. In other
words, the transition $s \lla{a}{A} s'$ is matched only by the empty
transition in $B$. Then, $g$ is such that every live execution of $(A,L)$ contains
an infinite number of transitions that are not always-silent.

\en
\ed

Clause~\ref{clause:live-fwd-sim-inv:init} is the usual condition of
a forward simulation requiring that every start state of $(A,L)$
be related to at least one start state of $(B,M)$.

Clause~\ref{clause:live-fwd-sim-inv:trans} is the condition of a
forward simulation which requires that every transition $s \lla{a}{A} s'$ of $(A,L)$ be
``simulated'' by an execution fragment $\al$ of $(B,M)$ which has the same trace.
We also require that every complemented-pair $q \in M$ is matched to a
complemented-pair $p \in \clos{L}$ by the mapping $h$ and that such
corresponding pairs impose a constraint on the transition $s \lla{a}{A} s'$ of $(A,L)$
and the simulating execution fragment $\al$ of $(B,M)$, as follows.
If $\al$ contains some $q.\R$ state, then at least one of $s, s'$ is a $p.\R$ state, and
if at least one of $s, s'$ is a $p.\G$ state, then $\al$ contains some $q.\G$ state.
This requirement thus enforces the matching discussed at the beginning
of this section, from which live trace inclusion follows.

Clause~\ref{clause:live-fwd-sim-inv:silent} is needed to ensure that a
live execution of $(A,L)$ has at least one corresponding \emph{infinite}
execution in $(B,M)$. This execution can then be shown, using
clause~\ref{clause:live-fwd-sim-inv:trans}, to be live (see
Lemma~\ref{lem:liveness} below). If $s \lla{a}{A} s'$ is always-silent, then
$a$ must be an internal action. Thus, in practice,
clause~\ref{clause:live-fwd-sim-inv:silent} holds, since executions
with an (infinite) suffix consisting solely of internal actions are
not usually considered to be live.
Clause~\ref{clause:live-fwd-sim-inv:silent} can itself be expressed as a
complemented-pair (which is added to $L$).  Call an action $a$ of $A$
\intrdef{non-always-silent} iff no transition arising from its execution is
always-silent. Thus, every transition arising from the execution of $a$ can be
matched with respect to $g$ by some nonempty execution fragment of $B$. It is also
possible that the transition can be matched by the empty fragment, but what is
important is that it is always possible to \emph{choose} a nonempty fragment to
match with. This means that we can always match a live execution $\al$ of $(A,L)$ with
some \emph{infinite} execution of $(B,M)$, by always matching the non-always-silent
transitions in $\al$ with nonempty execution fragments of $(B,M)$.

By definition, any external action of $A$ is non-always-silent. An internal
action of $A$ may or may not be non-always-silent.
We introduce an auxiliary boolean variable $\nonalwayssilent$
that is set to $\true$ each time a non-always-silent action of $A$ is executed, and is
set to $\false$ infinitely often by a new internal action of $A$
whose precondition is $\true$ and whose effect is $\nonalwayssilent := \false$
(every execution of this new action can be simulated by the empty
transition in $B$, since $\nonalwayssilent$ has no effect on any other state
component of $A$, nor on the execution of other actions in $A$).
Then the pair $\pair{\true}{\nonalwayssilent}$ expresses that a non-always-silent
action of $A$ is executed infinitely often, which implies that each
live execution of $(A,L)$ contains an infinite number of non-always-silent transitions.
The pair $\pair{\true}{\nonalwayssilent}$ can then be refined at the next
lower level of abstraction in exactly the same way as all the other pairs in $L$.
See Section~\ref{sec:example} for an example of this technique.

It is clear from the definitions that if $(g,h)$ is a
liveness-preserving forward simulation from $(A,L)$ to $(B,M)$
w.r.t. invariants, then $g$ is a forward simulation from $A$ to
$B$ w.r.t. the same invariants.  We write $(A,L) \lpfs (B,M)$ if there
exists a liveness-preserving forward simulation from $(A,L)$ to
$(B,M)$ w.r.t. invariants, and $(A,L) \lpfs (B,M)$ via $f$ if $f$
is a liveness-preserving forward simulation from $(A,L)$ to $(B,M)$
w.r.t. invariants.

\bd[Liveness-preserving Refinement Mapping w.r.t. Invariants]
\label{def:live-ref-inv}
Let $(A,L)$ and $(B,M)$ be live automata with the same external actions.
Let $I_A$, $I_B$ be invariants of $A$, $B$, respectively. 
Let $r = (g,h)$ where $g : \states(A) \mapsto \states(B)$
and $h: M \mapsto \clos{L}$ is a total mapping over $M$.
Then $r$ is a \intrdef{liveness-preserving refinement mapping} from $(A,L)$ to
$(B,M)$ with respect to $I_A$ and $I_B$ iff:

\bn

\item \label{clause:live-ref-inv:init} If $s \in \start(A)$, then
$g(s) \in \start(B)$.

\item \label{clause:live-ref-inv:trans}
If $s \lla{a}{A} s'$, $s \in I_A$, and $g(s) \in I_B$, then
there exists a finite execution fragment $\al$ of $B$ such that
$\fstate(\al) = g(s)$, 
$\lstate(\al) = g(s')$, and $\trace(\al) = \trace(a)$.
Furthermore, for all $q \in M$,
   \bn

   \item \label{clause:live-ref-inv:RED}
	 if $\al \in q.\R$ then	$s \in p.\R$ or $s' \in p.\R$, and

   \item \label{clause:live-ref-inv:GREEN}
	 if $s \in p.\G$ or $s' \in p.\G$ then $\al \in q.\G$,

   \en
where $p = h(q)$.

\item \label{clause:live-ref-inv:silent}
Call a transition $s \lla{a}{A} s'$ \intrdef{always-silent} iff
$s \in I_A$ and for every
finite execution fragment
$\al$ of $B$ such that
$\fstate(\al) = g(s)$, 
$\lstate(\al) = g(s')$, and $\trace(\al) = \trace(a)$,
we have $|\al| = 0$, i.e., $\al$ consists of a single state. In other
words, the transition $s \lla{a}{A} s'$ is matched only by the empty
transition in $B$. Then, $g$ is such that every live execution of $(A,L)$ contains
an infinite number of transitions that are not always-silent.

\en
\ed
We write $A \lprm B$ if there exists a liveness-preserving
refinement mapping from $A$ to $B$ w.r.t. invariants, and 
$A \lprm B$ via $r$ if $r$ is a liveness-preserving refinement
mapping from $A$ to $B$ w.r.t. invariants.
It is clear from the definitions that a liveness-preserving refinement mapping
is a special case of a liveness-preserving forward simulation.
Furthermore, if $(g,h)$ is a
liveness-preserving refinement mapping from $(A,L)$ to $(B,M)$ w.r.t. some invariants,
then $g$ is a refinement mapping from $A$ to $B$ w.r.t. the same invariants.

\bd[Liveness-preserving Backward Simulation w.r.t. Invariants]
\label{def:live-back-sim-inv}
Let $(A,L)$ and $(B,M)$ be live automata with the same external actions.
Let $I_A$, $I_B$ be invariants of $A$, $B$ respectively.
Let $b = (g,h)$ where $g \subseteq \states(A) \times \states(B)$
and $h: M \mapsto \clos{L}$ is a total mapping over $M$.
Then $b$ is a \intrdef{liveness-preserving backward simulation} from $(A,L)$ to
$(B,M)$ with respect to $I_A$ and $I_B$ iff:

\bn

\item \label{clause:live-back-sim-inv:nonempty} 
If $s \in I_A$, then $g[s] \ints I_B \ne \emptyset$.

\item \label{clause:live-back-sim-inv:init} 
If $s \in \start(A)$, then $g[s] \ints I_B \sub \start(B)$.

\item \label{clause:live-back-sim-inv:trans}
If $s \lla{a}{A} s'$, $s \in I_A$, and $u' \in g[s'] \ints I_B$, then
there exists a finite execution fragment $\al$ of $B$ such that
$\fstate(\al) \in g[s] \ints I_B$, 
$\lstate(\al) = u'$, and
$\trace(\al) = \trace(a)$.
Furthermore, for all $q \in M$,
   \bn

   \item \label{clause:live-back-sim-inv:RED}
	 if $\al \in q.\R$ then	$s \in p.\R$ or $s' \in p.\R$, and

   \item \label{clause:live-back-sim-inv:GREEN}
	 if $s \in p.\G$ or $s' \in p.\G$ then $\al \in q.\G$,

   \en
where $p = h(q)$.

\item \label{clause:live-back-sim-inv:silent}
Call a transition $s \lla{a}{A} s'$ \intrdef{sometimes-silent} iff $s \in I_A$ and for
some finite execution fragment
$\al$ of $B$ such that
$\fstate(\al) \in g[s] \ints I_B$, 
$\lstate(\al) \in g[s']$, and $\trace(\al) = \trace(a)$,
we have $|\al| = 0$, i.e., $\al$ consists of a single state. In other
words, the transition $s \lla{a}{A} s'$ can be matched by the empty
transition in $B$. Then, $g$ is such that every live execution of $(A,L)$ contains
an infinite number of transitions that are not sometimes-silent.

\en
\ed
Clauses~\ref{clause:live-back-sim-inv:nonempty} and
\ref{clause:live-back-sim-inv:init} are the usual conditions of a backward
simulation requiring that a state in the invariant $I_A$ of $(A,L)$ is related
to at least one state in the invariant $I_B$ of $(B,M)$, and that every start
state of $(A,L)$ is related only to start states of $(B,M)$, ignoring states not
in the invariant $I_B$. These clauses are needed due to the ``backwards'' nature
of the bisimulation, since, from a state $u'$ in the
invariant $I_B$ it is possible, when ``going backwards'' along a transition to
reach a state $u$ not in the invariant, i.e., $u \lla{a}{B} u'$, $u \notin I_B$,
and $u' \in I_B$ is possible.  Also, a start state in $(A,L)$ must \emph{always}
be matched by a start state in $(B,M)$, since the matching state in $(B,M)$
cannot be chosen initially: it is constrained by the succeeding transitions,
i.e., it is ``chosen'' last of all, and so the result must be an initial state
of $B$ regardless of the choice.

Clause~\ref{clause:live-back-sim-inv:trans} is the condition of a backward
simulation which requires that every transition $s \lla{a}{A} s'$ of $(A,L)$ be
``simulated'' by an execution fragment $\al$ of $(B,M)$, except that we also
require that every complemented-pair $q \in M$ is matched to a complemented-pair
$p \in \clos{L}$ by the mapping $h$ and that such corresponding pairs
impose a constraint on the transition $s \lla{a}{A} s'$ of $(A,L)$ and the
simulating execution fragment $\al$ of $(B,M)$, as follows.  If $\al$ contains
some $q.\R$ state, then at least one of $s, s'$ is a $p.\R$ state, and if 
at least one of $s, s'$
is a $p.\G$ state, then $\al$ contains some $q.\G$ state.  This requirement thus
enforces the matching discussed at the beginning of this section, from which
live trace inclusion follows.

Clause~\ref{clause:live-back-sim-inv:silent} is needed to ensure that a
live execution of $(A,L)$ has at least one corresponding \emph{infinite}
execution in $(B,M)$. This execution can then be shown, using
clause~\ref{clause:live-back-sim-inv:trans}, to be live (see
Lemma~\ref{lem:liveness} below). If $s \lla{a}{A} s'$ is sometimes-silent, then
$a$ must be an internal action. Thus, in practice,
clause~\ref{clause:live-back-sim-inv:silent} holds, since executions
with an (infinite) suffix consisting solely of internal actions are
not usually considered to be live.
Clause~\ref{clause:live-back-sim-inv:silent} can itself be expressed
as a complemented-pair (which is added to $L$), 
which can then be refined at the next lower level of abstraction.
Call an action $a$ of $A$
\intrdef{non-sometimes-silent} iff no transition arising from its execution is
sometimes-silent. Thus, every transition arising from the execution of $a$ must
\emph{always} be 
matched with respect to $g$ by some nonempty execution fragment of $B$.

We can now express the requirement that a non-sometimes-silent
action of $A$ is executed infinitely often, as a complemented-pair, and refine
this pair at the next lower level of abstraction.
The details are similar to those discussed above for
Clause~\ref{clause:live-fwd-sim-inv:silent} of
Definition~\ref{def:live-fwd-sim-inv}, and are omitted.
Note the difference with forward simulation; there, we only had to ensure that 
it was infinitely often possible to \emph{choose} a nonempty execution fragment
to match with. 
With backward simulations, we have to show that infinitely often, 
\emph{all} the matching execution fragments are nonempty.

It is clear from the definitions that if $(g,h)$ is a
liveness-preserving backward simulation from $(A,L)$ to $(B,M)$ w.r.t. some invariants,
then $g$ is a backward simulation from $A$ to $B$ w.r.t. the same invariants.
We write $A \lpbs B$ if there exists a liveness-preserving
backward simulation from $A$ to $B$ w.r.t. some invariants, and 
$A \lpbs B$ via $b$ if $b$ is a liveness-preserving backward
simulation from $A$ to $B$ w.r.t. some invariants.
If the backward simulation $g$ is
image-finite, then we write $A \ilpbs B$, $A \ilpbs B$ via $b$, respectively.

\bd[Liveness-preserving History Relation w.r.t. Invariants]
\label{def:live-hist-inv}
Let $(A,L)$ and \linebreak $(B,M)$ be live automata with the same external actions.
Let $I_A$, $I_B$ be invariants of $A$, $B$, respectively. A \intrdef{history relation}
from $A$ to $B$ with respect to $I_A$ and $I_B$
is a relation $hs$ over $\states(A) \times \states(B)$
that satisfies:
\bn

\item \label{clause:live-hist-inv:init}
$hs$ is a liveness-preserving forward simulation from $A$ to $B$
w.r.t. $I_A$ and $I_B$, and

\item \label{clause:live-hist-inv:trans}
$hs^{-1}$ is a refinement from $B$ to $A$ w.r.t. $I_B$ and $I_A$.

\en
\ed
We write $A \lphr B$ if there exists a liveness-preserving
history relation from $A$ to $B$ w.r.t. some invariants, and 
$A \lphr B$ via $h$ if $h$ is a liveness-preserving history relation
from $A$ to $B$ w.r.t. some invariants.

\bd[Liveness-preserving Prophecy Relation w.r.t. Invariants]
\label{def:live-proph-inv}
Let $(A,L)$ and $(B,M)$ be live automata with the same external actions.
Let $I_A$, $I_B$ be invariants of $A$, $B$, respectively. A \intrdef{prophecy relation}
from $A$ to $B$ with respect to $I_A$ and $I_B$
is a relation $p$ over $\states(A) \times \states(B)$
that satisfies:
\bn

\item \label{clause:live-proph-inv:init}
$p$ is a liveness-preserving backward simulation from $A$ to $B$
w.r.t. $I_A$ and $I_B$, and

\item \label{clause:live-proph-inv:trans}
$p^{-1}$ is a refinement from $B$ to $A$ w.r.t. $I_B$ and $I_A$.

\en
\ed
We write $A \lppr B$ if there exists a liveness-preserving
prophecy relation from $A$ to $B$ w.r.t. some invariants, and 
$A \lppr B$ via $p$ if $p$ is a liveness-preserving prophecy
relation from $A$ to $B$ w.r.t. some invariants.  If the liveness-preserving prophecy
relation is image-finite, then we write $A \ilppr B$,
$A \ilppr B$ via $p$, respectively.

We use $\l F, \l R, i \l B, \l H, i \l P$ to denote 
liveness-preserving forward simulation,
liveness-preserving refinement mapping,
image-finite liveness-preserving backward simulation,
liveness-preserving history relation,
image-finite liveness-preserving prophecy relation,
respectively. 
Thus, when we write $X \in \{\l F, \l R, i \l B, \l H, i \l P\}$, we mean that $X$ is
one of these relations.

Liveness-preserving simulation relations induce a correspondence
between the live executions of the concrete and the abstract automata. This
correspondence is captured by the notion of $R^\l$-relation.
We remind the reader of the definition
$\trace(\al',j,k) = \trace(b_j \cdots b_k)$ if $j \le k$, and
$= \lambda$ (the empty sequence) if $j > k$.

\bd[$R^\l$-relation and Live Index Mappings] Let $(A,L)$ and $(B,M)$ be
live automata with the same external actions. Let $R^\l = (R,H)$ where 
$R$ is a relation over $\states(A) \times \states(B)$ and
$H: M \mapsto \clos{L}$ is a total mapping over $M$.
Furthermore, let $\alpha$ and
$\alpha'$ be executions of $(A,L)$ and $(B,M)$, respectively: \ms\\
\ind $\alpha  = s_0 a_1 s_1 a_2 s_2 \cdots$ \\
\ind $\alpha' = u_0 b_1 u_1 b_2 u_2 \cdots$ \ms\\
Say that $\alpha$ and $\alpha'$ are \intrdef{$R^\l$-related}, written
$(\alpha, \alpha') \in R^\l$, if there exists a total, nondecreasing mapping
$m: \{0,1,\ldots,|\alpha|\} \mapsto \{0,1,\ldots,|\alpha'|\}$
such that:
\bn
\item \label{clause:live-index-mapping:init} $m(0) = 0$,

\item \label{clause:live-index-mapping:corr}
      $(s_i, u_{m(i)}) \in R$ for all $i$, $0 \leq i \leq |\alpha|$,

\item \label{clause:live-index-mapping:trace}
      $\trace(\al', {m(i-1)+1}, {m(i)}) = \trace(a_i)$ for all 
               $i$, $0 < i \leq |\alpha|$,

\item \label{clause:live-index-mapping:cofinal}
      for all $j, 0 \leq j \leq |\alpha'|$, there exists an $i$, 
                  $0 \leq i \leq |\alpha|$, such that $m(i) \geq j$, and

\item \label{clause:live-index-mapping:pairs} for all complemented-pairs
$q \in M$ and all $i$, $0 < i \leq |\alpha|:$
   \bn
   \item \label{clause:live-index-mapping:RED}
         if $(\ex j \in m(i-1) \ldots m(i) : u_j \in q.\R)$ then
		$s_{i-1} \in p.\R$ or $s_i \in p.\R$, and

   \item \label{clause:live-index-mapping:GREEN}
         if $s_{i-1} \in p.\G$ or $s_i \in p.\G$ then
		$(\ex j \in m(i-1) \ldots m(i) : u_j \in q.\G)$,
   \en
where $p = H(q)$.
\en
The mapping $m$ is referred to as a \intrdef{live index mapping} from
$\alpha$ to $\alpha'$ with respect to $R^\l$.  Write $((A,L),(B,M)) \in
R^\l$ if for every live execution $\alpha$ of $(A,L)$, there exists a
live execution $\alpha'$ of $(B,M)$ such that $(\alpha,\alpha') \in
R^\l$.
\label{def:live-index-mapping}
\ed

Note that $(\al,\al') \in R^\l$ does not require $\al, \al'$ to be live executions.
By Definitions~\ref{def:index-mapping} and
\ref{def:live-index-mapping}, it is clear that, if $R^\l = (R,H)$,
then $(\al, \al') \in R^\l$ implies $(\al, \al') \in R$.
The following lemma establishes a correspondence between the prefixes
of a live execution of the concrete automaton and an infinite family
of finite executions of the abstract automaton.

\bl
\label{lem:live-exec-forward}
Let $(A,L)$ and $(B,M)$ be live automata with the same external
actions, and such that  $(A,L) \lpfs (B,M)$ via $f$ for some $f = (g,h)$.
Let $\al$ be an arbitrary live execution of $(A,L)$. Then there exists
a collection $(\al_i', m_i)_{0 \leq i}$ of finite
executions of $(B,M)$ and mappings such that:
\bn
   \item \label{lem:live-exec-forward:mapping}
   $m_i$ is a live index mapping from $\al |_i$ to $\al'_i$ with
   respect to $f$, for all $i \geq 0$, and

   \item \label{lem:live-exec-forward:prefix}
   $\al'_{i-1} \leq \al'_i$ and $m_{i-1} = m_i \up
   \{0,\ldots,i-1\}$ for all $i > 0$, and

   \item \label{lem:live-exec-forward:infinite}
	 $\al'_{i-1} < \al'_i$ for infinitely many $i > 0$.
\en
\el
\bpr
Let $\al = s_0 a_1 s_1 a_2 s_2 \ldots$ and let $I_A$, $I_B$ be
invariants of $A$, $B$, respectively, such that $f$ is a
liveness-preserving forward simulation from $(A,L)$ to $(B,M)$ with
respect to $I_A$ and $I_B$.
We construct $\al'_i$ and $m_i$ by induction on $i$.

Since $s_0 \in \start(A)$, we have $(s_0, v_0) \in g$ and $v_0 \in
\start(B)$ for some state $v_0$, by
Definition~\ref{def:live-fwd-sim-inv},
clause~\ref{clause:live-fwd-sim-inv:init}.
Let $\al'_0 = v_0$ and let $m_0$ be the mapping that maps $0$ to
$0$. Then, $m_0$ is a live index mapping from $\al|_0$ to $\al'_0$ with
respect to $f$ (in particular,
clause~\ref{clause:live-index-mapping:pairs} of
Definition~\ref{def:live-index-mapping} holds vacuously, 
since $| \al |_0 | = 0$).

Now inductively assume that $m_{i-1}$ (for $i > 0$) is a
live index mapping from $\al|_{i-1}$ to $\al'_{i-1}$ with respect to
$f$. Let $u_0 = \lstate(\al'_{i-1})$. Then, by
clause~\ref{clause:live-index-mapping:cofinal} of
Definition~\ref{def:live-index-mapping} and the fact that $m_{i-1}$ is
nondecreasing, we have $m_{i-1}(i-1) = |\al'_{i-1}|$ and $(s_{i-1},
u_0) \in g$. Since $s_{i-1}$, $s_i$, and $u_0$ are reachable, by
definition, they satisfy their respective invariants. Hence, by
Definition~\ref{def:live-fwd-sim-inv},
clause~\ref{clause:live-fwd-sim-inv:trans}, there exists a
finite execution fragment
$u_0 \lla{b_1}{B} u_1 \lla{b_2}{B} \cdots \lla{b_{n}}{B} u_n$
of $B$ such that $u_n \in g[s_i]$, $\trace(b_1 \cdots b_{n}) = \trace(a_i)$, and 
for all complemented-pairs $q \in M$:
   \bn
   \item if $(\ex j \in 1 \ldots n : u_j \in q.\R)$ then
		$s_{i-1} \in p.\R$ or $s_{i} \in p.\R$, and

   \item if $s_{i-1} \in p.\G$ or $s_{i} \in p.\G$ then
		$(\ex j \in 1 \ldots n : u_j \in q.\G)$,
   \en
where $p = h(q)$.
Now define
$\al'_i = \al'_{i-1} \cat (u_0 \lla{b_1}{B} u_1 \lla{b_2}{B} \cdots \lla{b_{n}}{B} u_n)$,
and define $m_i$ to be the mapping such that $m_i(j) = m_{i-1}(j)$ for
all $j$, $0 \leq j \leq i-1$, and $m_i(i) = |\al'_i|$. We argue that $m_i$ is a
live index mapping from $\al|_i$ to $\al'_i$ with respect to
$f$, i.e., that all clauses of Definition~\ref{def:live-index-mapping}
hold.
Clause~\ref{clause:live-index-mapping:init} holds since $m_i(0) =
m_{i-1}(0)$ by definition, and $m_{i-1}(0) = 0$ by the inductive hypothesis.
Clause~\ref{clause:live-index-mapping:corr} holds by the inductive
hypothesis and $u_n \in g[s_i]$.
Clause~\ref{clause:live-index-mapping:trace} holds by the inductive
hypothesis and $\trace(b_1 \cdots b_{n}) = \trace(a_i)$.
Clause~\ref{clause:live-index-mapping:cofinal} holds since $m_i(|\,\al|_i\,|)
= m_i(i) = |\al'_i|$, by definition.
Finally, clause~\ref{clause:live-index-mapping:pairs} holds by the inductive
hypothesis and the conditions for all complemented-pairs $q \in M$
just established above w.r.t. $s_{i-1} \lla{a_i}{A} s_i$ and 
$u_0 \lla{b_1}{B} u_1 \lla{b_2}{B} \cdots \lla{b_{n}}{B} u_n$.
Having established that $m_i$ is a live index mapping from $\al|_i$ to
$\al'_i$ with respect to $f$, we conclude that
clause~\ref{lem:live-exec-forward:mapping} of the lemma holds.

Clause~\ref{lem:live-exec-forward:prefix} of the lemma holds by
construction of $\al'_i$ and $m_i$, since $\al'_i$ and $m_i$ are obtained by
extending $\al'_{i-1}$ and $m_{i-1}$, respectively.

By Definition~\ref{def:live-fwd-sim-inv},
clause~\ref{clause:live-fwd-sim-inv:silent}, for infinitely many $i > 0$,
we can select the execution fragment
$u_0 \lla{b_1}{B} u_1 \lla{b_2}{B} \cdots \lla{b_{n}}{B} u_n$
that matches $s_{i-1} \lla{a_i}{A} s_i$ 
so that $n > 0$. Hence, for infinitely many $i > 0$, we have
$\al'_{i-1} < \al'_i$.
Thus, clause~\ref{lem:live-exec-forward:infinite} of the lemma holds.
\epr

\bd[Induced Digraph]
\label{def:induced-digraph}
Let $(A,L)$ and $(B,M)$ be live automata with the same external
actions and assume
$A \ilpbs B$ via $b = (g,h)$ with respect to invariants $I_A$ and $I_B$. 
For any execution $\al = s_0 a_1 s_1 a_2 s_2 \ldots$ of $A$, let the 
\intrdef{digraph induced} by $\al$, $b$, $I_B$, $L$, and $M$ be the directed graph $G$
given as follows:
\bn

\item \label{def:induced-digraph:nodes}
The nodes of $G$ are the ordered pairs $(u,i)$ such that $0 \le i \le |\al|$,
      and $u \in g[s_i] \ints I_B$, and

\item \label{def:induced-digraph:edges}
there is an edge from $(u,i)$ to $(u',i')$ iff $i' = i+1$ and there exists
      a finite execution fragment $\al'$ of $B$ such that $\fstate(\al') = u$,
      $\lstate(\al') = u'$, $\trace(\al') = \trace(a_{i+1})$, and for all
      complemented-pairs $q \in M$:
      \bn
      \item if $\al' \in q.\R$ then
		$s_{i} \in p.\R$ or $s_{i+1} \in p.\R$, and

      \item if $s_{i} \in p.\G$ or $s_{i+1} \in p.\G$ then
		$\al' \in q.\G$,
      \en
      where $p = h(q)$.

\en
\ed

\bl
\label{lem:induced-digraph}
Let $(A,L)$ and $(B,M)$ be live automata with the same external
actions and assume
$A \ilpbs B$ via $b$ with respect to invariants $I_A$ and $I_B$. 
Let $\al$ be any execution of $A$. Then the digraph $G$ induced
by $\al$, $b$, $I_B$, $L$, and $M$ satisfies:
\bn

\item \label{lem:induced-digraph:nodes}
For each $i$, $0 \le i \le |\al|$, there is at least one node in $G$ of the form
      $(u,i)$.

\item \label{lem:induced-digraph:roots}
The roots of $G$ are exactly the nodes of the form $(u,0)$.

\item \label{lem:induced-digraph:finite-roots}
$G$ has a finite number of roots.

\item \label{lem:induced-digraph:outdegree}
Each node in $G$ has finite outdegree.

\item \label{lem:induced-digraph:reachable}
Each node of $G$ is reachable from some root of $G$.

\en
\el
\bpr
Let $b = (g,h)$. Then $g$ is an image-finite backward simulation from $A$ to $B$.
We deal with each clause in turn.
\bn

\item Each state $s_i$ of $\al$ is reachable, and so belongs to $I_A$.
Hence $g[s_i] \ints I_B \ne \emptyset$ by
Clause~\ref{clause:live-back-sim-inv:nonempty} of
Definition~\ref{def:live-back-sim-inv}.  
Hence by Definition~\ref{def:induced-digraph}, 
clause~\ref{def:induced-digraph:nodes}, there exist nodes of $G$ of
the form $(u,i)$. 

\item Every node $(u,0)$ is a root of $G$ (i.e., it has no incoming
edges). We now show that any node $(u,i)$ with $i > 0$ cannot be a
root. Now $u \in g[s_i] \ints I_B$ by 
Definition~\ref{def:induced-digraph}, 
clause~\ref{def:induced-digraph:nodes}. Also,
$s_{i-1} \in I_A$ and $s_{i-1} \lla{a_i}{A} s_i$ by assumption, hence
by Definition~\ref{def:live-back-sim-inv},
clause~\ref{clause:live-back-sim-inv:trans}, 
there exists a finite execution fragment $\al'$ of $B$ such that
$\fstate(\al') \in g[s_{i-1}] \ints I_B$, 
$\lstate(\al') = u$, 
$\trace(\al') = \trace(a_i)$, and, for all $q \in M$,
   \bn

   \item if $\al' \in q.\R$ then
		$s_{i-1} \in p.\R$ or $s_i \in p.\R$, and

   \item if $s_{i-1} \in p.\G$ or $s_i \in p.\G$ then
		$\al' \in q.\G$,

   \en
where $p = h(q)$.
Hence, by Definition~\ref{def:induced-digraph},
clause~\ref{def:induced-digraph:edges}, there exists an edge in $G$ from 
$(\fstate(\al'), i-1)$ to $(u,i)$.

\item Since $g$ is image-finite, the set $g[s_0] \ints I_B$ is finite.
By Definition~\ref{def:induced-digraph},
clause~\ref{def:induced-digraph:nodes}, all nodes of $G$ of the form $(u,0)$
must satisfy $u \in g[s_0] \ints I_B$. Hence, there are a finite
number of such nodes. By clause 2 of the lemma (which has already been
established), these nodes are exactly the roots of $G$. Hence, the
number of roots is finite.

\item Let $(u,i)$ be an arbitrary node of $G$.
By Definition~\ref{def:induced-digraph},
clause~\ref{def:induced-digraph:edges},
from any node of the form $(u,i)$, all outgoing edges are to
nodes of the form $(u',i+1)$. 
Since $g$ is image-finite, the set $g[s_{i+1}] \ints I_B$ is finite.
By Definition~\ref{def:induced-digraph},
clause~\ref{def:induced-digraph:nodes}, all nodes of $G$ of the form $(u,i+1)$
must satisfy $u \in g[s_{i+1}] \ints I_B$. Hence, there are a finite
number of such nodes. Hence, the outdegree of any node of $G$ of the form 
$(u,i)$ is finite. Since $(u,i)$ was chosen arbitrarily, the result follows.

\item We establish this by induction on the second component $i$ of
the nodes $(u,i)$ of $G$.
For the base case, $i=0$ and nodes $(u,0)$ are reachable by definition
since they are roots.
Assume the induction hypothesis that all nodes of the form $(u,i)$ are
reachable from some root of $G$, and consider an arbitrary node of the
form $(u,i+1)$.

Now $u \in g[s_{i+1}] \ints I_B$ by Definition~\ref{def:induced-digraph}, 
clause~\ref{def:induced-digraph:nodes}. Also,
$s_{i} \in I_A$ and $s_{i} \lla{a_{i+1}}{A} s_{i+1}$ by assumption, hence
by Definition~\ref{def:live-back-sim-inv},
clause~\ref{clause:live-back-sim-inv:trans}, 
there exists a finite execution fragment $\al'$ of $B$ such that
$\fstate(\al') \in g[s_{i}] \ints I_B$, 
$\lstate(\al') = u$, 
$\trace(\al') = \trace(a_{i+1})$, and, for all $q \in M$,
   \bn

   \item if $\al' \in q.\R$ then
		$s_{i} \in p.\R$ or $s_{i+1} \in p.\R$, and

   \item if $s_{i} \in p.\G$ or $s_{i+1} \in p.\G$ then
		$\al' \in q.\G$,

   \en
where $p = h(q)$.
Hence, by Definition~\ref{def:induced-digraph},
clause~\ref{def:induced-digraph:edges}, there exists an edge in $G$ from 
$(\fstate(\al'), i)$ to $(u,i+1)$.
By the induction hypothesis, 
$(\fstate(\al'), i)$ is reachable. Hence, so is $(u,i+1)$.
\en

Since all the clauses are established,
Lemma~\ref{lem:induced-digraph} holds.
\epr

\bl
\label{lem:live-exec-backward}
Let $(A,L)$ and $(B,M)$ be live automata with the same external
actions, and such that  $(A,L) \ilpbs (B,M)$ via $b$ for some $b = (g,h)$.
Let $\al$ be an arbitrary live execution of $(A,L)$. Then there exists
a collection $(\al_i', m_i)_{0 \leq i}$ of finite
executions of $(B,M)$ and mappings such that:
\bn
   \item \label{lem:live-exec-backward:mapping}
   $m_i$ is a live index mapping from $\al |_i$ to $\al'_i$ with
   respect to $b$, for all $i \geq 0$, and

   \item \label{lem:live-exec-backward:prefix}
   $\al'_{i-1} \leq \al'_i$ and $m_{i-1} = m_i \up
   \{0,\ldots,i-1\}$ for all $i > 0$, and

   \item \label{lem:live-exec-backward:infinite}
	 $\al'_{i-1} < \al'_i$ for infinitely many $i > 0$.
\en
\el
\bpr
Let $\al = s_0 a_1 s_1 a_2 s_2 \ldots$ and let $I_A$, $I_B$ be
invariants of $A$, $B$, respectively, such that $b$ is a image-finite
liveness-preserving backward simulation from $(A,L)$ to $(B,M)$ with
respect to $I_A$ and $I_B$.  Let $G$ be the digraph induced by $\al$,
$b$, $I_B$, $L$ and $M$. Since $\al$ is infinite (all live executions are
infinite, by Definition~\ref{def:live-automaton}), $G$ is infinite. 
Hence, by clauses~\ref{lem:induced-digraph:finite-roots} and
\ref{lem:induced-digraph:outdegree} of
Lemma~\ref{lem:induced-digraph}, and Konig's lemma, $G$
contains an infinite path. Fix $p = (u_0,0)(u_1,1),\ldots$ to be any
such path.
By Definition~\ref{def:induced-digraph}, clause~\ref{def:induced-digraph:nodes},
$u_i \in g[s_i] \ints I_B$ for all $i \ge 0$.
We now construct $\al'_i$ and $m_i$ by induction on $i$, with $\al'_i$ such that 
$\lstate(\al'_i) = u_i$.

Now $s_0 \in \start(A)$ since $\al$ is an execution of $A$.
Also, by Definition~\ref{def:induced-digraph}, $u_0 \in g[s_0] \ints I_B$.
Hence, by clause~\ref{clause:live-back-sim-inv:init} of 
Definition~\ref{def:live-back-sim-inv}, $u_0 \in \start(B)$.
Let $\al'_0 = u_0$ and let $m_0$ be the mapping that maps $0$ to
$0$. Then, $m_0$ is a live index mapping from $\al|_0$ to $\al'_0$ with
respect to $b$ (in particular,
clause~\ref{clause:live-index-mapping:pairs} of
Definition~\ref{def:live-index-mapping} holds vacuously, 
since $|\al |_0 | = 0$), and $\lstate(\al'_0) = u_0$.

Now inductively assume that $m_{i-1}$ (for $i > 0$) is a live index mapping from
$\al|_{i-1}$ to $\al'_{i-1}$ with respect to $b$, and that $\lstate(\al'_{i-1})
= u_{i-1}$. By construction of path $p$, there is an edge in $G$ from
$(u_{i-1},i-1)$ to $(u_i,i)$. Hence, by Definition~\ref{def:induced-digraph},
there exists a finite execution fragment $\al''$ such that 
$\fstate(\al'') = u_{i-1}$,
$\lstate(\al'') = u_i$, $\trace(\al'') = \trace(a_{i})$, and, for all 
complemented-pairs $q \in M$:
   \bn
   \item if $\al'' \in q.\R$ then
		$s_{i-1} \in p.\R$ or $s_{i} \in p.\R$, and

   \item if $s_{i-1} \in p.\G$ or $s_{i} \in p.\G$ then
		$\al'' \in q.\G$,
   \en
where $p = h(q)$.
Now define $\al'_i = \al'_{i-1}\cat\al''$,
and define $m_i$ to be the mapping such that $m_i(j) = m_{i-1}(j)$ for
all $j$, $0 \leq j \leq i-1$, and $m_i(i) = |\al'_i|$. We argue that $m_i$ is a
live index mapping from $\al|_i$ to $\al'_i$ with respect to
$b$, i.e., that all clauses of Definition~\ref{def:live-index-mapping}
hold, and that $\lstate(\al'_i) = u_i$.
Clause~\ref{clause:live-index-mapping:init} holds since $m_i(0) =
m_{i-1}(0)$ by definition, and $m_{i-1}(0) = 0$ by the inductive hypothesis.
Clause~\ref{clause:live-index-mapping:corr} holds by the inductive
hypothesis, $\lstate(\al'') = u_i$, and $u_i \in g[s_i]$ (which we established above).
Clause~\ref{clause:live-index-mapping:trace} holds by the inductive
hypothesis and $\trace(\al'') = \trace(a_i)$.
Clause~\ref{clause:live-index-mapping:cofinal} holds since $m_i(|\,\al|_i\,|)
= m_i(i) = |\al'_i|$, by definition.
Finally, clause~\ref{clause:live-index-mapping:pairs} holds by the inductive
hypothesis and the conditions for all complemented-pairs $q \in M$
established above w.r.t. $s_{i-1} \lla{a_i}{A} s_i$ and $\al''$.
Having established that $m_i$ is a live index mapping from $\al|_i$ to
$\al'_i$ with respect to $f$, we conclude that
clause~\ref{lem:live-exec-backward:mapping} of the lemma holds.
Also, $\lstate(\al'_i) = \lstate(\al'') = u_i$, as required for the induction
step to be valid.

Clause~\ref{lem:live-exec-backward:prefix} of the lemma holds by
construction of $\al'_i$ and $m_i$,  since $\al'_i$ and $m_i$ are obtained by
extending $\al'_{i-1}$ and $m_{i-1}$, respectively.

By Definition~\ref{def:live-back-sim-inv},
clause~\ref{clause:live-back-sim-inv:silent}, for infinitely many $i > 0$,
the execution fragment $\al''$  which matches $s_{i-1} \lla{a_i}{A} s_i$
must have length $|\al''| \ge 1$. Hence, for infinitely many $i > 0$, we have
$\al'_{i-1} < \al'_i$.
Thus, clause~\ref{lem:live-exec-backward:infinite} of the lemma holds.
\epr

Our next lemma shows that, if infinite concrete and abstract
executions correspond in the sense of $(\al,\al') \in R^\l$, and the
concrete execution is live, then so is the abstract execution.

\bl Let $(A,L)$ and $(B,M)$ be live automata with the same external
actions. Let $R^\l = (R,H)$ where $R$ is a relation over $\states(A)
\times \states(B)$ and $H: M \mapsto \clos{L}$ is a total mapping
over $M$. Let $\al, \al'$ be arbitrary infinite executions of $(A,L)$,
$(B,M)$ respectively. If $(\alpha,\alpha') \in R^\l$, then 
$\alpha \in \execs(A,L)$ implies $\alpha' \in \execs(B,M)$.
\label{lem:liveness}
\el
\bpr
We assume the antecedents of the lemma and establish
$\alpha' \not\in \execs(B,M)$ implies $\alpha \not\in \execs(A,L)$. 
Let: \ms\\
\ind $\alpha  = s_0 a_1 s_1 a_2 s_2 \cdots$ \\
\ind $\alpha' = u_0 b_1 u_1 b_2 u_2 \cdots$ \ms\\
Since $(\alpha,\alpha') \in R^\l$, there exists a live index mapping $m:
\{0,1,\ldots,|\alpha|\} \mapsto \{0,1,\ldots,|\alpha'|\}$ satisfying
the conditions in Definition~\ref{def:live-index-mapping}. Suppose $\alpha'
\not\in \execs(B,M)$. Then, by Definition~\ref{def:live-execution}, there exists
a complemented-pair $q \in M$ such that
	$\alpha' \sat \iof q.\R \land \ea \neg q.\G$.
Let
$p = H(q)$. We prove:
\bleqn{(*)}
	$\alpha \sat \iof p.\R \land \ea \neg p.\G$.
\eleqn
Since $\alpha' \sat \iof q.\R$, there exist an infinite number of pairs
of states $(u_{m(i-1)}, u_{m(i)})$ along $\alpha'$ that contain a
$q.\R$-state between
them (inclusive, i.e., the $q.\R$-state could be $u_{m(i-1)}$ or
$u_{m(i)}$).  By clauses~\ref{clause:live-index-mapping:corr} and
\ref{clause:live-index-mapping:trace}
of Definition~\ref{def:live-index-mapping}, for each
such pair there corresponds a pair of states $(s_{i-1}, s_i)$ along $\alpha$
such that
$(s_{i-1}, u_{m(i-1)}) \in R$ and $(s_i, u_{m(i)}) \in R$.
Also, by clause~\ref{clause:live-index-mapping:RED} of
Definition~\ref{def:live-index-mapping}, $s_{i-1} \in p.\R$ or $s_i \in
p.\R$. Since this holds for an infinite number of values of the index $i$,
we conclude
\bleqn{(a)}
	$\alpha \sat \iof p.\R$.
\eleqn
Since $\alpha' \sat \ea \neg q.\G$, there exists a state $u_g$
along $\alpha'$ such that
    $\fa \l \geq g : u_\l \not\in q.\G$.
Now assume that $\alpha \sat \iof p.\G$. Since $m$ is nondecreasing and cofinal
in $\{0,1,\ldots,|\alpha'|\}$ (clause~\ref{clause:live-index-mapping:cofinal},
Definition~\ref{def:live-index-mapping}),
there exists an $s_{i-1}$ along $\alpha$ such that $s_{i-1} \in p.\G$ and
$m(i-1) \geq g$. By clauses~\ref{clause:live-index-mapping:corr} and
\ref{clause:live-index-mapping:trace} of Definition~\ref{def:live-index-mapping},
$(s_{i-1}, u_{m(i-1)}) \in R$ and $(s_{i}, u_{m(i)}) \in R$.
Also, by clause~\ref{clause:live-index-mapping:GREEN} of 
Definition~\ref{def:live-index-mapping},
at least one of $u_{m(i-1)}, u_{m(i-1)+1}, \ldots, u_{m(i)}$ is a $q.\G$ state. 
Since $m(i-1) \geq g$, this contradicts
    $\fa \l \geq g : u_\l \not\in q.\G$
above. Hence the assumption $\alpha \sat \iof p.\G$ must be false, and so:
\bleqn{(b)}
	$\alpha \sat \ea \neg p.\G$.
\eleqn
From (a) and (b), we conclude (*). From (*), we have $\al \not\sat p$.
Now $p \in \clos{L}$, since $H: M \mapsto \clos{L}$.
Hence, $\al \not\in \execs(A,\clos{L})$ by Definition~\ref{def:live-execution}.
Hence, by Proposition~\ref{prop:execs-clos}, $\al \not\in \execs(A,{L})$.
\epr

We can now establish a correspondence theorem for live executions. Our theorem
states that, if a liveness-preserving simulation relation $S^\l$ is established from
a concrete automaton to an abstract automaton, then for every live execution
$\al$ of the concrete automaton, there exists a corresponding (in the sense of
$(\al,\al') \in S^\l$) live execution $\al'$ of the abstract automaton. Our proof
uses Lemmas~\ref{lem:live-exec-forward} and \ref{lem:live-exec-backward}
to establish the existence of an infinite
family of finite executions corresponding to prefixes of $\al$. We then construct
$\al'$ from this infinite family using the ``diagonalization'' technique of
\cite{GSSL93}. 
Finally, we invoke Lemma~\ref{lem:liveness} to show that $\al'$
is live, given that $\al$ is live.

\bt [Live Execution Correspondence Theorem]
Let $(A,L)$ and $(B,M)$ be live automata with the same external actions.
Suppose $(A,L) \simu_X (B,M)$ via $S^\l$, where 
$X \in \{\l F, \l R, i\l B,$ $\l H, i\l P\}$.
Then $((A,L),(B,M)) \in S^\l$.
\label{thm:live execution correspondence}
\et
\bpr
We proceed by cases on $X$.

\case{1}{$X = \l F$} So $S^\l$ is a liveness-preserving forward simulation 
$f = (g,h)$, and $(A,L) \lpfs (B,M)$ via $f$.
Let $\al = s_0 a_1 s_1 a_2 s_2 \ldots$ be an arbitrary live execution
of $(A,L)$, and let $(\al_i', m_i)_{0 \leq i}$ be a
collection of finite executions of $(B,M)$ and mappings
as given by Lemma~\ref{lem:live-exec-forward}.
By definition of $((A,L),(B,M)) \in f$, we
must show that there exists a live execution $\al'$ of $(B,M)$ such that
$(\al,\al') \in f$.

By Definition~\ref{def:live-execution}, $\al$ is infinite.
Let $m$ be the unique mapping over the natural
numbers defined by $m(i) = m_i(i)$, for all $i \geq 0$. Let $\al'$ be the limit of
$\al'_i$ under the prefix ordering, that is, $\al'$ is the unique execution of
$(B,M)$ defined by $\al' |_{m(i)} = \al'_i$ for all $i \ge 0$,
with the restriction that for any
index $j$ of $\al'$, there exists an $i$ such that $\al'|_j \leq
\al'_i$. By Lemma~\ref{lem:live-exec-forward},
clause~\ref{lem:live-exec-forward:infinite}, $\al'$ is infinite.

We now show that $m$ is a live index mapping from $\al$ to $\al'$ with
respect to $f$. The proof that $m$ is nondecreasing and total and
satisfies
clauses~\ref{clause:live-index-mapping:init}--\ref{clause:live-index-mapping:cofinal}
of Definition~\ref{def:live-index-mapping} proceeds in exactly the
same way that the proof of the corresponding assertions does in the proof of the
Execution Correspondence Theorem in \cite{GSSL93}.
We repeat the details for sake of completeness.

Suppose $m$ is not nondecreasing. Then there exists an
$i$ such that $m(i) < m(i-1)$. However, $m(i) = m_i(i)$ and $m(i-1) =
m_{i-1}(i-1) = m_i(i-1)$, so this contradicts the fact that $m_i$ is
an index mapping and is therefore nondecreasing. Likewise, we can see
that the range of $m$ is within $\{0,\ldots,|\al'|\}$.

Clause~\ref{clause:live-index-mapping:init} of
Definition~\ref{def:live-index-mapping} holds since $m_0$ is an index
mapping and therefore satisfies $m_0(0) =
0$. Hence $m(0) = m_0(0) = 0$.
Assume clauses~\ref{clause:live-index-mapping:corr} or
\ref{clause:live-index-mapping:trace} do not hold. Then, there must exist
an $i$ for which one of the clauses is invalidated. However, this
contradicts the fact that, for all $i$, $m_i$ is an index mapping from $\al|_i$ to
$\al'_i$ with respect to $f$.
Now assume that clause~\ref{clause:live-index-mapping:cofinal} does
not hold. Hence, there is an index $j$ in $\al'$ such that $m(i) < j$
for all $i$. By definition of $\al'$, there exists an $i$ such that
$\al'|_j \leq \al'_i$. Thus  $|\al'_i| \geq j$.
 Now Lemma~\ref{lem:live-exec-forward} gives us
$m_i(i) = |\al'_i|$. Hence $m(i) \geq j$, since $m(i) =
m_i(i)$. This contradicts $m(i) < j$.

Now assume that $m$ violates
clause~\ref{clause:live-index-mapping:pairs} of
Definition~\ref{def:live-index-mapping}. Then, there exists a pair $q
\in M$ and an $i > 0$ for which
clause~\ref{clause:live-index-mapping:pairs} is invalidated.  However,
this contradicts the fact that, for all $i > 0$, $m_i$ is a live index
mapping from $\al|_i$ to $\al'_i$ with respect to $f$
(Lemma~\ref{lem:live-exec-forward}, clause~\ref{lem:live-exec-forward:mapping}).
Hence $m$ satisfies clause~\ref{clause:live-index-mapping:pairs} of
Definition~\ref{def:live-index-mapping}.  Since $m$ satisfies all
clauses of Definition~\ref{def:live-index-mapping}, $m$ is a live
index mapping from $\al$ to $\al'$ with respect to $f$, and so
$(\al,\al') \in f$.
Since $\al \in \execs(A,L)$, $(\al,\al') \in f$, and $\al, \al'$ are both
infinite, we can apply Lemma~\ref{lem:liveness} to conclude $\al' \in
\execs(B,M)$, i.e., $\al'$ is a live execution of $(B,M)$, which
establishes the theorem in this case.

\case{2}{$X = \l R$} So $S^\l$ is a liveness-preserving refinement mapping 
$r = (g,h)$ and $(A,L) \lprm (B,M)$ via $r$.
Since a liveness-preserving refinement mapping is a liveness-preserving forward
simulation, the result follows from Case 1.

\case{3}{$X = i \l B$} So $S^\l$ is an image-finite 
liveness-preserving backward simulation 
$b = (g,h)$, and $(A,L) \ilpbs (B,M)$ via $b$.
The argument is identical to that of Case 1, except that we invoke 
Lemma~\ref{lem:live-exec-backward} instead of Lemma~\ref{lem:live-exec-forward}.

\case{4}{$X = \l H$} So $S^\l$ is a
liveness-preserving history relation $hs$
and $(A,L) \lphr (B,M)$ via $hs$.
From Definition~\ref{def:live-hist-inv}, $hs$ is a 
liveness-preserving forward simulation from $A$ to $B$.
Hence, the argument of Case 1 applies.

\case{5}{$X = i \l P$} So $S^\l$ is an image-finite 
liveness-preserving prophecy relation
$p = (g,h)$, and $(A,L) \ilppr (B,M)$ via $p$.
From Definition~\ref{def:live-proph-inv}, $p$ is an 
image-finite liveness-preserving backward simulation from $A$ to $B$.
Hence, the argument of Case 3 applies.

Since all cases of $X$ have been dealt with, the theorem is established.
\epr

We now establish our main result: liveness-preserving 
simulation relations imply the live preorder.

\bt [Liveness]
Let $(A,L)$ and $(B,M)$ be live automata with the same external actions.
Suppose $(A,L) \simu_X (B,M)$, where 
$X \in \{\l F, \l R, i\l B, \l H, i\l P\}$.
Then $(A, L) \lpreorder (B, M)$.
\label{thm:liveness}
\et
\bpr
From $(A,L) \simu_X (B,M)$, we have $(A,L) \simu_X (B,M)$ via $S^\l$ for some
$S^\l = (g,h)$.  We establish 
	$\traces(\execs(A,L)) \sub \traces(\execs(B,M))$,
which, by Definition~\ref{def:preorders}, proves the theorem.
Let $\beta$ be an arbitrary trace in $\traces(\execs(A,L))$. By
definition, $\beta = \trace(\alpha)$ for some live execution
$\al \in \execs(A,L)$.
By the Live Execution Correspondence Theorem~(\ref{thm:live
execution correspondence}), there exists a live execution 
$\al' \in \execs(B,M)$
such that $(\al, \al') \in S^\l$. Since $(\al, \al') \in S^\l$, we have
$(\al, \al') \in g$ by Definitions~\ref{def:index-mapping} and
\ref{def:live-index-mapping}.  Hence, by Lemma~\ref{lem:traces}, $\trace(\al)
= \trace(\al')$.  Hence $\beta = \trace(\al')$, and so $\beta \in
\traces(\execs(B,M))$, since $\al' \in \execs(B,M)$.
Since $\beta$ was chosen arbitrarily, we conclude
$\traces(\execs(A,L)) \sub \traces(\execs(B,M))$, as desired.
\epr

\section{Refining Liveness Properties Within the Same Level of Abstraction}
\label{sec:lattices}

The previous section showed how to refine an abstract liveness
condition $M$ to a concrete liveness condition $L$: every pair $q \in M$ is
mapped into some pair $p$ in the semantic closure $\clos{L}$ of $L$, and
then a liveness-preserving simulation relation that relates the
$\R$ and $\G$ sets of $p, q$ appropriately is devised.
We assume that the liveness properties $L$, $M$ are directly
specified, and so the pairs in $M$ and in $L$ are easy to
identify.\footnote{For example, if we were attempting to mechanize our
method, we would assume that $M$, $L$ are recursive sets.}
However, pairs in $\clos{L} - L$ are not directly specified, but only
given implicitly by $A$, $L$, and Definition~\ref{def:closure}.
Thus, the question arises, given a pair $q \in M$ that is mapped to
some pair $p$, how do we establish $p \in \clos{L}$?
We do so as follows.

Given such a pair $p$, we refine it 
 into a finite ``lattice'' of
pairs that are already known to be in $\clos{L}$. Let $P$ be a finite subset of
$\clos{L}$, and let $\lprec$ be an irreflexive partial order
over $P$\footnote{Following convention, we shall refer to this
ordered set simply as $P$ when no confusion arises.}. If $r \in P$, define
$\suc(r) = \{w \in P ~|~ r \lprec w \land \fa w': r \lpreceq w' \lprec w \imp
r = w' \}$, where $r \lpreceq w \df r \lprec w$ or $r = w$.
Thus, $\suc(r)$ is the set of all ``immediate successors''
of $r$ in $(P,\lprec)$. 
We now impose two technical conditions on $P$:
(1) for every pair $r$, the $\G$ set of $r$ must be a subset of the
union of the $\R$ sets of all the immediate successors of $r$,
i.e., $r.\G \sub \UN_{w \in \suc(r)} w.\R$, 
and
(2) $P$ has a single $\lprec$-minimum element $\bottom(P)$, and a
single $\lprec$-maximum element $\topp(P)$, and $\bottom(P).\R = p.\R$ and
$\topp(P).\G = p.\G$.

Now let $\al$ be an arbitrary live execution of $(A,L)$. Then, $\al \sat \iof r.\R
\imp \iof r.\G$ and $\al \sat \iof w.\R \imp \iof w.\G$, for all $w \in
\suc(r)$.  Since $\suc(r)$ is finite and $r.\G \sub \UN_{w \in \suc(r)} w.\R$, 
it follows that, if $r.\G$ holds
infinitely often in $\al$, then $w.\R$ holds infinitely often in
$\al$, for some $w \in \suc(r)$.  Hence, by ``chaining'' the above implications,
we get  $\al \sat \iof r.\R \imp \iof \UN_{w \in \suc(r)} w.\G$.
Thus, $\pair{r.\R}{\UN_{w \in \suc(r)} w.\G} \in \clos{L}$ by Definition~\ref{def:closure}.
Thus, the
$\lprec$ ordering provides a way of relating the complemented-pairs of
$P$ so that the complemented-pairs property (infinitely often $\R$
implies infinitely often $\G$) can be generalized to encompass a pair
and its immediate successor pairs.  By starting with the
$\lprec$-minimum pair $\bottom(P)$, and applying the above argument
inductively (using $\lprec$ as the underlying ordering), we can
establish the complemented-pairs property for 
$\pair{\bottom(P).\R}{\topp(P).\G}$, i.e., $\al \sat \iof \bottom(P).\R \imp \iof
\topp(P).\G$, and so $\pair{\bottom(P).\R}{\topp(P).\G} \in \clos{L}$.
Since we require $\bottom(P).\R = p.\R$ and $\topp(P).\G = p.\G$, we obtain the
desired result that $p \in \clos{L}$.

\bd[Complemented-pairs Lattice]
Let $(A,L)$ be a live automaton.
Then $(P, \lprec)$ is a 
\intrdef{complemented-pairs lattice over $\clos{L}$} iff\/\footnote{Note that we use the
term ``lattice'' in an informal sense, since our complemented-pairs lattices
do not satisfy the mathematical definition of a lattice.}
\bn
   \item $P$ is a finite subset of $\clos{L}$,

   \item $\lprec$ is an irreflexive partial order over $P$,

   \item $P$ contains an element $\topp(P)$ which satisfies
         	$\fa r \in P: r \lpreceq \topp(P)$,
	and an element $\bottom(P)$ which satisfies
		$\fa r \in P: \bottom(P) \lpreceq r$, and

   \item \label{def:lattice:succ}
         $\fa r \in P - \{\topp(P)\}: r.\G \subseteq \UN_{w \in \suc(r)} w.\R$.
\en
\label{def:lattice}
\ed

\noindent
The elements $\topp(P)$ and $\bottom(P)$ are necessarily unique, since
$\lprec$ is a partial order.
Let $\lattices(\clos{L})$ denote the set
of all complemented-pairs lattices over $\clos{L}$.

\bl
Let $(A,L)$ be a live automaton, $(P,\lprec) \in \lattices(\clos{L})$,
$\bot = \bottom(P)$, and $\top = \topp(P)$. Then 
$\pair{\bot.\R}{\top.\G} \in \clos{L}$.
\label{lem:lattice}
\el
\bpr Let $\alpha$ be an arbitrary live execution of $(A,L)$.
We show $\al \sat \iof \bot.\R \imp \iof \top.\G$.
By Definition~\ref{def:closure}, this establishes the lemma.

We assume $\alpha \sat \iof \bot.\R$ and establish $\alpha \sat \iof \top.\G$.
First, we establish:
\bleqn{(*)}
	If $r \in P$, $r \neq \top$, and $\alpha \sat \iof r.\R$, then
        $\alpha \sat \iof w.\R$
         for some $w \in \suc(r)$.
\eleqn
Proof of (*): Assume the antecedent of (*). Since $\alpha$ is live
and $r \in \clos{L}$, we have
$\alpha \sat \iof r.\R \imp \iof r.\G$ by Definition~\ref{def:closure}.
Hence $\alpha \sat \iof r.\G$.
By Definition~\ref{def:lattice}, $r.\G \subseteq \UN_{w \in \suc(r)} w.\R$.
Hence $\alpha \sat \iof \UN_{w \in \suc(r)} w.\R$. Since $P$
is finite, $\suc(r)$ is finite. It follows that 
$\alpha \sat \iof w.\R$ for some $w \in \suc(r)$.
(End of proof of (*).)

We now construct a sequence $r_1, r_2, \ldots, r_i, \ldots$ of pairs in $P$ such
that $\fa i \geq 1: \alpha \sat \iof r_i.\R$.  We let $r_1 = \bot$, noting
that $\alpha \sat \iof \bot.\R$ by assumption.  We derive $r_{i+1}$ by applying
(*) to $r_i$. It follows by induction on the length of the derived sequence that 
$\alpha \sat \iof r_{i+1}.\R$ ($r_1 = \bot$ supplies the base case).
Now suppose $\top$ is not in $r_1, r_2, \ldots$  Then (*) can be
applied indefinitely. Since $r_{i+1} \in \suc(r_i)$, it follows that 
$r_j \lprec r_{i+1}$ for all $j \in 1..i$. Hence $r_j \neq r_{i+1}$ for all $j \in 1..i$.
Thus $r_1, r_2, \ldots$ is an infinite sequence of pairwise different
complemented-pairs in
$P$. But this is impossible, since $P$ is finite. Hence the assumption that
$\top$ is not in $r_1, r_2, \ldots$ is false. It follows that 
$r_1, r_2, \ldots$ is a finite sequence of 
pairwise different complemented-pairs, with $\top$ as its last member. 
Hence 
$\alpha \sat \iof \top.\R$. Since $\alpha$ is live and $\top \in \clos{L}$,
$\alpha \sat \iof \top.\R \imp \iof \top.\G$. Hence $\alpha \sat \iof \top.\G$,
as desired.
\epr

\vspace{1.0ex}

We remark that when constructing a lattice to refine a complemented-pair, we can use
requirement~\ref{def:lattice:succ} of Definition~\ref{def:lattice}
$(r.\G \subseteq \UN_{w \in \suc(r)} w.\R)$ as a
constraint that suggests how to order the complemented-pairs of the
lattice.
Also, while Lemma~\ref{lem:lattice} presents one method of establishing the
membership of complemented-pairs in $\clos{L}$, our overall
methodology is not restricted to this particular method. Any
appropriate deductive technique that suffices can be used, for example that of 
\cite{MP93}, which is based on linear temporal logic.
This provides a way of using deductive methods generally, 
and those based on temporal logic in particular, within a
framework which accommodates the refinement of liveness properties 
across multiple levels of abstraction.

\section{Example---The Eventually Serializable Data Service}
\label{sec:example}

The eventually-serializable data service (ESDS) of \cite{FGLLS99,LLSG92} is a
replicated, distributed data service that trades off immediate
consistency for improved efficiency. A shared data object is
replicated, and the response to an operation at a particular replica
may be out of date, i.e., not reflecting the effects of other
operations that have not yet been received by that replica. Thus,
operations may be reordered \emph{after} the response is issued.
Replicas communicate amongst each other the operations they receive,
so that eventually every operation ``stabilizes,'' i.e., its ordering
is fixed with respect to all other operations. Clients may require an operation
to be \intr{strict}, i.e., stable at the time of response, and so
it cannot be reordered after the response is issued. Clients may also
specify, in an operation $x$, a set $x.\prev$ of other
operations that should precede $x$ (client-specified constraints, $\CSC$).
We let $\Op$ be the (countable) set of all operations on the data object,
and $V$ be the set of all possible results of operations in $\Op$.
$\reps$ is the set of all replicas, and
$\client(x)$ is the client issuing operation $x$.
We use $x,y$ to index over operations, 
$c$ to index over clients, and $r,r',i$ to index over replicas.
Each operation $x$ has a unique identifier $\idof{x}$. 
$\idset$ is the set of identifiers of operations in $\Op$.

In Appendix~\ref{app:esds}, we give the I/O automata code (in
``precondition-effect'' style) from \cite{FGLLS99}.  I/O automata
\cite{LT89} add an input/output distinction to the external actions,
i.e, all external actions of an automaton are either input actions
(which must furthermore be enabled in all states), or output
actions. This is needed to define a parallel composition operator
$\pl$ with good compositional properties.
Figure~\ref{fig:users} gives the environment of the ESDS system: a set
of users, or clients, which output requests $\request(x)$
to perform operations $x$,
and input responses $\response(x, v)$ to the requests, with returned value $v$.
Figure~\ref{fig:ESDSI} presents the specification $\ESDSI$. As a
high-level specification,  $\ESDSI$ is a single automaton, and
therefore it does not address issues of concurrency and
distribution. The only concern is to specify the set of correct
traces, which are by definition the traces of $\ESDSI$.
$\ESDSI$ inputs requests $\request(x)$, and outputs 
responses $\response(x, v)$ to the requests, with returned value $v$.
Once $\request(x)$ has been received, it is ``entered'' into the
current partial order $\po$, via internal action
$\enter(x, \newpo)$, which updates the value of
$\po$ to that given by $\newpo$.
This new value must include all operations in $\prevof{x}$,
and all operations that have stabilized, as preceding $x$.
Note that $\spn{R} = \{x ~|~ x R y \lor y R x\}$, where $R$ is a
binary relation.
At any time, it is permissible to impose new ordering constraints,
which is done by internal action $\addcons(\newpo)$.
The $\stabilize(x)$ internal action checks that $x$ is totally ordered
with respect to all other operations
($\forall y \in \ops$, $y \preceq_{\po} x \logicor x \preceq_{\po} y$),
and that all operations that precede $x$ have already stabilized
($\ops |_{\prec_{\po } x} \subseteq \stabilized$).
In this case, $x$ itself can be stabilized. 
The $\calc(x, v)$ internal action computes a return value $v$ for the
operation $x$. If $x$ is strict, then  $\calc(x, v)$ checks (in its
precondition) that $x$ has stabilized. 
The $\valset(x, \ops, \prec_{\po})$ function returns the set of all
values for $x$ which are consistent with the set $\ops$ of all
operations that have been entered, and the partial order $\prec_{\po}$ 
defined by $\po$. The actual value returned is then chosen
nondeterministically from this set.

As an intermediate step, we refine $\ESDSI$ to a
second level specification $\ESDSII$.  This refinement consists only
of changing some of the transitions. The state space and the
signature remain the same.  Figure~\ref{fig:ESDSII} presents these
changes, as changes to the ``precondition-effect'' definitions of some
of the actions in the action list.
The main difference with $\ESDSI$ is that the precondition to
stabilize an operation $x$ is relaxed: now, all operations that
precede $x$ are not required to be stable themselves, but are only
required to be totally ordered with respect to all other entered operations
($\prec_{\po}$ totally orders $\ops |_{\prec_{\po } x}$).
This intermediate version $\ESDSII$ is useful, as it is easier to construct a
simulation from the implementation to $\ESDSII$, and another
simulation from $\ESDSII$ to $\ESDSI$, 
than it is to construct a simulation from the 
implementation directly to $\ESDSI$.

The implementation consists of front-ends, replicas, and channels.
Each client $c$ has a front-end $\Frontend(c)$, see 
Figure~\ref{fig:front-end}, which inputs requests $\request(x)$, and
relays them onto one or more of the replicas $\Replica(r)$, via output
action $\send_{cr}(\reqmsg{x})$.  $\Frontend(c)$ receives
responses from the replicas via input action
$\receive_{rc}(\respmsg{x,v})$, and relays the response onto the
client via output action $\response(x,v)$.
While the frontend can receive several replies for $x$ from various
replicas, it only relays one of these onto the client.
A replica $r$ (Figure~\ref{fig:replica}) receives requests to perform
operation $x$ via input action $\receive_{cr}(\reqmsg{x})$.
It queues received operations into a set $\pending_r$ of pending
operations.
A pending operation $x$ can be ``performed'' by the internal action
$\doit_r(x, l)$ if all operations in $\prevof{x}$ have been
performed. In this case, $x$ is assigned a ``label'' $l$ larger than
the labels of all operations known to be done at replica $r$. This
label determines the values that can be returned for $x$, using the
$\valset$ function.
Once $x$ has been processed by $\doit_r(x, l)$, a value $v$ for $x$
can be returned by the output action $\send_{rc}(\respmsg{x, v})$. 
$v$ is nondeterministically chosen from among the set returned by 
$\valset(x, \done_r[r], \prec_{\LC_r})$, which computes all values for
$x$ that are consistent with the set $\done_r[r]$ of operations done
at replica $r$, and the partial order $\prec_{\LC_r}$ on operations
that is determined by the labels assigned to each operation.
In addition, replicas ``gossip'' amongst each
other, by means of the actions 
$\send_{rr'}(\gossipmsg{R, D, L, S})$ and
$\receive_{r'r}(\gossipmsg{R, D, L, S})$.
The purpose of gossiping is 
to bring each other up to date on the operations that they have
executed.
All communication between the front-ends and the replicas is by means
of reliable asynchronous channels.  Figure~\ref{fig:channel}
shows a channel from process $i$ to process $j$ with messages drawn
from some set $\msgset$.

We will use $\ESDSAlg$ to refer to the parallel composition of all
replicas, front-ends, and channels, with all $\send$ and $\receive$
actions hidden\footnote{I/O automata composed in parallel synchronize
on actions with the same name, and otherwise execute independently.
An action is hidden by removing it from the set of output actions and
adding it to the set of internal actions. We refer the reader to
\cite[section 3]{FGLLS99} for formal definitions of parallel
composition and hiding.}.  
Since the users must be taken into account, the first-level
specification, second-level specification, and implementation are the
I/O automata $\ESDSI \pl \Users$, $\ESDSII \pl \Users$, 
and $\ESDSAlg \pl \Users$, respectively.
We refer the reader to \cite{FGLLS99} for a complete description of
the ESDS system.

The liveness condition used in (the conference version of)
\cite{FGLLS99} is that every request should eventually receive a
response, and every operation should stabilize. We express this as the
following complemented-pairs liveness condition $\MI$ for the
specification $\ESDSI \pl \Users$:\footnote{Throughout this section,
our notation is consistent with \cite{FGLLS99}.}
\be

\item $\{ \pair{x \in \wait}{x \not\in \wait} ~|~ x \in \Op \}$,
i.e.,  every request eventually receives a response.

\item $\{ \pair{x \in \wait}{x \in \stabilized} ~|~ x \in \Op \}$,
i.e., every operation eventually stabilizes.

\ee
Because the number of submitted operations $x$ in general grows without
bound with time, a countably infinite number of pairs is needed to express
this liveness condition in the natural manner illustrated above. Note
that we use predicates to denote sets of states.

\subsection{Refinement from ${\ESDSI \pl \Users}$ to $\ESDSII \pl \Users$}

The top-level specification $\ESDSI \pl \Users$ and second-level
specification $\ESDSII \pl \Users$ have the same state-space, they
only differ in some actions, as shown in Figure~\ref{fig:ESDSII}.
Hence, we let the liveness condition $\MII$ of $\ESDSII \pl \Users$
consist of the same complemented-pairs as those in $\MI$,
and we map each pair of $\MI$ into the same pair of $\MII$.

\bfg

\horline

$G$ is a relation between states in $\ESDSII \pl \Users$ and $\ESDSI \pl \Users$,
such that
$(s,u) \in G$ if and only if               
$s \in \states(\ESDSII \pl \Users)$, $u \in \states(\ESDSI \pl \Users)$, and:
\be

\item $u.\wait = s.\wait$

\item $u.\rept = s.\rept$

\item $u.\ops = s.\ops$

\item $u.\po = s.\po$

\item $u.\stabilized \supseteq s.\stabilized$

\ee

\horline
\caption{Forward Simulation from $\ESDSII \pl \Users$ to $\ESDSI \pl \Users$}
\label{fig:sim-ESDSII-to-ESDSI}
\efg

In \cite{FGLLS99}, it is shown that the relation $G$ given in
Figure~\ref{fig:sim-ESDSII-to-ESDSI} %
is a forward simulation relation from $\ESDSII \pl \Users$ to
$\ESDSI \pl \Users$.  We show that $G$ is also a liveness-preserving
forward simulation.
For the pair $\pair{x \in \wait}{x \not\in \wait}$ it is clear that
$G$ satisfies clause~\ref{clause:live-fwd-sim-inv:trans} of
Definition~\ref{def:live-fwd-sim-inv}, since $G$ only relates states
that agree on the value of $\wait$.
For the pair $\pair{x \in \wait}{x \in \stabilized}$, we see from
Figure~\ref{fig:sim-ESDSII-to-ESDSI} that if $s \in \states(\ESDSII
\pl \Users)$ and $u \in \states(\ESDSI \pl \Users)$ are related by $G$,
and $s$ satisfies ${x \in \stabilized}$, then $u$ also satisfies
${x \in \stabilized}$,
since $s.\stabilized \sub u.\stabilized$. Since $s$ and $u$ agree on
the value of $\wait$, we conclude that $G$ satisfies
clause~\ref{clause:live-fwd-sim-inv:trans} of
Definition~\ref{def:live-fwd-sim-inv}, for this pair too.

By inspection, we verify that, in every live execution of $\ESDSII$, there is
an infinite number of executions of non-$\stabilize$ actions.  Now according to
the definition of $G$ in Figure~\ref{fig:sim-ESDSII-to-ESDSI}, every action in
$\ESDSII$ is simulated by the same action in $\ESDSI$, except for the
$\stabilize$ action; a single $\stabilize(x)$ action in $\ESDSII$ can be
simulated by a possibly empty sequence of $\stabilize$ actions in $\ESDSI$.
Hence, any transition generated by executing any action other than $\stabilize$
is not always-silent, by clause~\ref{clause:live-fwd-sim-inv:silent} of
Definition~\ref{def:live-fwd-sim-inv}.  Since every live execution of $\ESDSII$
contains an infinite number of these transitions,
clause~\ref{clause:live-fwd-sim-inv:silent} of
Definition~\ref{def:live-fwd-sim-inv} is satisfied.

Since each pair of $\MI$ is mapped into a pair of $\MII$ itself, rather
than the semantic closure $\clos{\MII}$ of $\MII$, we are done (i.e.,
there is no need to construct complemented-pairs lattices for these
pairs).

Since Definition~\ref{def:live-fwd-sim-inv} is now satisfied, we have
established $(\ESDSII \pl \Users, \MII) \lpsim (\ESDSI \pl \Users,
\MI)$.  Hence, applying Theorem~\ref{thm:liveness}, we conclude
$(\ESDSII \pl \Users, \MII) \lpreorder (\ESDSI \pl \Users, \MI)$.

\subsection{Refinement from $\ESDSII \pl \Users$ to $\ESDSAlg \pl \Users$}

Let $L$ be the liveness condition of $\ESDSAlg \pl \Users$.  Since
$\ESDSAlg \pl \Users$ is an implementation, we take $L$ to be the
following: every action that is continuously enabled from some point
onwards is eventually executed (fair scheduling), and every message
that is sent is eventually received (fair polling of channels).
These are reasonable liveness properties to expect of an implementation.

We map the pair 
$\pair{x \in \wait}{x \not\in \wait}$ of $\MII$ into the pair 
$\pair{x \in \wait_c}{x \not\in \wait_c}$, where 
$c = \client(x)$ is the client that requests operation $x$.
We map the pair
$\pair{x \in \wait}{x \in \stabilized}$ of $\MII$ into the pair 
$\pair{x \in \wait_c}{x \in \INT_i \stable_i[i]}$.

The proof obligations are then to exhibit a liveness-preserving
forward simulation for this choice of pair-mapping, and to show that
the pairs 
$\pair{x \in \wait_c}{x \not\in \wait_c}$ and 
$\pair{x \in \wait_c}{x \in \INT_i \stable_i[i]}$
are members of $\clos{L}$, since they are not members of $L$.

\subsubsection{Establishing a Liveness-preserving Forward Simulation}

In \cite{FGLLS99}, it is shown that the relation $F$ given in
Figure~\ref{fig:sim-ESDSAlg-to-ESDSII} %
is a forward simulation relation from $\ESDSAlg \pl \Users$ to
$\ESDSII \pl \Users$.  We establish that $F$ is also a liveness-preserving
forward simulation.
We first 

By Definition~\ref{def:fwd-sim-inv}, $F$ already satisfies
clause~\ref{clause:live-fwd-sim-inv:init} of
Definition~\ref{def:live-fwd-sim-inv}.  We argue that $F$ also satisfies
clauses~\ref{clause:live-fwd-sim-inv:trans} and
\ref{clause:live-fwd-sim-inv:silent}.  Let
$\SpReq = \pair{x \in \wait}{x \not\in \wait}$, 
$\ImpReq = \pair{x \in \wait_c}{x \not\in \wait_c}$,
$\SpStab = \pair{x \in \wait}{x \in \stabilized}$,
$\ImpStab = \pair{x \in \wait_c}{x \in \INT_i \stable_i[i]}$.
Let $B = \ESDSII \pl \Users$, and $A = \ESDSAlg \pl \Users$.
Let $s, u$ range over the states of $\ESDSAlg \pl \Users$, $\ESDSII
\pl \Users$ respectively. 
We use the notation $s.v$ to denote the value of state variable $v$ in
state $s$, and likewise for $u.v$.

\bfg

\horline

$F$ is a relation between states in $\ESDSAlg \pl \Users$ 
and $\ESDSII \pl \Users$, i.e.,
$F \sub \states(\ESDSAlg \pl \Users) \times
        \states(\ESDSII \pl \Users)$, such that
$(s,u) \in F$ if and only if:              %
\be

\item $u.\requested = s.\requested$

\item $u.\responded = s.\responded$

\item $u.\wait = \UN_c s.\wait_c$

\item $u.\rept = \UN_c s.\rept_c  \un s.\mathit{potential\_rept}_c$

\item $u.\ops = s.\ops = \UN_r s.\done_r[r]$

\item $u.\po \sub s.\po$

\item $u.\stabilized = \INT_r s.\stable_r[r]$

\ee

where
$s.\mathit{potential\_rept}_c = 
\{ (x,v) ~|~ \respmsg{x, v} \in \UN_r s.\channel_{rc} \land s.\wait_c \}$
is the set of responses en route to $\Frontend(c)$, and
$u.\po$ is the partial order induced by the various operation
constraints in the implementation. See \cite{FGLLS99} for details.

\horline
\caption{Forward simulation from $\ESDSAlg \pl \Users$ 
         to $\ESDSII \pl \Users$}
\label{fig:sim-ESDSAlg-to-ESDSII}
\efg
\paragraph{Establishing clause~\ref{clause:live-fwd-sim-inv:trans} of
    Definition~\ref{def:live-fwd-sim-inv} for the pairs 
    ${\SpReq =  \pair{x \in \wait}{x \not\in \wait} \in \MI}$ and
    $\ImpReq = \pair{x \in \wait_c}{x \not\in \wait_c}$.}
$F$ relates states $s$ and $u$ only if $u.wait = \UN_c s.wait_c$.
Hence $x \in u.\wait$ iff $x \in s.\wait_c$, where $c = \client(x)$.
Thus $u$ is a $\SpReq.\R$ state iff $s$ is a $\ImpReq.\R$ state, and
$u$ is a $\SpReq.\G$ state iff $s$ is a $\ImpReq.\G$ state.

Let $s \lla{a}{A} s'$ and consider all possibilities for $a$.
If $a$ is one of $\send$ (along any channel), $\receive$ (from any
channel), or $\doit_r$ (for any replica $r$), then $a$ does not change
$\wait_c$ (for any client $c$), and the actions of $\ESDSII
\pl \Users$ that simulate $a$ do not change $\wait$. Hence if 
$u_0 \lla{b_1}{B} u_1 \lla{b_2}{B} u_2 \lla{b_3}{B} \cdots \lla{b_{n}}{B} u_n$ is
the simulating execution fragment of $\ESDSII \pl \Users$,
corresponding to $s \lla{a}{A} s'$ for the aforementioned cases of $a$,
then we immediately conclude that (1) all $u_i, i \in 0 \ldots n$ have the same
value of $\wait$, and (2) $s$ and $s'$ have the same value of $\UN_c
\wait_c$. Together with $u_0.\wait = \UN_c s.\wait_c$, this allows us
to conclude
$(\ex i \in  0 \ldots n: u_i \in \SpReq.\R)$ iff
$s \in \ImpReq.\R$ or $s' \in \ImpReq.\R$, and
$s \in \ImpReq.\G$ or $s' \in \ImpReq.\G$ iff
$(\ex i \in 0 \ldots n : u_i \in \SpReq.\G)$.
Thus clause~\ref{clause:live-fwd-sim-inv:trans} of
Definition~\ref{def:live-fwd-sim-inv} is satisfied in this case.

If $a$ is $\request(x)$, this is simulated by the same action in $\ESDSII
\pl \Users$. $\request(x)$ adds $x$ to $\wait_c$ in $\ESDSAlg \pl
\Users$, and adds $x$ to $\wait$ in $\ESDSII \pl \Users$. Hence, using
similar reasoning as above, we
easily verify that clause~\ref{clause:live-fwd-sim-inv:trans} of
Definition~\ref{def:live-fwd-sim-inv} is satisfied in this case. The
argument for $a = \response(x,v)$ is similar. This concludes our
argument that clause~\ref{clause:live-fwd-sim-inv:trans} of
Definition~\ref{def:live-fwd-sim-inv} holds for the pairs 
$\SpReq$ and $\ImpReq$.

\paragraph{Establishing clause~\ref{clause:live-fwd-sim-inv:trans} of
   Definition~\ref{def:live-fwd-sim-inv} for the pairs
   $\SpStab = \pair{x \in \wait}{x \in \stabilized} \in \MI$ and 
   $\ImpStab = \pair{x \in \wait_c}{x \in \INT_i \stable_i[i]}$.}
$F$ relates states $s$ and $u$ only if $u.wait = \UN_c s.wait_c$ and
$u.\stabilized = \INT_i s.\stable_i[i]$
(definition of $F$ in \cite{FGLLS99}, and Figure~\ref{fig:sim-ESDSAlg-to-ESDSII}).
Hence $x \in u.\wait$ iff $x \in s.\wait_c$, where $c = \client(x)$, and
$x \in u.\stabilized$ iff  $x \in \INT_i s.\stable_i[i]$.
Thus $u \in \SpStab.\R$ iff $s \in \ImpStab.\R$, and
$u \in \SpStab.\G$ iff $s \in \ImpStab.\G$.

Let $s \lla{a}{A} s'$ and let 
$u_0 \lla{b_1}{B} u_1 \lla{b_2}{B} u_2 \lla{b_3}{B} \cdots \lla{b_{n}}{B} u_n$
be the execution fragment of $\ESDSII \pl \Users$ that simulates $s \lla{a}{A} s'$.
Given the previous remarks,
we conclude immediately that clause~\ref{clause:live-fwd-sim-inv:trans} of
Definition~\ref{def:live-fwd-sim-inv} is satisfied when
$u_1, \ldots, u_{n-1}$ are not present, i.e., the simulating fragment
consists of either a single state or a single transition.

The only case where 
$u_0 \lla{b_1}{B} u_1 \lla{b_2}{B} u_2 \lla{b_3}{B} \cdots \lla{b_{n}}{B} u_n$
consists of more than one transition is when $a = \receive_{rr'}(m)$ .
In this case, the actions $b_1, \ldots, b_{n}$ are 
$\addcons(s'.po), \stabilize(x_1), \ldots,$ $\stabilize(x_k)$, where 
$\set{x_1,\ldots,x_k} = \INT_i s'.\stable_i[i]$
(see \cite{FGLLS99}, Section 8).
Now $\INT_i s.\stable_i[i] \sub \INT_i s'.\stable_i[i]$ by inspection
of the $\receive_{rr'}(m)$ action in Figure~\ref{fig:replica}.
Also, 
$u_0.\stabilized = \INT_i s.\stable_i[i]$, and
$u_n.\stabilized = \INT_i s'.\stable_i[i] = \set{x_1,\ldots,x_k}$, by
definition of $F$ and ${x_1,\ldots,x_k}$.

Now $\receive_{rr'}(m)$ does not affect $\wait_c$, and
$\addcons(s'.po), \stabilize(x_1), \ldots,$ $\stabilize(x_k)$
do not affect $\wait$. Hence,
$(\ex i \in 0 \ldots n : u_i \in \SpStab.\R)$ iff
$s \in \ImpStab.\R$ or $s' \in \ImpStab.\R$.
Also, suppose 
$s \in \ImpStab.\G$ or $s' \in \ImpStab.\G$, i.e.,
$x \in \INT_i s.\stable_i[i]$ or $x \in \INT_i s'.\stable_i[i]$.
Hence $x \in \INT_i s'.\stable_i[i]$ since 
$\INT_i s.\stable_i[i] \sub \INT_i s'.\stable_i[i]$.
Since 
$u_n.\stabilized = \INT_i s'.\stable_i[i]$, we have $x \in u_n.\stabilized$.
Hence $u_n \in \SpStab.\G$. Hence $(\ex i \in 0 \ldots n : u_i \in \SpStab.\G)$.

We have thus established  clause~\ref{clause:live-fwd-sim-inv:trans} of
Definition~\ref{def:live-fwd-sim-inv} for the pairs $\SpStab$ and $\ImpStab$.

\paragraph{Establishing clause~\ref{clause:live-fwd-sim-inv:silent} of
Definition~\ref{def:live-fwd-sim-inv}.}
From Figure~\ref{fig:replica}, it is clear that the action
$\send_{rr'}(m)$ (for some $m$) is continuously enabled, and hence
executed infinitely often in any live execution of $\ESDSAlg \pl
\Users$.  Hence, the action $\receive_{rr'}(m)$ is also executed
infinitely often. Now, according to the definition of $F$ (see
\cite{FGLLS99}, Section 8), $\receive_{rr'}(m)$ is simulated by
the sequence of actions
$\addcons(s'.po)$, $\stabilize(x_1)$, $\ldots$, $\stabilize(x_k)$, where
$\set{x_1,\ldots,x_k} = \INT_i s'.\stable_i[i]$, and $s'$ is the state
of $\ESDSAlg \pl \Users$
resulting from the execution of $\receive_{rr'}(m)$.  Thus,
$\receive_{rr'}(m)$ is always matched by at least one action, namely
$\addcons(s'.po)$. Hence, any transition generated by executing
$\receive_{rr'}(m)$ is not always-silent, by
clause~\ref{clause:live-fwd-sim-inv:silent} of
Definition~\ref{def:live-fwd-sim-inv}.  Since every live execution
of $\ESDSAlg \pl \Users$ contains an infinite number of these transitions,
clause~\ref{clause:live-fwd-sim-inv:silent} of
Definition~\ref{def:live-fwd-sim-inv} is satisfied.

\subsubsection{Establishing Membership in $\clos{L}$}

\paragraph{Establishing $\pair{x \in \wait_c}{x \not\in \wait_c} \in \clos{L}$.}

We use a complemented-pairs lattice over $\clos{L}$, together with
Lemma~\ref{lem:lattice}, to establish 
$\pair{x \in \wait_c}{x \not\in \wait_c} \in \clos{L}$.
Recall that $L$ is the
complemented-pairs liveness condition for the implementation
$\ESDSAlg \pl \Users$. At the implementation level, the natural
liveness hypothesis is that each continuously enabled action is
eventually executed, and each message in transit eventually arrives.
We use this hypothesis to justify the pairs in $L$ (which are also in
$\clos{L}$, by definition).
Figure~\ref{fig:lattice-req} shows the complemented-pairs lattice that
we use. $c = \client(x)$ is the
client that invoked operation $x$. We display the portion of the
lattice corresponding to a single replica $r$. The $\vdots$
indicate where isomorphic copies corresponding to the other replicas
occur (the number of replicas is finite).
Let $L$ consist of all the pairs in Figure~\ref{fig:lattice-req}. 
It is straightforward to verify that Figure~\ref{fig:lattice-req}
satisfies all the conditions of Definition~\ref{def:lattice}. 
\bfg
\begin{center}
\horline\\[1.5ex]
\begin{picture}(0,0)%
\includegraphics{lattice-req.pstex}%
\end{picture}%
\setlength{\unitlength}{3947sp}%
\begingroup\makeatletter\ifx\SetFigFont\undefined%
\gdef\SetFigFont#1#2#3#4#5{%
  \reset@font\fontsize{#1}{#2pt}%
  \fontfamily{#3}\fontseries{#4}\fontshape{#5}%
  \selectfont}%
\fi\endgroup%
\begin{picture}(7302,5011)(511,-5450)
\put(1201,-1411){\makebox(0,0)[lb]{\smash{{\SetFigFont{12}{14.4}{\rmdefault}{\mddefault}{\updefault}.}}}}
\put(1201,-1636){\makebox(0,0)[lb]{\smash{{\SetFigFont{12}{14.4}{\rmdefault}{\mddefault}{\updefault}.}}}}
\put(1201,-1861){\makebox(0,0)[lb]{\smash{{\SetFigFont{12}{14.4}{\rmdefault}{\mddefault}{\updefault}.}}}}
\put(7726,-1411){\makebox(0,0)[lb]{\smash{{\SetFigFont{12}{14.4}{\rmdefault}{\mddefault}{\updefault}.}}}}
\put(7726,-1636){\makebox(0,0)[lb]{\smash{{\SetFigFont{12}{14.4}{\rmdefault}{\mddefault}{\updefault}.}}}}
\put(7726,-1861){\makebox(0,0)[lb]{\smash{{\SetFigFont{12}{14.4}{\rmdefault}{\mddefault}{\updefault}.}}}}
\put(7726,-4036){\makebox(0,0)[lb]{\smash{{\SetFigFont{12}{14.4}{\rmdefault}{\mddefault}{\updefault}.}}}}
\put(7726,-4261){\makebox(0,0)[lb]{\smash{{\SetFigFont{12}{14.4}{\rmdefault}{\mddefault}{\updefault}.}}}}
\put(7726,-4486){\makebox(0,0)[lb]{\smash{{\SetFigFont{12}{14.4}{\rmdefault}{\mddefault}{\updefault}.}}}}
\put(1201,-4036){\makebox(0,0)[lb]{\smash{{\SetFigFont{12}{14.4}{\rmdefault}{\mddefault}{\updefault}.}}}}
\put(1201,-4261){\makebox(0,0)[lb]{\smash{{\SetFigFont{12}{14.4}{\rmdefault}{\mddefault}{\updefault}.}}}}
\put(1201,-4486){\makebox(0,0)[lb]{\smash{{\SetFigFont{12}{14.4}{\rmdefault}{\mddefault}{\updefault}.}}}}
\put(2776,-3586){\makebox(0,0)[lb]{\smash{{\SetFigFont{10}{12.0}{\rmdefault}{\mddefault}{\updefault}$\pair{x \in \pndg_r \ints \rcvd_r}{x \in \pndg_r \ints \done_r[r]}$}}}}
\put(2776,-1486){\makebox(0,0)[lb]{\smash{{\SetFigFont{10}{12.0}{\rmdefault}{\mddefault}{\updefault}$\pair{<\resp,x,v> \in \chan_{rc}}{(x,v) \in \rept_c}$}}}}
\put(526,-2386){\makebox(0,0)[lb]{\smash{{\SetFigFont{10}{12.0}{\rmdefault}{\mddefault}{\updefault}$\lpb x \in \pndg_r \ints \done_r[r] \land x.\strict,$}}}}
\put(601,-2581){\makebox(0,0)[lb]{\smash{{\SetFigFont{10}{12.0}{\rmdefault}{\mddefault}{\updefault}$<\resp,x,v> \in \chan_{rc}$}}}}
\put(601,-2776){\makebox(0,0)[lb]{\smash{{\SetFigFont{10}{12.0}{\rmdefault}{\mddefault}{\updefault}$\rpb$}}}}
\put(5026,-2386){\makebox(0,0)[lb]{\smash{{\SetFigFont{10}{12.0}{\rmdefault}{\mddefault}{\updefault}$\lpb x \in \pndg_r \ints \done_r[r] \land \neg x.\strict,$}}}}
\put(5026,-2776){\makebox(0,0)[lb]{\smash{{\SetFigFont{10}{12.0}{\rmdefault}{\mddefault}{\updefault}$\rpb$}}}}
\put(5101,-2581){\makebox(0,0)[lb]{\smash{{\SetFigFont{10}{12.0}{\rmdefault}{\mddefault}{\updefault}$<\resp,x,v> \in \chan_{rc}$}}}}
\put(2626,-4486){\makebox(0,0)[lb]{\smash{{\SetFigFont{10}{12.0}{\rmdefault}{\mddefault}{\updefault}$\pair{<\req, x> \in \chan_{cr}}{x \in \pndg_r \ints \rcvd_r}$}}}}
\put(2851,-5386){\makebox(0,0)[lb]{\smash{{\SetFigFont{10}{12.0}{\rmdefault}{\mddefault}{\updefault}$\pair{x \in \wait_c}{\ex r: <\req, x> \in \chan_{cr}}$}}}}
\put(3376,-586){\makebox(0,0)[lb]{\smash{{\SetFigFont{10}{12.0}{\rmdefault}{\mddefault}{\updefault}$\pair{(x,v) \in \rept_c}{x \not\in \wait_c}$}}}}
\end{picture}%

\end{center}
\horline
\caption{Complemented-pairs lattice that establishes
         $\pair{x \in \wait_c}{x \not\in \wait_c} \in \clos{L}$
         ($c = \client(x)$).}
\label{fig:lattice-req}
\efg
We justify the complemented-pairs in Figure~\ref{fig:lattice-req} as
follows:
\bn

\item $\pair{x \in \wait_c}{\ex r : <\req, x> \in \chan_{cr}}$.\\
$send_{cr}$ is continuously enabled and eventually happens, for at
least one replica $r$.

\item $\pair{<\req, x> \in \chan_{cr}}{x \in \pndg_r \ints \rcvd_r}$.\\
Liveness of $\chan_{cr}$, and the definition of action $\receive_{cr}$
in Figure~\ref{fig:replica}.

\item $\pair{x \in \pndg_r \ints \rcvd_r}{x \in \pndg_r \ints \done_r[r]}$.\\
If $x.\prev \subseteq \done_r[r]$ holds continuously, then
either $\doit_r$ is continuously enabled and eventually happens (making $x \in \done_r[r]$ true),
or $\doit_r$ is disabled because $x \in \done_r[r]$ becomes true due to a gossip message.
Establishing $x.\prev \subseteq \done_r[r]$ essentially requires a
``sublattice'' for each $x' \in x.\prev$. This sublattice is a
``chain'' consisting of three pairs, with the ordering (a) $\lprec$
(b) $\lprec$ (c):
   \bn

   \item $\pair{x \in \pndg_r \ints \rcvd_r}{x' \in \pndg_{r'} \ints \rcvd_{r'}}$
   is the bottom element. It is justified since each client
   includes in $x.\prev$ only operations that have already been requested. Thus
   $x' \in x.\prev$ is eventually received by some replica $r'$, at which point 
   $x' \in \pndg_{r'} \ints \rcvd_{r'}$ holds.

   \item $\pair{x' \in \pndg_{r'} \ints \rcvd_{r'}}
               {x' \in \pndg_{r'} \ints \done_{r'}[r']}$
   is the middle element.
   It is justified ``inductively,'' i.e., it can be expanded into a sublattice in
   exactly the same way as 
   $\pair{x \in \pndg_r \ints \rcvd_r}{x \in \pndg_r \ints \done_r[r]}$.
   This ``nested'' expansion is guaranteed to terminate however, since $x.\prev$
   is finite, for all $x$.

   \item $\pair{x' \in \done_{r'}[r']}{x' \in \done_r[r]}$ is the top
   element. It is justified since $r'$ eventually sends a gossip message to $r$.

   \en
By applying Lemma~\ref{lem:lattice} to this sublattice, we conclude 
$\pair{x \in \pndg_r \ints \rcvd_r}{x' \in \done_r[r]} \in \clos{L}$.
Now $\done_r[r]$ increases monotonically, $x' \in \done_r[r]$ is
stable---once true, it remains true.
Hence, from the aforementioned pair for each 
$x' \in x.\prev$, we conclude that $x.\prev \sub \done_r[r]$
eventually holds, and remains true subsequently, as required.

Note that the condition $l > label_r(y.id)$ does not need to be verified as
eventually holding, since it merely expresses a constraint on the value of the
``action parameter'' $l$, i.e., the only instances of $\doit_r(x,l)$ which are
enabled are those having values of $l$ that satisfy $l > label_r(y.id)$. That
is, $l$ is properly regarded as part of the ``name'' of the action
$\doit_r(x,l)$.

\item  $\pair{x \in \pndg_r \ints \done_r[r] \land x.\strict}{<\resp, x, v> \in \chan_{rc}}$.\\
\label{pairs-list:strict}
This is justified by the following sublattice, where the ordering
relation is (a) $\lprec$ (b) $\lprec$ (c) $\lprec$ (d) $\lprec$ (e).
$x \in \pndg_r$, $x.\strict$, 
are implicit conjuncts of all the predicates in the sublattice, except
the $\GREEN$ predicate of pair (e), and are omitted for clarity.
   \bn

   \item $\pair{x \in \ints \done_r[r]}{x \in \ints_{r'} \done_{r'}[r']}$.
   Justified since $r$ sends
   gossip messages to every other replica $r'$.

   \item $\pair{x \in \ints_{r'} \done_{r'}[r']}{x \in \stable_r[r]}$.
   Justified since each $r'$ sends gossip messages to $r$.

   \item $\pair{x \in \stable_r[r]}{x \in \ints_{r'} \stable_{r'}[r']}$.
   Justified since $r$ sends gossip messages to every other replica $r'$.

   \item $\pair{x \in \ints_{r'} \stable_{r'}[r']}
               {x \in \ints_{r'} \stable_{r}[r']}$.
   Justified since each $r'$ sends gossip messages to $r$.

   \item $\pair{x \in \ints_{r'} \stable_{r}[r']}
               {<\resp, x, v> \in \chan_{rc}}$.
   Justified since  $x \in \pndg_r$, 
   $x \in \done_{r}[r]$, and
   $x \in \ints_{r'} \stable_{r}[r']$ all hold continuously, since
   $\done_{r}[r]$ and $\stable_{r}[r']$ grow monotonically.
   Hence $send_{rc}(<\resp, x, v>)$ is continuously enabled, and so is
   eventually executed.

   \en

\item $\pair{x \in \pndg_r \ints \done_r[r] \land \neg x.\strict}
            {<\resp, x, v> \in \chan_{rc}}$.\\
$send_{rc}(<\resp, x, v>)$ is continuously enabled and eventually happens.

\item $\pair{<\resp, x, v> \in \chan_{rc}}{(x,v) \in \rept_c}$.\\
Liveness of $\chan_{rc}$, and the definition of action $\receive_{rc}$
in Figure~\ref{fig:front-end}.

\item $\pair{(x,v) \in \rept_c}{x \not\in \wait_c}$.\\
$\response(x,v)$ is continuously enabled and eventually happens.
\en

\paragraph{Establishing $\pair{x \in \wait_c}{x \in \INT_i \stable_i[i]} \in \clos{L}$.}

We use the complemented-pairs lattice over $\clos{L}$ given in 
Figure~\ref{fig:lattice-stab} together with
Lemma~\ref{lem:lattice}.
\bfg
\begin{center}
\horline\\[1.5ex]
\begin{picture}(0,0)%
\includegraphics{lattice-stab.pstex}%
\end{picture}%
\setlength{\unitlength}{3947sp}%
\begingroup\makeatletter\ifx\SetFigFont\undefined%
\gdef\SetFigFont#1#2#3#4#5{%
  \reset@font\fontsize{#1}{#2pt}%
  \fontfamily{#3}\fontseries{#4}\fontshape{#5}%
  \selectfont}%
\fi\endgroup%
\begin{picture}(6627,4786)(1186,-5450)
\put(7726,-4036){\makebox(0,0)[lb]{\smash{{\SetFigFont{12}{14.4}{\rmdefault}{\mddefault}{\updefault}.}}}}
\put(7726,-4261){\makebox(0,0)[lb]{\smash{{\SetFigFont{12}{14.4}{\rmdefault}{\mddefault}{\updefault}.}}}}
\put(7726,-4486){\makebox(0,0)[lb]{\smash{{\SetFigFont{12}{14.4}{\rmdefault}{\mddefault}{\updefault}.}}}}
\put(1201,-4036){\makebox(0,0)[lb]{\smash{{\SetFigFont{12}{14.4}{\rmdefault}{\mddefault}{\updefault}.}}}}
\put(1201,-4261){\makebox(0,0)[lb]{\smash{{\SetFigFont{12}{14.4}{\rmdefault}{\mddefault}{\updefault}.}}}}
\put(1201,-4486){\makebox(0,0)[lb]{\smash{{\SetFigFont{12}{14.4}{\rmdefault}{\mddefault}{\updefault}.}}}}
\put(1201,-1711){\makebox(0,0)[lb]{\smash{{\SetFigFont{12}{14.4}{\rmdefault}{\mddefault}{\updefault}.}}}}
\put(1201,-1936){\makebox(0,0)[lb]{\smash{{\SetFigFont{12}{14.4}{\rmdefault}{\mddefault}{\updefault}.}}}}
\put(1201,-2161){\makebox(0,0)[lb]{\smash{{\SetFigFont{12}{14.4}{\rmdefault}{\mddefault}{\updefault}.}}}}
\put(7726,-1711){\makebox(0,0)[lb]{\smash{{\SetFigFont{12}{14.4}{\rmdefault}{\mddefault}{\updefault}.}}}}
\put(7726,-1936){\makebox(0,0)[lb]{\smash{{\SetFigFont{12}{14.4}{\rmdefault}{\mddefault}{\updefault}.}}}}
\put(7726,-2161){\makebox(0,0)[lb]{\smash{{\SetFigFont{12}{14.4}{\rmdefault}{\mddefault}{\updefault}.}}}}
\put(2776,-3586){\makebox(0,0)[lb]{\smash{{\SetFigFont{10}{12.0}{\rmdefault}{\mddefault}{\updefault}$\pair{x \in \pndg_r \ints \rcvd_r}{x \in \pndg_r \ints \done_r[r]}$}}}}
\put(2626,-4486){\makebox(0,0)[lb]{\smash{{\SetFigFont{10}{12.0}{\rmdefault}{\mddefault}{\updefault}$\pair{<\req, x> \in \chan_{cr}}{x \in \pndg_r \ints \rcvd_r}$}}}}
\put(2851,-5386){\makebox(0,0)[lb]{\smash{{\SetFigFont{10}{12.0}{\rmdefault}{\mddefault}{\updefault}$\pair{x \in \wait_c}{\ex r: <\req, x> \in \chan_{cr}}$}}}}
\put(3301,-2611){\makebox(0,0)[lb]{\smash{{\SetFigFont{10}{12.0}{\rmdefault}{\mddefault}{\updefault}$\pair{x \in \done_r[r]}{x \in \INT_i \done_i[i]}$}}}}
\put(3226,-811){\makebox(0,0)[lb]{\smash{{\SetFigFont{10}{12.0}{\rmdefault}{\mddefault}{\updefault}$\pair{x \in \stable_r[r]}{x \in \INT_i \stable_i[i]}$}}}}
\put(3271,-1711){\makebox(0,0)[lb]{\smash{{\SetFigFont{10}{12.0}{\rmdefault}{\mddefault}{\updefault}$\pair{x \in \INT_i \done_i[i]}{x \in \stable_r[r]}$}}}}
\end{picture}%

\end{center}
\horline
\caption{Complemented-pairs lattice that establishes
         $\pair{x \in \wait_c}{x \in \INT_i \stable_i[i]} \in \clos{L}$
         ($c = \client(x)$).}
\label{fig:lattice-stab}
\efg
The bottom three complemented-pairs in Figure~\ref{fig:lattice-stab}
also occur in Figure~\ref{fig:lattice-req}, and have therefore already
been justified. We justify the remaining pairs as follows.
\bn

\item $\pair{x \in \done_r[r]}
            {x \in \INT_{i} \done_{i}[i]}$.
Justified since $r$ sends
gossip messages to every other replica.

\item $\pair{x \in \INT_{i} \done_{i}[i]}
            {x \in \stable_r[r]}$.
Justified since each $i$ sends gossip messages to $r$.

\item $\pair{x \in \stable_r[r]}
            {x \in \INT_{i} \stable_{i}[i]}$.
Justified since $r$ sends gossip messages to every other replica.

\en

Since Definition~\ref{def:live-fwd-sim-inv} is now satisfied, we have established
$(\ESDSAlg \pl \Users, L) \lpsim (\ESDSII \pl \Users, \MII)$.  Hence, applying
Theorem~\ref{thm:liveness}, we conclude $(\ESDSAlg \pl \Users, L) \lpreorder
(\ESDSII \pl \Users, \MII)$.
Together with 
$(\ESDSII \pl \Users, \MII) \lpreorder (\ESDSI \pl \Users, \MI)$
established above, we have
$(\ESDSAlg \pl \Users, L) \lpreorder
(\ESDSI \pl \Users, \MI)$, as desired.

We have illustrated three levels of abstraction, and two
liveness-preserving forward simulations, between the top and middle,
and middle and bottom levels. It is straightforward to continue this
process. For example, an actual implementation would not simply route
a request to any replica, but would select the replica according to
certain criteria, for example load balancing/performance \cite{MDZ99},
or distance from the client \cite{PK00}.
Thus, the front-ends
and replicas would be refined to incorporate a
load-balancing/anycast/replica (or mirror) location ``service'' which,
given a request from a client $c$, assigns some replica $r$ to service
that request. We then map the complemented-pair 
$\pair{x \in \wait_c}{\ex r : <\req, x> \in \chan_{cr}}$
into a pair at the next lower level
which expresses the liveness of the service: the service
eventually assigns some replica $r$ to every request $x$.  This pair could then
be justified by constructing a lattice whose elements are the
specified or derived liveness properties of the service.

\section{Discussion}
\label{sec:discussion}

\subsection{Alternative Choices for Specifying Liveness Properties}
\label{sec:alternatives}

We have used the complemented-pairs acceptance condition to specify
liveness properties. There are other acceptance conditions 
for finite automata over infinite strings that we could have
chosen: Buchi, generalized-Buchi, Rabin, and Muller. We briefly
discuss each in turn.

A Buchi condition is a single set $\GREEN$ of states, and the
computation must contain an infinite number of states from $\GREEN$.
This can be expressed as a single complemented
pair $\pair{\true}{\GREEN}$, and so is subsumed by complemented-pairs.
A generalized-Buchi condition  is a set 
$\{ \GREEN_i ~|~ i \in \eta \}$ 
of sets of states, and for each $\GREEN_i$, the computation should
contain an infinite number of states from $\GREEN_i$.
This can be expressed as the set of complemented-pairs
$\{ \pair{\true}{\GREEN_i} ~|~ i \in \eta \}$
and so is also subsumed by complemented-pairs.

The Rabin condition is a set 
$\{ \pair{\true}{\GREEN_i} ~|~ i \in \eta \}$
of pairs, however the acceptance condition is different.
A computation $\al$ is accepted iff for some pair
$\pair{\RED_i}{\GREEN_i}$, 
$\al$ does not contain an infinite number of states in $\RED_i$, and
$\al$ does contain an infinite number of states in $\GREEN_i$.
This condition is a ``disjunctive'' one, it constrains a computation
only with respect to any one of the pairs, not all of them at once.
Since, in writing specifications, conjunction is far more useful than
disjunction, i.e., we typically list some properties \emph{all} of which
must be satisfied, we feel that this condition would not be useful in
practice.

The Muller condition is a set 
$\{ \GREEN_i ~|~ i \in \eta \}$
of sets of states, and, the set of states that
occur infinitely often along the computation should be exactly one of
the $\GREEN_i$. This condition is not very suitable for an
infinite-state model, since it is possible (and indeed, often the
case) that an infinite computation does not contain any particular state
that recurs infinitely often, since the model usually contains
unbounded data, such as integers, reals, sequences, or sets.
Thus, the set of states \emph{each of which} 
occurs infinitely often along the computation, is usually empty.

Finally, we consider the ``temporal leads-to'' property.  Roughly, $p$
leads-to $q$ means that, whenever $p$ holds, then $q$ subsequently 
holds. In our framework, leads-to properties can be expressed and
verified by using history variables. 
Let $\flag_p$ be a boolean history variable that is initially false,
is set whenever $p \land \neg q$ holds, and reset whenever $q$ holds.
Then, the complemented-pair $\pair{\flag_p}{q}$ expresses ``$p$ leads-to $q$.''
Since $\flag_p$ is not used to affect control flow, it
does not need to be ``implemented.'' Thus, the issue of
atomically detecting the values of $p$ and $q$ at run time and updating
$\flag_p$, does not arise.

\remove{
\subsection{Open Problems for Infinite State Acceptance Conditions}
\label{sec:open-problems}

...
}

\subsection{Application to Fault-tolerance}
\label{sec:fault-tolerance}

Our method can be applied to the verification of \emph{fault
tolerance} properties. We consider situations in which the occurrence
of a fault can cause the system to enter a ``bad'' state, i.e., one
that is unreachable under normal execution \cite{AAE98}.
Let $\good$ denote the set of states that are reachable under normal system
execution from a start state, and
let $\fault$ denote the set of states that result immediately
after a fault occurs, i.e., the post-states of faults (the faults can
occur in any state, good or bad). 
If follows that, under normal execution (no faults) only good states
are reachable from good states.

\remove{
Thus, $\good \un \bad$ is the set of states reachable when the system
execution is subject to faults.
}

We are interested in ``nonmasking'' fault-tolerance properties of the type: once
faults stop occurring, the system will eventually recover to a good state (and
therefore remain forever after in good states, since only good states
are reachable from good states in the absence of faults). 
Expressed in temporal logic, this is $(\ea \neg fault \implies \iof good)$.
This is logically equivalent to $\iof (fault \lor good)$. We can
express this as the complemented pair $\pair{\true}{fault \lor good}$.

\remove{
The constraint that eventually
faults stop occurring means that the number of faults that occur is finite. This
can be expressed as the pair $\pair{\fault}{\emptyset}$. The recovery property
can be expressed as the pair $\pair{\bad}{\good}$.  Since $\bad \un \good$ exhausts
all the states reachable under fault-prone execution, this means that every bad
state is eventually followed by a good state, or equivalently, a good state
eventually occurs, i.e., the pair $\pair{\true}{\good}$.
}

Hence, the liveness condition $\pair{\true}{fault \lor good}$ defines
the set of ``live'' executions to be either 
(1) those along which an
infinite number of faults occur (in which case we have no obligation
to recover to a good state) or 
(2) those along which an infinite number of good
states occur. In the latter case, we may also assume that faults stop
occuring, since the negation of this is covered by case (1).
Since only good states are reachable from good states, it follows that
there is some suffix consisting entirely of good states, and so the
system has recovered.

\remove{
only a finite number of faults occur, and furthermore, every bad state
is eventually followed by a good state.  In the suffix of an execution
in which faults no longer occur, the property of being in a good state
is stable: once the system is in a good state, it remains in a good
state.  Thus, the live executions are those in which only a finite
number of faults occur, and eventually the execution reaches a good
state and remains in good states forever after. This is precisely the
fault-tolerance property we wish to verify.
}

Thus, ``live'' executions are those in which the system exhibits the
desired fault-tolerance property.  The trace of such an execution is
then an ``external fault-tolerant behavior.''

We can now refine such nonmasking fault-tolerance properties, i.e., to
establish that the external fault-tolerant behaviors of an
implementation are included in those of the specification.  Our
framework thus can take the place of theories that are specialized to
dealing with nonmasking fault-tolerance, e.g., \cite{DA02},
which we have shown is just a particular kind of liveness property.

\subsection{Mechanization Of Our Method}
\label{sec:mechanization}

Our method imposes the following proof obligations:
\bn

\item Devise an appropriate liveness-preserving simulation and check
that it satsifies all of the conditions of its definition
(one of Definitions~\ref{def:live-fwd-sim-inv}--\ref{def:live-proph-inv}).

\item For each derived pair, devise a complemented-pairs lattice and check
  that it satisfies the conditions of Definition~\ref{def:lattice}.

\en

These conditions can be formalized in a first-order assertion language
with interpreted symbols. We refer the reader to \cite{GL00,GV04} for
details. The conditions can be verified using theorem provers
such as PVS \cite{OSR92}. For lack of space, we omit an extended
discussion of these issues, which can be found, for example, in
\cite{GV04}.
That paper presents \empi{normed simulations}, where the 
existence of a finite execution fragment at the abstract level that
matches a concrete transition is replaced by the existence 
of either a single matching transition, or an internal transition that
decreases a supplied norm (a function over a well-founded domain).
It should be possible to extend the ideas in this paper to normed
simulations. For example, if the concrete transition contains a $\RED$
state, then we require that, by the time that either the matching
abstract transition has been generated, or the norm function has
decreased to minimum, that a corresponding $\RED$ state has appeared
at the abstract level. We leave the details to another occasion.

\section{Expressive Completeness of Complemented-pairs Liveness Conditions}
\label{sec:expressiveness}

We now investigate the expressiveness of complemented-pairs: what 
are the live execution properties which can be expressed by
complemented-pairs conditions? First, we make this notion precise.

\bd
Let $A$ be an automaton and 
let $\varphi$ be a live execution property for $A$. Then we say that
a liveness condition $L$ \intrdef{expresses} $\varphi$ if and only if 
$(A,L)$ is a live automaton and $\lexecs(A,L) = \varphi$.
\ed

The use of (complemented-pairs) liveness conditions to specify
liveness means that the liveness of an execution depends only on the
set of states which occur in that execution, and not on their
ordering. This is necessary, to satisfy the machine closure condition,
since ordering is a safety property: once an ordering is violated
along a finite execution, no extension can then satisfy the ordering.

In Section~\ref{sec:completeness}, we show that, under some
assumptions that are natural for infinite-state systems, that the
generalized Buchi condition is expressively complete, i.e., they can
express any live execution property. Since complemented pairs subsumes
generalized Buchi, the result then carries over to our framework.

In Section~\ref{sec:forest}, we show that complemented pairs are
expressively complete if history variables can be used.

\remove{
In a strict sense, this implies a loss of expressive power.  This is
however an unavoidable consequence of reducing reasoning about
liveness from reasoning over entire executions to reasoning over
individual states and finite execution fragments. 
Moreover, this expressiveness can
be regained by using history variables, which record some aspect of
the execution up to the current state. At the extreme, we can use a
history variable which maintains a complete record of the execution so
far, and in this case, as we show below (Section~\ref{sec:forest}),
any live execution property
is expressible by some complemented-pairs condition. Using such a
history variable is tantamount to introducing aspects of reasoning
about finite executions into our method, in the guise of reasoning
about individual states/transitions.
Thus, by using history variables that record only a few aspects of the
current execution, versus history variables that record everything
which has occurred so far, we can trade off expressiveness 
versus an increased proof burden. Furthermore, in all cases, our
method avoids reasoning about \emph{infinite} executions, which, for
example, methods based on fairness must do.
}

\subsection{Relative Expressive Completeness of Complemented-pairs
  Liveness Conditions}
\label{sec:completeness}

\remove{
We now show that, under some assumptions that are natural for
infinite-state systems, that complemented-pairs conditions are
expressively complete, i.e, they can express any live execution
property.
}

In infinite-state systems, it is often the case that the occurrence of
``significant'' events is permanently recorded by changes to the state.
For instance, in the eventually-serializable data service of
Section~\ref{sec:example}, the execution of every operation on the
data results
in a permanent record of that operation's unique identifier. 
Any database system which maintains logs is also an example of this.
So is a real-time system in which clocks maintain the time,
if we consider the passage of time to be a
significant event.
This large class of systems justifies the assumption that a particular
state cannot repeat infinitely often along a live execution, since we
expect that significant events (e.g., operation execution, transaction
commit, time passage) occur infinitely often along a live execution.
Thus, we assume the following condition in this section:

\begin{assumption}[No infinite repetition]
\label{ass:no-infinite-repetition}
Let $(A,L)$ be a live automaton, and
$\al = s_0 a_1 s_1 \ldots$ be a live execution of $(A,L)$. Then, there
is no state $s$ such that $s = s_i$ for an infinite number of values
for the index $i$.  
That is, no state occurs infinitely often along $\al$.
\end{assumption}

Since a generalized-Buchi condition depends only on the set of states
which occur in that execution, we take it as reasonable that if one
execution contains ``more'' states than another, and the latter
execution is live, then the former execution should also be live. 
In this section, we restrict attention to liveness properties which
satisfy this condition, which we call \emph{robust} properties.
Our notion of one execution containing ``more'' states than another 
is captured by a relation $\subs$ between executions.

\bd[$\subs$]
\label{def:subsumption}
Let $\al = s_0 a_1 s_1 \ldots$ and $\ga$ be infinite executions of automaton $A$.
Then $\ga \subs \al$ iff there exists a suffix $\ga' = u_0 b_1 u_1 \ldots$ of $\ga$ 
and a mapping $m: \{0,1,\ldots\} \mapsto \{0,1,\ldots\}$ such that 
\bn
\item $\fa i \ge 0: s_{m(i)} = u_i$, and 
\item $\fa i \ge 0: m^{-1}(i)$ is a finite set.
\en
\ed

Thus, $\ga \subs \al$ iff $\ga$ has some suffix $\ga'$ which can be
put into a correspondence with $\al$ as follows. If a state $s$
occurs some finite $(> 0)$ number of times in $\ga'$, 
then state $s$ also occurs some finite number of times in $\al$.
If $s$ occurs infinitely often in $\ga'$, then $s$ also occurs
infinitely often in $\al$. Note that Assumption~\ref{ass:no-infinite-repetition}
does not rule this out, since it applies only to live executions.
$\subs$ is clearly reflexive and transitive, and so is a preorder.
We formalize the condition discussed above as the class of 
\intr{robust} live execution properties.

\bd[Robust Live Execution Property]
\label{def:robust-live-execution-property}
Let $\varphi$ be a live execution property for automaton $A$.
Then, $\varphi$ is \emph{robust for $A$} if and only if:\\
\ind for all $\ga, \al \in \iexecs(A)$, 
       if $\ga \subs \al$ and $\ga \in \varphi$ then $\al \in \varphi$.
\ed

Our robustness condition corresponds more closely to using a
generalized-Buchi acceptance condition than a complemented-pairs
acceptance condition (see Section~\ref{sec:alternatives} above). Since
complemented-pairs subsume generalized-Buchi, this is still within our
framework, and also allows for a simpler technical development.
The definition of live trace properties corresponding to robust live
execution properties is straightforward.

\bd[Robust Live Trace Property]
\label{def:robust-live-trace-property}
Let $A$ be an automaton, and $\psi \sub \traces(A)$. Then,
$\psi$ is a \emph{robust live trace property} for $A$ if and only if
there exists a robust live execution property $\varphi$ for $A$ such that 
$\psi = \traces(\varphi)$.
\ed

\remove{
We can now establish our relative completeness result: if $\varphi$ is
a robust live execution property, then there exists a
complemented-pairs liveness condition $L$ which expresses $\varphi$, that
is, the executions which satisfy $L$ are exactly those in $\varphi$.
}

We now show that an execution in $\varphi$ can be distinguished
from an execution outside $\varphi$ by means of a simple Buchi
acceptance condition.
For an execution $\al$, define $\states(\al) = \{ s ~|~ s
\mbox{ occurs along } \al \}$.

\bp
\label{prop:distinguish-two-execs}
Let $A$ be an automaton, and let $\varphi$ be an arbitrary
robust live execution property for $A$.
Let  $\ga, \al \in \iexecs(A)$ be such that 
$\ga \in \varphi$ and $\al \not\in \varphi$. 
Then there exists a set $\G_{\al,\ga} \sub \states(A)$ such that
$\ga \sat \iof \G_{\al,\ga}$ and $\al \sat \always \neg \G_{\al,\ga}$.
\ep
\bpr
Since $\ga$ is an infinite execution, we have by 
Assumption~\ref{ass:no-infinite-repetition} that $\states(\ga)$ is an infinite set. 
Now suppose that $\states(\ga) - \states(\al)$ is a finite set.
Then, by Assumption~\ref{ass:no-infinite-repetition}, there exists a suffix 
$\ga'$ of $\ga$ which contains no state in $\states(\ga) - \states(\al)$.
Hence $\states(\ga') \sub \states(\al)$. 
By Assumption~\ref{ass:no-infinite-repetition}
each state along $\ga'$ repeats only a finite number of times. 
Hence we have $\ga' \subs \al$ by Definition~\ref{def:subsumption}.
Hence $\ga \subs \al$, again by Definition~\ref{def:subsumption}.
Thus by Definition~\ref{def:robust-live-execution-property}, $\al \in \varphi$,
contrary to assumption. 
We conclude that $\states(\ga) - \states(\al)$ is an infinite set.
Thus $\ga \sat \iof (\states(\ga) - \states(\al))$.
Also, $\al \sat \always \neg (\states(\ga) - \states(\al))$, by definition.
So, letting $\G_{\al,\ga} = \states(\ga) - \states(\al)$ establishes the proposition.
\epr

We next show that an execution outside $\varphi$ can be distinguished
from every execution inside $\varphi$ by means of a simple Buchi
acceptance condition.

\bp
\label{prop:distinguish-nonlive-exec}
Let $A$ be an automaton, and let $\varphi$ be an arbitrary
robust live execution property for $A$.
Let  $\al \in \iexecs(A)$ be such that $\al \not\in \varphi$. 
Then there exists a set $\G_{\al} \sub \states(A)$ such that
$\al \sat \Box \neg \G_{\al}$ and
$\fa \ga \in \varphi: \ga \sat \iof \G_{\al}$.
\ep
\bpr
Let $\ga$ be an arbitrary execution in $\varphi$, and let $\G_{\al,\ga}$ be the set
given by Proposition~\ref{prop:distinguish-two-execs} for $\al$, $\ga$.
Then $\ga \sat \iof \G_{\al,\ga}$ and $\al \sat \Box \neg \G_{\al,\ga}$.
Let $\G_{\al} = \UN_{\ga \in \varphi} \G_{\al,\ga}$. Then,
$\fa \ga \in \varphi: \ga \sat \iof \G_{\al}$, since $\G_{\al,\ga} \sub \G_{\al}$.
Also, 
$\al \sat \Box \neg \G_{\al}$ since $\al \sat \Box \neg \G_{\al,\ga}$ for every 
$\G_{\al,\ga}$, $\ga \in \varphi$. 
\epr

We now present the relative completeness result: every 
execution outside $\varphi$ can be distinguished
from every execution inside $\varphi$ by means of a 
generalized-Buchi acceptance condition.

\remove{
We now present the relative completeness result: every 
execution outside $\varphi$ can be distinguished
from every execution inside $\varphi$ by means of a 
complemented-pairs liveness condition. In fact, since the condition we
construct has $\true$ for the $\RED$ sets of all its pairs, it is
also a generalized-Buchi acceptance condition.
}

\bt[Relative Expressive Completeness of Generalized-Buchi]
\label{thm:execs-relative-completeness}
Let $A$ be an automaton, and let $\varphi$ be an arbitrary
robust live execution property for $A$. 
Then there exists a generalized-Buchi condition 
$L = \{ \G_i ~|~ i \in \eta \}$ 
over $A$ such that                     %
$\varphi = \{ \ga ~|~ \fa {i \in \eta}:  \ga \sat \iof \G_i \}$.
\et
\bpr
If $\varphi = \iexecs(A)$ then letting $L = \{ \true \}$
establishes the theorem. Hence we assume that
$\varphi$ is a proper subset of $\iexecs(A)$ for the rest of the proof.
Let $\al$ be an arbitrary execution in $\iexecs(A) - \varphi$, and let
$\G_{\al}$ be as given in Proposition~\ref{prop:distinguish-nonlive-exec} for $\al$.
Let $L = \{ \G_{\al} ~|~ \al \in \iexecs(A) - \varphi\}$.
Define $\execs(A,L) = 
\{ \ga ~|~ \fa {\al \in \iexecs(A) - \varphi}: \ga \sat \iof \G_{\al} \}$.
We show that $\varphi = \execs(A,L)$.
The proof is by double-containment.

\noindent
{$\execs(A,L) \sub \varphi$}:
Choose arbitrarily $\al \not\in \varphi$.
So $\al \in \iexecs(A) - \varphi$.
Hence $\al \sat \Box \neg \G_{\al}$ by 
Proposition~\ref{prop:distinguish-nonlive-exec}, and so 
$\al \not\sat \iof \G_{\al}$.
Thus $\al \not\in \execs(A,L)$ by definition of $\execs(A,L)$.
Taking the contrapositive yields 
$\al \in \execs(A,L)$ implies $\al \in \varphi$, i.e., $\execs(A,L) \sub \varphi$.

\noindent
{$\varphi \sub \execs(A,L)$}:
Choose arbitrarily $\ga \in \varphi$ and $\al \in \iexecs(A) - \varphi$.
Hence $\ga \sat \iof \G_{\al}$ by
Proposition~\ref{prop:distinguish-nonlive-exec}. 
Hence $\fa \al \in \iexecs(A) - \varphi:  \ga \sat \iof \G_{\al}$.
Hence $\ga \in \execs(A,L)$ by definition of $\execs(A,L)$.
Thus $\varphi \sub \execs(A,L)$.
\remove{
Since $\execs(A,L) = \varphi$, we have by Definition~\ref{def:liveness-property}
that $L$ satisfies the constraint of Definition~\ref{def:live-automaton}.
Hence $L$ is a liveness condition over $A$. This concludes the proof.}
\epr

\bco[Relative Expressive Completeness of Complemented-pairs]
\label{thm:traces-relative-completeness}
Let $A$ be an automaton, and let $\psi$ be an arbitrary
robust live trace property for $A$. Then there exists a 
complemented-pairs liveness condition
$L$ over $A$ such that $\traces(\execs(A,L)) = \psi$.
\eco
\bpr
Let $\psi$ be an arbitrary robust live trace property for $A$.
By Definition~\ref{def:robust-live-trace-property}, 
there exists a robust live execution property $\varphi$ for $A$ such that 
$\psi = \traces(\varphi)$.
By Theorem~\ref{thm:execs-relative-completeness}, there exists a 
generalized-Buchi condition $\{ \G_i ~|~ i \in \eta \}$ over $A$
such that 
$\varphi = \{ \ga ~|~ \fa {i \in \eta} : \ga \sat \iof \G_i \}$.
Let $L = \{ \pair{\true}{\G_i} ~|~ i \in \eta \}$. Then 
$\execs(A,L) = \varphi$.
Hence there exists a complemented-pairs liveness condition
$L$ over $A$ such that $\traces(\execs(A,L)) = \psi$.
\epr

\subsection{Expressive Completeness of Complemented-pairs for Liveness Properties of 
            Forest Automata}
\label{sec:forest}

An automaton $A$ is a \intrdef{forest automaton} iff for each reachable state
$s$ of $A$, there is exactly one (finite) execution of $A$ with last
state $s$. 
Thus, if $\al, \al'$ are arbitrary different infinite executions of $A$, then
they have only a finite number of states in common.
Any automaton can be turned into a forest automaton by
adding a history variable which records the execution up to the current
state. While this is obviously impractical for a real implementation,
such a variable is only needed for modeling and analysis purposes; it
does not have to be implemented since it does not affect the actual 
execution of the automaton.\footnote{The terms ``ghost variable'' and
``auxiliary variable'' have been used in the literature for this notion.}

Let $\al$ be an arbitrary infinite execution of $A$.
Define $\pr(\al) = \pair{\states(\al)}{\emptyset}$.

\bp
\label{prop:exec-pair}
Let $A$ be a forest automaton. Then
$\fa \al, \al' \in \iexecs(A): \al' \ne \al \mbox{~iff~} \al' \sat \pr(\al)$.
\ep
\bpr
Let $\al, \al'$ be arbitrary elements of $\iexecs(A)$.
If $\al' \ne \al$, then $\al' \sat \ea \neg \states(\al)$,
since $\al, \al'$ have only a finite number of states in common.
Hence $\al' \not\sat \iof \states(\al)$, and so $\al' \sat \pr(\al)$.
If $\al' = \al$, then $\al' \sat \iof \states(\al)$, and so 
      $\al' \not\sat \pr(\al)$.
\epr

We show that, if $\varphi$ is a live execution
property for automaton $A$, then there exists a liveness
condition which expresses $\varphi$, i.e. such that an execution
satisfies every complemented-pair in the condition iff it is a member
of $\varphi$.

\bt[Expressive Completeness of Complemented-pairs for Forest Automata]
\label{thm:execs-forest-completeness}
Let $A$ be a forest automaton, and let $\varphi$ be an arbitrary
live execution property for $A$.
Then there exists a complemented-pairs liveness condition
$L$ over $A$ such that $\execs(A,L) = \varphi$.
\et
\bpr
If $\varphi = \iexecs(A)$ then letting $L = \{ \pair{\true}{\true} \}$
establishes the theorem. Hence we assume that
$\varphi$ is a proper subset of $\iexecs(A)$ for the rest of the proof.
Let $L = \{\pr(\al) ~|~ \al \in \iexecs(A) - \varphi\}$.
We show that $\execs(A,L) = \varphi$. The proof is by double-containment.

\noindent
{$\execs(A,L) \sub \varphi$}:
Choose arbitrarily $\al' \in \execs(A,L)$ and $\al \in \iexecs(A) - \varphi$.
Now $\execs(A,L) \sub \iexecs(A)$ by definition, and so $\al' \in \iexecs(A)$.
From the definition of $L$, we have $\al' \sat \pr(\al)$.
Hence, by Proposition~\ref{prop:exec-pair}, $\al \ne \al'$.
Since $\al$ was chosen arbitrarily from 
$\iexecs(A) - \varphi$, we conclude $\al' \not\in \iexecs(A) - \varphi$.
Hence  $\al' \in \varphi$, since $\al' \in \iexecs(A)$. 

\noindent
{$\varphi \sub \execs(A,L)$}:
Choose arbitrarily $\al' \in \varphi$ and $\al \in \iexecs(A) - \varphi$.
Hence $\al \ne \al'$.
Hence, by Proposition~\ref{prop:exec-pair}, $\al' \sat \pr(\al)$.
Since $\al$ was chosen arbitrarily from 
$\iexecs(A) - \varphi$, we conclude, from the definition of $L$, that 
$\al' \in \lexecs(A,L)$.
\remove{
Since $\execs(A,L) = \varphi$, we have by Definition~\ref{def:liveness-property}
that $L$ satisfies the constraint of Definition~\ref{def:live-automaton}.
Hence $L$ is a liveness condition over $A$. This concludes the proof.}
\epr

\bco[Expressive Completeness of Complemented-pairs for Forest Automata]
\label{thm:traces-forest-completeness}
Let $A$ be a forest automaton, and let $\psi$ be an arbitrary
live trace property for $A$. Then there exists a complemented-pairs liveness condition
$L$ over $A$ such that $\traces(\execs(A,L)) = \psi$.
\eco
\bpr
Let $\psi$ be an arbitrary live trace property for $A$.
By Definition~\ref{def:live-trace-property}, 
there exists a live execution property $\varphi$ for $A$ such that 
$\psi = \traces(\varphi)$.
By Theorem~\ref{thm:execs-forest-completeness}, there exists a liveness condition
$L$ over $A$ such that $\execs(A,L) = \varphi$.
Hence there exists a liveness condition
$L$ over $A$ such that $\traces(\execs(A,L)) = \psi$.
\epr

\section{Related Work}
\label{sec:related}

The use of an infinite number of complemented pairs was proposed by
Vardi \cite{Var91}, which defines a recursive Streett automaton to be
one whose transition relation is recursive, and whose complemented
pairs are defined by recursive sets. Recursive Buchi automata are
defined similarly. Recursive Wolper automata are those with a
recursive transition relation and no acceptance conditions. Every
infinite run of the Wolper automaton is accepting.
The paper shows that Recursive Wolper, Buchi, and Street
automata all accept the same set of languages, namely $\Sigma_1^1$.
In our approach, we make no restrictions on the set of complemented
pairs. For example, we allow uncountable sets of pairs, which could be
useful for specifications over uncountable domains, e.g., the reals.

The safety-liveness classification was first proposed in \cite{Lam77}.
Formal characterizations of safety and liveness, variously based on Buchi
automata, temporal logic, or the Borel hierarchy, 
were given in \cite{AS87,MP90,Sis94}. 
Many researchers have proposed deductive systems for proving
properties of infinite-state reactive and distributed systems, including liveness
properties, e.g., \cite{AS89,Lam94,Lam02,MP91}.
Some of the methods proposed to date incorporate diagrammatic
techniques, similar in spirit to our complemented-pairs
lattices. 
In particular, Owicki and Lamport \cite{OL82} propose \emph{proof lattices},
and Manna and Pnueli \cite{MP83,MP93} propose
\emph{proof diagrams}, both for establishing liveness properties of
concurrent programs. 
In \cite{MP94}, Manna and Pnueli propose three different kinds of 
\emph{verification diagrams}, two for safety properties, and one for
liveness properties of the form 
$\always(U \implies \eventually V)$, where $U,V$ are state-assertions,
that is, temporal leads-to properties. Nodes in this diagram are
labeled with state-assertions, and directed edges between nodes
represent program transitions. 
Some of these edges correspond to ``helpful'' transitions, which are
guaranteed to occur (using fairness) if execution enters their source
node, and whose occurrence makes progress towards making $V$ true.
Browne et. al. \cite{BMS95} and Manna et. al. \cite{MBSU98} present
\emph{generalized verification diagrams}, 
which can be used to establish arbitrary temporal properties of
programs, including liveness properties. These are a particular kind
of $\omega$-automaton (``formula automata'').
These methods relate a program, expressed in an operational notation,
to a property expressed in temporal logic, i.e., they relate two
artifacts expressed in very different notations.  Thus, they cannot be
used to refine liveness properties in a multi-stage stepwise
refinement method that, starting with a high-level specification,
expressed in a particular (operational) notation, 
constructs a sequence of artifacts, all expressed in the \emph{same} notation, 
and each a refinement of the previous one, and ending with the
detailed implementation.

Our complemented-pairs lattices relate a liveness property of an
automaton, to a liveness property of a lower level automaton, i.e.,
the relationship is between two artifacts expressed in the same
notation. This forms the basis for a multi-stage proof
technique that refines high-level liveness properties down to the
liveness properties of an implementation in several manageable
steps (our use of ``sublattices'' in Section~\ref{sec:example} is an
example of this). Furthermore, each indivudual refinement step is
itself decomposed into the tasks involved in constructing lattices and 
discharging the associated ``verification conditions.''
We feel that this ability to decompose a liveness proof into
multiple stages directly attacks the scalability problem,
and is one of our main contributions.
\remove{
[[[removed in response to referees]]]
In addition, all of the above discussed proof systems rely on either
well-foundedness arguments or on fairness, to establish liveness
properties. Thus, they inherently require reasoning over entire executions.
}
UNITY \cite{CM88} provides a framework in which a subclass of general
liveness properties, namely ``leads-to'' can be verified and
refined. The approach is proof theoretic, and also relies on
fairness. We showed in Section~\ref{sec:alternatives} above how to
deal with leads-to properties in our framework.
All of the aforementioned methods operate only at the level of
executions, and do not provide a notion of external behavior, such as
a set of traces.

Gawlick et. al. \cite{GSSL93,GSSL98} presents a proof method
for liveness properties.  In that paper, a
liveness property of an automaton $A$ is modeled as a subset $L$ of
the executions of $A$.\footnote{$L$ must satisfy the machine closure
constraint of Definition~\ref{def:live-automaton}.}
However, the method presented there imposes a proof obligation
concerning the liveness of individual executions, without providing
any rule or method for discharging this obligation. Specifically, 
in addition to establishing a simulation, we have to
show that if an execution $\alpha$ of the implementation $A$
corresponds to an execution $\alpha'$ of the specification $B$,
and $\alpha$ is live (i.e., $\alpha$ is a member of the liveness
property), then $\alpha'$ is also live\footnote{See \cite{GSSL93}, page 89.}.
Merely establishing a simulation between $A$ and $B$
is insufficient to show this, since the simulation relation makes no
reference to the liveness conditions of $A$ and $B$.
The main concern in \cite{GSSL93} is the interaction between
liveness properties and parallel composition; a
notion of ``environment-freedom'' is introduced which 
enables the use of compositional verification for liveness.
The published version \cite{GSSL98} omits the proof method.

Likewise, Jensen~\cite{Jen99} presents simulation relations for 
proving liveness properties, and also requires that an ``inclusion''
condition be verified. A difference is that the live executions are
exactly the fair executions, and so the inclusion property becomes:
if an execution $\alpha$ of the implementation $A$
corresponds to an execution $\alpha'$ of the specification $B$,
and $\alpha$ is fair, then $\alpha'$ is also fair 
(Theorems 2.9 and 2.10 in \cite{Jen99}).

Sogaard-Andersen, Lynch, and Lampson \cite{SLL93} presents a similar
method, with the main difference being that the liveness property is
given by a linear temporal logic formula. Now, the proof obligation is
that if an execution $\alpha$ of the implementation $A$ corresponds to
an execution $\alpha'$ of the specification $B$, and $\alpha$
satisfies the liveness formula for $A$, then $\alpha'$ satisfies the
liveness formula for $B$.

Henzinger et.~al.~\cite{HKR97} presents various extensions of
simulation that take fairness into account.  Fairness is expressed
using either Buchi or Streett (i.e., complemented-pairs) acceptance
conditions. However, the fair simulation notions are defined using a
game-theoretic semantics, and require a priori that fair executions of
the concrete automaton have matching fair executions in the abstract
automaton. There is no method of matching the $\RED$ and $\GREEN$
states in the concrete and abstract automata to assure fair trace
containment. Also, the setting is finite state, and the paper
concentrates on algorithms for checking fair simulation.

Alur and Henzinger \cite{AH95} proposes the use of complemented-pairs acceptance
conditions to define liveness properties. However it restricts the
conditions to contain only a finite number of pairs. As our example in
Section~\ref{sec:example} shows, it is very convenient to be able to
specify an infinite number of pairs---in this case, we were able to
use two pairs for each operation $x$ submitted to the data service,
one pair to check for response, and the other to check for
stabilization. It would be quite difficult to specify the liveness
properties of the data service using only a finite number of pairs.
If however, the system being considered is finite-state, then we
remark that much of the work on temporal logic model
checking seems applicable. For example, the algorithm
of Emerson and Lei~\cite{EL85} for model checking under fairness assumptions can
handle the complemented-pairs acceptance condition.
While \cite{AH95} gives rules for compositional and modular
reasoning, it does not provide a method for refining liveness
properties. As stated above, we believe this is a crucial aspect of a
successful methodology for dealing with liveness. It should be clear
that Figure~\ref{fig:lattice-req} provides a very succinct
presentation for the refinement of the liveness property expressed by
$\pair{x \in \wait}{x \not\in \wait}$, namely that every request
eventually receives a response.

Our work is in the linear-time setting, where the external behavior is
a set of traces. In the branching-time setting, the external behavior
can be given as a ``trace-tree'' \cite{HKR97}, i.e., a tree whose
branches are traces. Our liveness-preserving
simulation relations should imply an appropriate containment notion
between ``live-trace-trees,'' i.e., a tree whose branches are live
traces. However we point out technical differences between our setting
and \cite{AH95,HKR97}: we abstract away states and internal actions to
obtain traces, whereas in \cite{AH95,HKR97} an execution is a sequence
of states (actions are not named), and a 
trace is obtained by applying an ``observation function'' to each
state along the execution.

Kesten, Pnueli, and Vardi~\cite{KP00,KPV01} present a method of
\empi{finitary abstraction}: construct a finite-state abstraction (``abstract
system'') of an
infinite-state ``concrete'' system, and model check this abstraction for the
required properties. The method deals with properties expressed in
full linear time temporal logic, (and so handles both safety and
liveness), and is complete, i.e., a suitable finite state abstraction
can always be constructed.
The semantics of the concrete system is given by a \empi{Fair Discrete
System} (FDS), which consists of
(1) a finite set of typed system variables, containing the data and
control state (the \empi{concrete} variables),
(2) a predicate giving the set of initial states, 
(3) a predicate giving the transition relation, 
(4) a \empi{justice condition};  a finite set of predicates $\{J_1,...,J_k\}$,
where each $J_i$ must hold infinitely often along a computation, and
(5) a \empi{compassion condition}, 
a finite set of pairs of predicates $\{<p_1,q_1>,....,<p_n,q_n>\}$;
along a computation, if $p_i$ holds infinitely often, then
$q_i$ must hold infinitely often.
The {justice} and {compassion} conditions ensure that the concrete
system satisfies liveness properties by restricting attention to
``fair'' computations.
For a given concrete system, a finite-state abstract system is
specified syntactically, by giving a set of abstract variables (with
finite domains), and for each abstract variable, giving its value as
an expression over the concrete variables. This implicitly defines a
mapping from concrete to abstract states, and gives rise to two
abstraction operators on concrete predicates: 
(1) a universal (contracting) abstraction, that holds in an abstract
state iff the concrete predicate holds in all corresponding concrete
states, and 
(2) an existential (expanding) abstraction, that holds in an abstract
state iff the concrete predicate holds in some corresponding concrete
state.
The (concrete) temporal properties to be verified are abstracted
by distributing these operators through temporal modalities (nexttime,
until) and disjunction. Distribution through negation converts a 
universal abstraction into an existential one, and vice-versa.
The abstract system is obtained by applying existential abstraction to
the initial state predicate and each justice predicate. The transition 
relation is abstracted by ``lifting'' it to the abstract level using
the definitions of the abstract variables in terms of the concrete
variables.
The compassion pairs \mbox{$<p_i,q_i>$} are abstracted by applying
universal abstraction to $p_i$ and existential abstraction to $q_i$.
\remove{
This is analogous to our requirement in
Definition~\ref{def:live-fwd-sim-inv} that a $\RED$ state in an
abstract execution fragment implies the existence of a $\RED$ state in
the corresponding concrete transition, while a $\GREEN$ state in the
concrete transition implies the existence of a $\GREEN$ state in the
corresponding abstract execution fragment.
}
A main result is that if the abstracted system satisfies the
abstracted property, then the concrete system satisfies the concrete
property.  Another main result is that the method is complete: if the
concrete system satsifies the property, then there exists a
corresponding finite state abstract system and abstracted property
such that the abstract system satsifies the abstract property.  To
obtain completeness, the concrete system must be ``augmented'' by
composing it (synchronously) with a ``ranking monitor,'' which tracks
the difference in successive values of a variant function (``progress
measure'' in the paper) that decreases with progress towards
satisfying the liveness property, and is defined over a well-founded
domain. The reason for incompleteness of the unaugmented method is
liveness properties.
A major difference with our approach is that the number of
complemented pairs is finite, whereas we allow an infinite set.
Furthermore, the abstract system in our approach is not necessarily
finite state. Verification in our approach is by manually devising a
liveness-preserving simulation relation, and the needed complemented
pairs lattices, and then checking the conditions in the corresponding
definitions, possibly with mechanization via theorem proving (see
Section~\ref{sec:mechanization}).  Verification in \cite{KP00,KPV01}
is by manually devising the finitary abstraction mapping and the
ranking monitors, and then model-checking the resulting abstracted
system against the abstracted property. There is no method for
deriving a liveness property at one level from other liveness
properties at the same level, like our complemented-pairs lattices
provide.

In \cite{Ur98}, a method of abstraction based on Galois theory is presented. 
This is based on extensions of the framework of abstract
interpretation \cite{CC77} to temporal properties. 
Again, there are two abstraction notions: under-approximation and
over-approximation.
In \cite{DGG00}, the interaction between abstraction and model
checking under fairness is discussed.  
It is pointed out that abstraction really requires three-valued logic,
since, e.g., a proposition that is true in one concrete state and
false in another has ``unknown'' value in an abstract state
that represents both concrete states.
To handle fairness properly, two abstractions of the transition
relation are introduced, called the free and constrained transition
relations.

\section{Conclusions and Further Work}
\label{sec:conc}

We have presented five liveness-preserving simulation relations that
allow us to refine the liveness properties of infinite-state
distributed systems.
Our method for refining liveness requires reasoning only over
individual states and finite execution fragments, rather than
reasoning over entire executions.  We believe that the use of
simulation-based refinement together with complemented-pairs lattices
for expressing and combining liveness properties provides a
powerful and general framework for refining liveness properties.
In particular, our approach facilitates the decomposition of the
refinement task at each level into simpler subtasks: devise the
liveness-preserving simulation relation, and devise the
complemented-pairs lattices. Since the lattices are a kind of diagram,
they also facilitate the decomposition of proofs and the separation of
concerns, which contributes to scalability of the method.

The general approach and techniques used in this paper do not depend
intimately on the particular automaton model that we used.  Thus, for
example, our approach can be applied to labeled transition systems,
which are used to define operational semantics for process algebras
such as Algebra of Communicating Processes \cite{BW90},
Communicating Sequential Processes \cite{Hoa85},
Calculus of Communicating Systems \cite{Mil89}, and the $\pi$-calculus \cite{Mil99}.
Our approach can also be extended in a
straightforward way to formalisms with unlabeled actions, such as
(finite or infinite) Kripke structures, 
since the fact that actions are named is not used in any essential
way, it just contributes to the ``matching'' condition in simulation
relations, and to the definition of external behavior (trace).

We showed that the Streett acceptance condition (generalized to
arbitrary cardinality) is expressive enough to define any liveness
property, provided that it satisfies a notion of robustness, or
provided that history variables can be used.  \remove{ This suggests a
new direction of research: investigating such infinite-state
generalizations of Buchi, Rabin, Muller, and Streett automata.  }

Simulation relations as a proof method for refinement have been widely
studied. One major impediment to their widespread adoption in practice
is the absence of efficient methodologies for establishing simulation
relations. Doing so usually requires long proofs, with many invariants,
etc. Some of the ideas in this paper may be applicable to
decomposing and simplifying the task of establishing simulation
relations in the first place. For example, it may be possible to apply
our approach to refining the invariants that are used in such
proofs. Another potential application is to models of computation for
dynamic \cite{ALy01}, real-time \cite{LSVK03}, hybrid \cite{LSV03}, and
probabilistic \cite{S95}  systems.
For example, a real-time analogue of a complemented-pair condition
would be: if a $\RED$ state occurs, then a $\GREEN$ state must occur
within $t$ time units. A complemented-pairs lattice that refines a
complemented-pair would then have to satisfy, in addition to the
current requirements of Definition~\ref{def:lattice}, a condition for
the time bounds: every path from the bottom element to the top element
should have a ``total'' time bound matching the pair being refined.
In \cite{ALy01}, we present an automata-theoretic model for dynamic
computation, in which individual processes (automata) that constitute
a system can be created and destroyed, and can dynamically change
their action signature. Since the techniques of this paper assume only
a generic automaton structure, they are applicable to the model of 
\cite{ALy01}. Combining these two pieces of work will result in a
comprehensive method for verifying the liveness properties of
dynamic systems.

\clearpage
\bibliographystyle{plain}
\bibliography{bibfiles/ABBREV,bibfiles/DIST,bibfiles/IOAUT,bibfiles/MODEL,bibfiles/SYNTH,bibfiles/LOGIC,bibfiles/MOBILE,bibfiles/SENG}

\clearpage
\appendix

\section{Simulation Relations}
\label{app:simulation-relations}

We present here five simulation relations, using the definitions of \cite{LV95}.

\bd[Forward Simulation] Let $A$ and $B$ be automata with the same
external actions. A \intrdef{forward simulation} from $A$ to $B$
is a relation $f$ over $\states(A) \times \states(B)$ that satisfies:
\bn
   \item If $s \in \start(A)$, then $f[s] \ints \start(B) \neq \emptyset$.
   \item If $s \lla{a}{A} s'$ and $u \in f[s]$, then there exists a
finite execution fragment $\al$ of $B$ such that $\fstate(\al) = u$, 
$\lstate(\al) \in f[s']$, and $\trace(\al) = \trace(a)$.
\en
\ed

Simulation based proof methods typically use \intr{invariants} to
restrict the steps that have to be considered. An invariant of an
automaton is a predicate that holds in all of its reachable states, or
alternatively, is a superset of the reachable states.

\bd[Forward Simulation w.r.t. Invariants]
Let $A$ and $B$ be automata with the same external actions and with
invariants $I_A$, $I_B$, respectively. A \intrdef{forward simulation}
from $A$ to $B$ with respect to $I_A$ and $I_B$
is a relation $f$ over $\states(A) \times \states(B)$
that satisfies:
\bn

\item \label{clause:fwd-sim-inv:init} If $s \in \start(A)$, then
$f[s] \ints \start(B) \neq \emptyset$.

\item \label{clause:fwd-sim-inv:trans}
If $s \lla{a}{A} s'$, $s \in I_A$, and $u \in f[s] \ints I_B$, then
there exists a finite execution fragment $\al$ of $B$ such that
$\fstate(\al) = u$, 
$\lstate(\al) \in f[s']$, and $\trace(\al) = \trace(a)$.

\en
\label{def:fwd-sim-inv}
\ed
We write $A \simu_F B$ if there exists a forward simulation from $A$
to $B$ w.r.t. some invariants, and 
$A \simu_F B$ via $f$ if $f$ is a forward simulation from
$A$ to $B$ w.r.t. some invariants.

\bd[Refinement Mapping w.r.t. Invariants]
Let $A$ and $B$ be automata with the same external actions and with
invariants $I_A$, $I_B$, respectively. A \intrdef{refinement mapping}
from $A$ to $B$ with respect to $I_A$ and $I_B$
is a function $r$ from $\states(A)$ to $\states(B)$
that satisfies:
\bn

\item \label{clause:ref-inv:init} If $s \in \start(A)$, then
$r(s) \in \start(B)$.

\item \label{clause:ref-inv:trans}
If $s \lla{a}{A} s'$, $s \in I_A$, and $r(s) \in I_B$, then
there exists a finite execution fragment $\al$ of $B$ such that
$\fstate(\al) = r(s)$, 
$\lstate(\al) = r(s')$, and $\trace(\al) = \trace(a)$.

\en
\label{def:ref-inv}
\ed
We write $A \simu_R B$ if there exists a refinement mapping from $A$
to $B$ w.r.t. some invariants, and 
$A \simu_R B$ via $r$ if $r$ is a refinement mapping from
$A$ to $B$ w.r.t. some invariants.

\bd[Backward Simulation w.r.t. Invariants]
Let $A$ and $B$ be automata with the same external actions and with
invariants $I_A$, $I_B$, respectively. A \intrdef{backward simulation}
from $A$ to $B$ with respect to $I_A$ and $I_B$
is a relation $b$ over $\states(A) \times \states(B)$
that satisfies:
\bn

\item If $s \in I_A$, then $b[s] \ints I_B \ne \emptyset$.

\item \label{clause:back-sim-inv:init} 
If $s \in \start(A)$, then $b[s] \ints I_B \sub \start(B)$.

\item \label{clause:back-sim-inv:trans}
If $s \lla{a}{A} s'$, $s \in I_A$, and $u' \in b[s'] \ints I_B$, then
there exists a finite execution fragment $\al$ of $B$ such that
$\fstate(\al) \in b[s] \ints I_B$, 
$\lstate(\al) = u'$, and
$\trace(\al) = \trace(a)$.

\en
\label{def:back-sim-inv}
\ed
A backward simulation $b$ w.r.t. invariants is \intr{image-finite} iff
for each $s \in \states(A)$, $b[s]$ is a finite set.
We write $A \simu_B B$ if there exists a backward simulation from $A$
to $B$ w.r.t. some invariants, and 
$A \simu_B B$ via $b$ if $b$ is a backward simulation from
$A$ to $B$ w.r.t. some invariants.
If the backward simulation is image-finite, then we write 
$A \simu_{iB} B$, $A \simu_{iB} B$ via $b$, respectively.

\bd[History Relation w.r.t. Invariants]
Let $A$ and $B$ be automata with the same external actions and with
invariants $I_A$, $I_B$, respectively. A \intrdef{history relation}
from $A$ to $B$ with respect to $I_A$ and $I_B$
is a relation $h$ over $\states(A) \times \states(B)$
that satisfies:
\bn

\item \label{clause:hist-inv:init}
$h$ is a forward simulation from $A$ to $B$ w.r.t. $I_A$ and $I_B$.

\item \label{clause:hist-inv:trans}
$h^{-1}$ is a refinement from $B$ to $A$ w.r.t. $I_B$ and $I_A$.

\en
\label{def:hist-inv}
\ed
We write $A \simu_H B$ if there exists a history relation from $A$
to $B$ w.r.t. some invariants, and 
$A \simu_H B$ via $h$ if $h$ is a history relation from
$A$ to $B$ w.r.t. some invariants.

\bd[Prophecy Relation w.r.t. Invariants]
Let $A$ and $B$ be automata with the same external actions and with
invariants $I_A$, $I_B$, respectively. A \intrdef{prophecy relation}
from $A$ to $B$ with respect to $I_A$ and $I_B$
is a relation $p$ over $\states(A) \times \states(B)$
that satisfies:
\bn

\item \label{clause:proph-inv:init}
$p$ is a backward simulation from $A$ to $B$ w.r.t. $I_A$ and $I_B$.

\item \label{clause:proph-inv:trans}
$p^{-1}$ is a refinement from $B$ to $A$ w.r.t. $I_B$ and $I_A$.

\en
\label{def:proph-inv}
\ed
A prophecy relation $p$ w.r.t. invariants is \intr{image-finite} iff
for each $s \in \states(A)$, $p[s]$ is a finite set.
We write $A \simu_P B$ if there exists a prophecy relation from $A$
to $B$ w.r.t. some invariants, and 
$A \simu_P B$ via $p$ if $p$ is a prophecy relation from
$A$ to $B$ w.r.t. some invariants.
If the prophecy relation is image-finite, then we write 
$A \simu_{iP} B$, $A \simu_{iP} B$ via $p$, respectively.

\newpage
\section{Linear-time Temporal Logic}
\label{app:TL}

We define the syntax and semantics of the temporal logic that we use as
follows. This is essentially linear-time temporal logic
without the until and nexttime operators.

\bd[Syntax of Linear-time Temporal Logic]
\label{def:TL-syntax}
The syntax of a linear-time temporal logic formula is given inductively 
as follows, where $f,g$ are sub-formulae, and $U$ is a set of states
$($which defines a state-assertion$)$\textup{:}
\be

\item Each of $U,\ f \land g$ and $\neg f$ is a formula

\item $\always  f$ is a formula which intuitively means that $f$
      holds in every state of the execution being considered

\item $\eventually  f$ is a formula which intuitively means that $f$
      holds in some state of the execution being considered

\ee
\ed

Formally, we define the semantics of linear-time temporal logic
formulae with respect to an infinite execution, that is, an infinite
sequence of states.

\bd[Semantics of Linear-time Temporal Logic]
We use the usual notation to indicate truth:
$\al \sat f$ means that $f$ is true of execution $\al$.
We define $\sat$ inductively, where $\al = s_0 s_1 s_2 \ldots$
is an infinite sequence of states, and
$\al^i = s_i s_{i+1} \ldots$ is the suffix of $\al$ starting in $s_i$.

\begin{tabbing}
aaa\=              aaaaaaaaaaaa\= aaaaaaa\=        \kill
   \>$\al \sat U$              \>iff     \>$s_0 \in U$\\
   \>$\al \sat \neg f$         \>iff     \>it is not the case that $\al \sat f$\\
   \>$\al \sat f \land g$      \>iff     \>$\al \sat f$ and $\al \sat g$\\
   \>$\al \sat \always f$      \>iff     \>for all $i \ge 0$, $\al^i \sat f$\\
   \>$\al \sat \eventually f$  \>iff     \>for some $i \ge 0$, $\al^i \sat f$
\end{tabbing}
\ed
In particular, $\al \sat \iof f$ meains that $\al^i \sat f$ for an
infinite number of values of $i$.

\newpage
\section{I/O Automaton Code for the ESDS Example, from \protect \cite{FGLLS99}}
\label{app:esds}

\begin{figure}[hb]

\horline
\ioautomatontitle{$\mathit{Users}$}

\begin{signature}
\io{%
$\response(x,v)$, 
	where $x \in \opdescs$ and $v \in \datavals$
}{%
$\request(x)$, 
	where $x \in \opdescs$
}
\end{signature}

\begin{statevarlist}

\item $\requested$, a subset of $\opdescs$, initially empty

\end{statevarlist}

\begin{actionlist}

\iocode{%

\prcef{\textbf{Output} $\request(x)$}
{%
  $\idof{x} \notin \idof{\requested}$ \\
  $\prevof{x} \subseteq \idof{\requested}$
}{%
  $\setinsert{\requested}{x}$
}

}{%

\ef{\textbf{Input} $\response(x, v)$}
{%
  None
}

}

\end{actionlist}

\horline
\caption{The Users Automaton}
\label{fig:users}
\efg

\begin{figure}[hb]

\horline
\ioautomatontitle{$\ESDSI$}

\begin{signature}

\ioi{%
$\request(x)$, 
	where $x \in \opdescs$
}{%
$\response(x,v)$, 
	where $x \in \opdescs$ and $v \in \datavals$
}{%
$\enter(x, \newpo)$, 
	where $x \in \opdescs$ 
	and $\newpo$ is a strict partial order on $\idset$ \\
$\stabilize(x)$, 
	where $x \in \opdescs$ \\
$\calc(x, v)$, 
	where $x \in \opdescs$ and $v \in \datavals$ \\ 
$\addcons(\newpo)$, 
	where $\newpo$ is a partial order on $\idset$
}
\end{signature}

\begin{statevarlist}

\item $\wait$, a subset of $\opdescs$, initially empty; 
	the operations requested but not yet responded to
\item $\rept$, a subset of $\opdescs \times \datavals$, initially empty; 
	operations and responses that may be returned to clients
\item $\ops$, a subset of $\opdescs$, initially empty; 
	the set of all operations that have ever been entered
\item $\po$, a partial order on $\idset$, initially empty; 
	constraints on the order operations in $\ops$ are applied
\item $\stabilized$, a subset of $\opdescs$, initially empty;
	the set of stable operations

\end{statevarlist}

\begin{actionlist}

\iocode{%
\ef{\textbf{Input} $\request(x)$}
{%
  $\setinsert{\wait}{x}$
}

\prcef{\textbf{Internal} $\enter(x, \newpo)$}
{%
  $x \in \wait$ \\
  $x \notin \ops$ \\
  $\prevof{x} \subseteq \idof{\ops}$ \\
  $\spn{\newpo} \subseteq \idof{\ops} \union \set{\idof{x}}$ \\
  $\po \subseteq \newpo$ \\
  $\CSC(\set{x}) \subseteq \newpo$ \\
  $\set{(\idof{y}, \idof{x}): y \in \stabilized} \subseteq \newpo$
}{%
  $\setinsert{\ops}{x}$ \\
  $\po \gets \newpo$
}

\prcef{\textbf{Internal} $\addcons(\newpo)$}
{%
  $\spn{\newpo} \subseteq \idof{\ops}$ \\
  $\po \subseteq \newpo$
}{%
  $\po \gets \newpo$
}

}{%

\prcef{\textbf{Internal} $\stabilize(x)$}
{%
  $x \in \ops$ \\
  $x \notin \stabilized$ \\
  $\forall y \in \ops$, 
	$y \preceq_{\po} x \logicor x \preceq_{\po} y$ \\
  $\ops |_{\prec_{\po } x} \subseteq \stabilized$
}{%
  $\setinsert{\stabilized}{x}$
}

\prcef{\textbf{Internal} $\calc(x, v)$}
{%
  $x \in \ops$ \\
  $\strictof{x} \imp x \in \stabilized$ \\
  $v \in \valset(x, \ops, \prec_{\po})$
}{%
   if $x \in \wait$ then $\setinsert{\rept}{(x, v)}$ 
}

\prcef{\textbf{Output} $\response(x, v)$}
{%
  $(x, v) \in \rept$ \\
  $x \in wait$
}{%
  $\setdelete{\wait}{x}$ \\
  $\rept \gets \rept - \set{ (x, v'): (x, v') \in rept}$ 
}

}

\end{actionlist}
\horline
\caption{The Specification ESDS-I}
\label{fig:ESDSI}
\efg

\bfg

\horline

\iocode{

\prcef{\textbf{Internal} $\enter(x, \newpo)$}
{%
  $x \in \wait$ \\
  $\prevof{x} \subseteq \idof{\ops}$ \\
  $\spn{\newpo} \subseteq \idof{\ops} \union \set{\idof{x}}$ \\
  $\po \subseteq \newpo$ \\
  $\CSC(\set{x}) \subseteq \newpo$ \\
  $\set{(\idof{y}, \idof{x}): y \in \stabilized} \subseteq \newpo$
}{%
  $\setinsert{\ops}{x}$ \\
  $\po \gets \newpo$
}

}{

\prcef{\textbf{Internal} $\stabilize(x)$}
{%
  $x \in \ops$ \\
  $\forall y \in \ops$, 
	$y \preceq_{\po} x \logicor x \preceq_{\po} y$ \\
  $\prec_{\po}$ totally orders $\ops |_{\prec_{\po } x}$
}{%
  $\setinsert{\stabilized}{x}$
}

}

\horline

\caption{The Specification ESDS-II. Only differences with ESDS-I are shown.}
\label{fig:ESDSII}
\efg

\bfg

\horline

\ioautomatontitle{$\Frontend(c)$}

\begin{signature}

\io{%
$\request(x)$, 
	where $x \in \opdescs$ and $c = \client({x})$ \\
$\receive_{rc}(m)$, 
	where $r$ is a replica and $m \in \respmsgset$ 
}{%
$\response(x,v)$, 
	where $x \in \opdescs$, $c = \client({x})$, and $v \in \datavals$ \\ 
$\send_{cr}(m)$, 
	where $r$ is a replica and $m \in \reqmsgset$
}
\end{signature}

\begin{statevarlist}

\item $\wait_c$, a subset of $\opdescs$, initially empty 
\item $\rept_c$, a subset of $\opdescs \times \datavals$, initially empty 

\end{statevarlist}

\begin{actionlist}

\iocode{%
\ef{\textbf{Input} $\request(x)$}
{%
  $\setinsert{\wait_c}{x}$
}

\prcef{\textbf{Output} $\send_{cr}(\reqmsg{x})$}
{%
  $x \in \wait_c$
}{%
  None
}

}{%

\ef{\textbf{Input} $\receive_{rc}(\respmsg{x,v})$}
{%
  if $x \in \wait_c$ then $\setinsert{\rept_c}{(x,v)}$
}

\prcef{\textbf{Output} $\response(x, v)$}
{%
  $(x, v) \in \rept_c$ \\
  $x \in \wait_c$
}{%
  $\setdelete{\wait_c}{x}$ \\
  $\setdelete{\rept_c}{(x, v'):(x, v') \in \rept_c}$
}
}

\end{actionlist}

\horline
\caption{The Automaton for the front end of client $c$}
\label{fig:front-end}
\efg

\bfg

\horline
\ioautomatontitle{$\Replica(r)$}

\begin{signature}

\ioi{%
$\receive_{cr}(m)$, 
	where $c$ is a client and $m \in \reqmsgset$ \\
$\receive_{r'r}(m)$, 
	where $r' \neq r$ is a replica and $m \in \gossipmsgset$ 
}{%
$\send_{rc}(m)$, 
	where $c$ is a client and $m \in \respmsgset$ \\
$\send_{rr'}(m)$, 
	where $r' \neq r$ is a replica and $m \in \gossipmsgset$ 
}{%
$\doit_r(x, l)$, 
	where $x \in \opdescs$ and $l \in \labels_r$
}
\end{signature}

\begin{statevarlist}

\item $\pending_r$, a subset of $\opdescs$, initially empty; 
	the messages that require a response
\item $\rcvd_r$, a subset of $\opdescs$, initially empty; 
	the operations that have been received
\item $\done_r[i]$ for each replica $i$, 
	a subset of $\opdescs$, initially empty; 
	the operations $r$ knows are done at $i$
\item $\stable_r[i]$ for each replica $i$, 
	a subset of $\opdescs$, initially empty; 
	the operations $r$ knows are stable at $i$ 
\item $\lbl_r: \idset \to \labels \union \set{\infty}$, 
	initially all $\infty$; 
	the minimum label $r$ has seen for $\id \in \idset$ 
\item Derived variable: 
	$\LC_r = \set{(\id, \id'): \lbl_r(\id) < \lbl_r(\id')}$, 
	a strict partial order on $\idset$; 
	the local constraints at $r$

\end{statevarlist}

\begin{actionlist}

\iocode{%

\ef{\textbf{Input} $\receive_{cr}(\reqmsg{x})$}
{%
  $\setinsert{\pending_r}{x}$ \\
  $\setinsert{\rcvd_r}{x}$
}

\prcef{\textbf{Internal} $\doit_r(x, l)$}
{%
  $x \in \rcvd_r - \done_r[r]$ \\
  $\prevof{x} \subseteq \idof{\done_r[r]}$ \\
  $l > \lbl_r(\idof{y})$ for all $y \in \done_r[r]$
}{%
  $\setinsert{\done_r[r]}{x}$ \\
  $\lbl_r(\idof{x}) \gets l$ 
}

\prcef{\textbf{Output} $\send_{rc}(\respmsg{x, v})$}
{%
  $x \in \pending_r \intersect \done_r[r]$ \\
  $\strictof{x} \imp x \in \Intersect_i \stable_r[i]$ \\
  $v \in \valset(x, \done_r[r], \prec_{\LC_r})$ \\ 
  $c = \client({x})$ 
}{%
  $\setdelete{\pending_r}{x}$
}

}{%

\prc{\textbf{Output} $\send_{rr'}(\gossipmsg{R, D, L, S})$}
{%
  $R = \rcvd_r$; $D = \done_r[r]$; \\ 
  $L = \lbl_r$; $S = \stable_r[r]$
}

\ef{\textbf{Input} $\receive_{r'r}(\gossipmsg{R, D, L, S})$}
{%
  $\rcvd_r \gets \rcvd_r \union R$ \\
  $\done_r[r'] \gets \done_r[r'] \union D \union S$ \\
  $\done_r[r] \gets \done_r[r] \union D \union S$ \\
  $\done_r[i] \gets \done_r[i] \union S$ for all $i \neq r, r'$ \\
  $\lbl_r \gets \min(\lbl_r, L)$ \\
  $\stable_r[r'] \gets \stable_r[r'] \union S$ \\
  $\stable_r[r] \gets \stable_r[r] \union S \union (\Intersect_i \done_r[i])$
}
}

\end{actionlist}

\horline
\caption{Automaton for replica $r$}
\label{fig:replica}
\efg

\bfg

\horline

\ioautomatontitle{$\Channel(i,j,\msgset)$}

\begin{signature}

\io{%
$\send_{ij}(m)$, 
	where $m \in \msgset$
}{%
$\receive_{ij}(m)$, 
	where $m \in \msgset$ 
}
\end{signature}

\begin{statevarlist}

\item $\channel_{ij}$, a multiset of messages, (taken from $\msgset$),
	initially empty
\end{statevarlist}

\begin{actionlist}

\iocode{%
\ef{\textbf{Input} $\send_{ij}(m)$}
{%
  $\setinsert{\channel_{ij}}{m}$
}

}{%

\prcef{\textbf{Output} $\receive_{ij}(m)$}
{%
  $m \in \channel_{ij}$
}{%
  $\setdelete{\channel_{ij}}{m}$
}
}

\end{actionlist}

\horline
\caption{The Channel Automaton}
\label{fig:channel}
\efg

\end{document}